\documentclass[10pt]{article}
\usepackage[margin=1in]{geometry}
\usepackage{amsfonts,amsmath,amssymb}
\usepackage[none]{hyphenat}
\usepackage{fancyhdr}
\usepackage{graphicx}
\usepackage{float}
\usepackage[nottoc,notlot,notlof]{tocbibind}

\usepackage{amssymb,amsmath,tikz}
\usepackage{pgfplots} 
\usepackage{amsthm}
\usepackage{mathrsfs}
\usepackage{pifont}
\usepackage{slashed}
\usepackage{mathtools}
\usepackage{cancel}

\showoutput
\DeclareMathOperator*{\p}{p}

\DeclareMathOperator*{\y}{y}
\DeclareMathOperator*{\q}{k}
\DeclareMathOperator*{\n}{n}

\newcommand*{\ud}{\mathrm{\,d}} 

\renewcommand{\qedsymbol}{$\blacksquare$}

\theoremstyle{plain}

\newtheorem*{twr*}{Theorem}
\newtheorem*{lem*}{Lemma}
\newtheorem{twr}{Theorem}
\newtheorem{lem}{Lemma}

\newtheorem*{defin*}{Definition}

\newtheorem*{rem*}{Remark}

\newtheorem{cor*}{Corollary}

\newtheorem*{notn*}{Notation}
\newtheorem*{wiener-ito*}{Wiener-It\^o-Segal Decomposition}
\newtheorem*{prop*}{Proposition}

\DeclareMathAlphabet{\mathpzc}{OT1}{pzc}{m}{it}

\begin{document}

\begin{titlepage}
\begin{center}
\vspace*{1cm}
\large{\textbf{BOGOLIUBOV'S CAUSAL PERTURBATIVE QED AND WHITE NOISE.}}\\
\large{\textbf{INTERACTING FIELDS}}\\
\vspace*{2cm}
\small{JAROS{\L}AW WAWRZYCKI}\\[1mm]
\tiny{Bogoliubov Labolatory of Theoretical Physics}
\\
\tiny{Joint Institute of Nuclear Research, 141980 Dubna, Russia}
\\
\vspace*{1cm}
\tiny{\today}\\
\vfill
\begin{abstract}
We present the Bogoliubov's causal perturbative QFT, which includes 
only one refinement: the creation-annihilation operators at a point, \emph{i.e.} 
for a specific momentum, are mathematically 
interpreted as the Hida operators from the white noise analysis. 
We leave the rest of the theory completely unchanged. This allows avoiding 
infrared-- and ultraviolet -- divergences in the transition 
to the adiabatic limit for interacting fields. We present here existence proof
of the adiabatic limit for interacting fields in causal QED with Hida operators.  
This limit exists if and only if the normalization in the Epstein-Glaser splitting
of the causal distributions, in the construction of the scattering operator, 
is ``natural'', and thus eliminates arbitrariness in the choice of the splitting 
making the theory definite, with its predictive power considerably strengthened. 
For example, we present a charge-mass relation which can be proved within this theory, 
and which is confirmed experimentally. 
\end{abstract}
\vspace*{0.5cm}
\tiny{{\bf Keywords}: scattering operator, causal perturbative method in QFT, interacting fields, \\
white noise, Hida operators, integral kernel operators, Fock expansion}
\end{center}
\vfill
\end{titlepage}

\tableofcontents
\thispagestyle{empty}
\clearpage

\setcounter{page}{1}

\section{The main idea and the main goal}

In the monograph \cite{Bogoliubov_Shirkov}, a way to the exact formulation of the renormalization method in perturbative QFT was initiated. It was there that the causal axioms (I) - (IV) (see below) were formulated clearly enough for the perturbative series for the scattering operator $S$, which allowed the perturbative QFT to be slowly transferred to the axiomatic path, and it was there that the idea of a strict mathematical construction of higher-order contributions $S_n$ to $S$ first appeared, which uses only the axioms (I) - (IV) (plus, possibly, Ward's identities), cf. \cite{Bogoliubov_Shirkov}, \S 29.2, and the works by Stepanov cited there. 

There was, however, some freedom as to the strict mathematical interpretation of the free-field operators and the higher-order contributions $S_n$ to $S$, with only the general proviso that they should be some kind of generalized distribution-type operators. The class of operators permitted here for $S_n$ may be represented by Wick polynomials in free fields with scalar coefficients, which may be general translationally invariant tempered distributions. 
As noted in \cite{Bogoliubov_Shirkov} the higher order
contributions $S_n$ are not uniquely determined by the causality axioms (I)-(IV), 
but only up to a quasi local generalized operators $\Lambda_n$ supported at the full diagonal. However,
concerning the rigorous treatment of $S_n$ (determined up to the terms supported at the diagonal) 
it was indicated in \S 29.2 only the proof of existence of $S_n$, without explicit mathematically rigorous construction of $S_n$. 
Instead, it was provided in \cite{Bogoliubov_Shirkov} only the practical computation
of the scalar coefficients in each $S_n$, which is based on the renormalization technique and which illustrates
the relation of the renormalization method with the construction of most general $S_n$ which are in accordance with the causality axioms (I)-(IV). 
A mathematically rigorous and \emph{explicit construction} of $S_n$, based solely on the Bogoliubov's axioms (I) - (IV), and which gives the same scalar coefficients 
(and Green functions) as the renormalization method, was given later by Epstein and Glaser \cite{Epstein-Glaser}. 

It should be emphasized, however, that the content of the axioms (I) - (IV) for the $S$ operator will depend on how we mathematically interpret the free field operators, their Wick products, and the higher-order contributions $S_n$ themselves in (I)-(IV). Thus, before giving a mathematically rigorous construction
of $S_n$, based on (I)-(IV), one has to fix the mathematical meaning of these generalized operators. Epstein and Glaser \cite{Epstein-Glaser} assumed that these operators are operator valued distributions precisely in the Wightman sense \cite{wig}, so in particular they assumed that the free field operators and 
$S_n$ are Wightman's operator valued distributions. Using this interpretation for the generalized operators, it was shown in \cite{Epstein-Glaser}
that the standard results of the renormalization technique for the computation of $S_n$ (or rather for the scalar coefficients in $S_n$, 
in particular for the Green functions) can be rigorously reconstructed from the axioms  (I) - (IV), if we add the assumption (V) of preservation of the
singularity degree at zero of the causal coefficient distributions during the computation of their splitting into the retarded and advanced parts,
thus arriving at a mathematically consistent formulation of perturbative QFT, without any UV divergences. The work \cite{Epstein-Glaser} 
is in fact a further development of the ideas outlined in \cite{Bogoliubov_Shirkov}.

In the rigorous construction of $S_n$ due to Bogoliubov, Epstein and Glaser, instead of the formal (divergent) multiplication by the step theta function
in the chronological product and then renormalization removing singular parts in the ill-defined products of distributions by the step function,  
one uses the inductive step of Epstein-Glaser in the perturbative approach, which is carried out solely within the axioms (I)-(V)  
in a mathematically consistent manner, and which reduces the whole problem of construction of
the higher order contributions $S_n$ to the problem of
splitting of causal numerical tempered distributions into their retarded and advanced parts. 
The splitting of a causal distribution is not unique and depends on the so-called singularity degree at zero (in space-time coordinates) 
of the splitted distribution -- freedom corresponding to the non-uniqueness in the ordinary renormalization procedure, and associated with the 
renormalization group freedom in the approach based on renormalization.
This method is known for fifty 
years and now we know that it gives the same results for the Green functions and generally for the scalar coefficients in 
$S_n$ as the standard renormalization technique \cite{Scharf}.  

However, some IR divergences remained for those QFT with infinite range of interaction (like QED), 
whenever one wanted to pass to the limit of the intensity of interaction function
$g$ to the constant function $1$, in order to get results with the physical interaction also in the remote part of space-time. 
This is the celebrated \emph{adiabatic limit problem}.
In some problems of QFT with infinite range of interaction (like QED), passing to this limit is unavoidable.      

We propose to improve this Bogoliubov-Epstein-Glaser perturbative QFT by reinterpreting mathematically the generalized operators (the free field
operators, their Wick products and the higher order contributions to the scattering operator) and regard them not as the operator valued distributions
in the Wightman sense, but as the integral kernel operators in the sense of the white noise calculus \cite{obataJFA}. This allows us to solve the
adiabatic limit problem, which was impossible when the generalized 
operators were interpreted as operator valued distributions in the Wightman sense \cite{wig}.

Namely, we are using the Hida white noise operators 
(here $\boldsymbol{\p}$ subsumes the spatial momenta components and the corresponding discrete spin
components in order to simplify notation) 
\[
\partial_{\boldsymbol{\p}}^{*}, \,\,\, \partial_{\boldsymbol{\p}}
\]
which respect the canonical commutation or anticommutation relations
\[
\big[\partial_{\boldsymbol{\p}}, \partial_{\boldsymbol{\q}}^{*}\big]_{{}_{\mp}} = \delta(\boldsymbol{\p}-\boldsymbol{\q}),
\]
as the creation-annihilation operators 
\[
a(\boldsymbol{\p})^{+}, \,\,\, a(\boldsymbol{\p}),
\]
of the free fields in the causal perturbative Bogoliubov-Epstein-Glaser QFT, \cite{Bogoliubov_Shirkov}, \cite{Epstein-Glaser}, 
leaving all the rest of the axioms (I)-(V) of the theory completely unchanged.
\emph{I.e.}
we are using the standard Gelfand triple
\[
\left. \begin{array}{ccccc} 
E & \subset & \mathcal{H} & \subset & E^* 
\end{array}\right.,
\]
over the single particle Hilbert space $\mathcal{H}$ of the total system of free fields 
determined by the corresponding standard self-adjoint operator $A$ in $\mathcal{H}$ (with some positive power $A^r$ being nuclear), 
and its lifting to the standard Gelfand triple 
\[
\left. \begin{array}{ccccc} 
(E) & \subset & \Gamma(\mathcal{H}) & \subset & (E)^* 
\end{array}\right.,
\]
over the total Fock space $\Gamma(\mathcal{H})$ of the total system of free fields underlying the actual QFT, e.g. QED, 
which is naturally determined by the self-adjoint and standard operator $\Gamma(A)$, and
with the nuclear Hida test space $(E)$ and its strong dual $(E)^*$.

This converts the free fields $\mathbb{A}$ and the $n$-th order contributions 
\[
S_n(g^{\otimes \, n}) \,\, 
\,\,\,\,\,\,\,\,\,\,\,\,\,\,\,\,\,\,\,\,\,
\textrm{and}
\,\,\,\,\,\,\,\,\,\,\,\,\,\,\,\,\,\,\,\,\,
\mathbb{A}_{{}_{\textrm{int}}}^{(n)}(g^{\otimes \, n},\phi) 
\]
written frequently as
\[
S_n(g) \,\,
\,\,\,\,\,\,\,\,\,\,\,\,\,\,\,\,\,\,\,\,\,
\textrm{and}
\,\,\,\,\,\,\,\,\,\,\,\,\,\,\,\,\,\,\,\,\,
\mathbb{A}_{{}_{\textrm{int}}}^{(n)}(g,\phi) ,
\]
to the scattering operator
\begin{gather*}
S(g) = \boldsymbol{1} + \sum\limits_{n=1}^{\infty} {\textstyle\frac{1}{n!}} S_n(g^{\otimes \, n}),
\,\,\,\,
S(g)^{-1} = \boldsymbol{1} + \sum\limits_{n=1}^{\infty} {\textstyle\frac{1}{n!}} \overline{S_n}(g^{\otimes \, n}),
\\
\textrm{or}
\,\,\,\,
S(g,h) = \boldsymbol{1} 
+ \sum\limits_{n=1}^{\infty} \sum\limits_{p=0}^{n} {\textstyle\frac{1}{n!}} S_{n-p,p}(g^{\otimes \, (n-p)} \otimes h^{\otimes \, p})
\end{gather*}
(denoted also by $S(g\mathcal{L})$, $S(g\mathcal{L})^{-1}$ or, respectively, $S(g\mathcal{L}+h\mathbb{A})$) and to the interacting fields 
\[
\mathbb{A}_{{}_{\textrm{int}}}(g,\phi) = \int{\textstyle\frac{i\delta}{\delta h(x)}}S(g\mathcal{L}+h\mathbb{A})^{-1}S(g\mathcal{L})\Big|_{{}_{h=0}} \phi(x) dx,
\]
into the finite sums of generalized integral kernel operators
\[
\mathbb{A}(\phi) = \int \kappa_{0,1}(\boldsymbol{\p};\phi) \, \partial_{\boldsymbol{\p}} \, d\boldsymbol{\p} 
+ \int \kappa_{1,0}(\boldsymbol{\p};\phi) \, \partial_{\boldsymbol{\p}}^{*} \, d\boldsymbol{\p}, 
\]
\begin{multline*}
S_n(g^{\otimes \, n}) = \sum\limits_{\mathpzc{l},\mathpzc{m}}
\int \kappa_{\mathpzc{l}\mathpzc{m}}\big(\boldsymbol{\p}_1, \ldots, \boldsymbol{\p}_\mathpzc{l},
\boldsymbol{\q}_1, \ldots, \boldsymbol{\q}_\mathpzc{m};  g^{\otimes \, n} \big) \,\,
\partial_{\boldsymbol{\p}_1}^{*} \ldots \partial_{\boldsymbol{\p}_\mathpzc{l}}^{*} 
\partial_{\boldsymbol{\q}_1} \ldots \partial_{\boldsymbol{\q}_\mathpzc{m}} 
d\boldsymbol{\p}_1 \ldots d\boldsymbol{\p}_\mathpzc{l}
d\boldsymbol{\q}_1 \ldots d\boldsymbol{\q}_\mathpzc{m}
\\
=
\int \ud^4 x_1 \ldots \ud^4x_n \, S_n(x_1, \ldots, x_n) \, g(x_1) \ldots g(x_n),
\end{multline*}
\begin{multline*}
\overline{S_n}(g^{\otimes \, n}) = \sum\limits_{\mathpzc{l},\mathpzc{m}}
\int \kappa_{\mathpzc{l}\mathpzc{m}}\big(\boldsymbol{\p}_1, \ldots, \boldsymbol{\p}_\mathpzc{l},
\boldsymbol{\q}_1, \ldots, \boldsymbol{\q}_\mathpzc{m};  g^{\otimes \, n} \big) \,\,
\partial_{\boldsymbol{\p}_1}^{*} \ldots \partial_{\boldsymbol{\p}_\mathpzc{l}}^{*} 
\partial_{\boldsymbol{\q}_1} \ldots \partial_{\boldsymbol{\q}_\mathpzc{m}} 
d\boldsymbol{\p}_1 \ldots d\boldsymbol{\p}_\mathpzc{l}
d\boldsymbol{\q}_1 \ldots d\boldsymbol{\q}_\mathpzc{m}
\\
=
\int \ud^4 x_1 \ldots \ud^4x_n \, \overline{S_n}(x_1, \ldots, x_n) \, g(x_1) \ldots g(x_n),
\end{multline*}
\begin{multline*}
S_{n-p,p}(g^{\otimes \, (n-p)} \otimes h^{\otimes \, p}) = \sum\limits_{\mathpzc{l},\mathpzc{m}}
\int \kappa_{\mathpzc{l}\mathpzc{m}}\big(\boldsymbol{\p}_1, \ldots, \boldsymbol{\p}_\mathpzc{l},
\boldsymbol{\q}_1, \ldots, \boldsymbol{\q}_\mathpzc{m}; g^{\otimes \, (n-p)} \otimes h^{\otimes \, p} \big) \,\,
\partial_{\boldsymbol{\p}_1}^{*} \ldots \partial_{\boldsymbol{\p}_\mathpzc{l}}^{*} 
\partial_{\boldsymbol{\q}_1} \ldots \partial_{\boldsymbol{\q}_\mathpzc{m}} 
d\boldsymbol{\p}_1 \ldots d\boldsymbol{\p}_\mathpzc{l}
d\boldsymbol{\q}_1 \ldots d\boldsymbol{\q}_\mathpzc{m}
\\
=
\int \ud^4 x_1 \ldots \ud^4y_p \, S_n(x_1, \ldots, x_n, y_1, \ldots y_p) \, g(x_1) \ldots g(x_{n-p}) h(y_1) \ldots h(y_{p}),
\end{multline*}
and
\begin{multline*}
\mathbb{A}_{{}_{\textrm{int}}}^{(n)}(g^{\otimes \, n},\phi) = \sum\limits_{\mathpzc{l},\mathpzc{m}}
\int \kappa_{\mathpzc{l}\mathpzc{m}}\big(\boldsymbol{\p}_1, \ldots, \boldsymbol{\p}_\mathpzc{l},
\boldsymbol{\q}_1, \ldots, \boldsymbol{\q}_\mathpzc{m};  g^{\otimes \, n}, \phi \big) \,\,
\partial_{\boldsymbol{\p}_1}^{*} \ldots \partial_{\boldsymbol{\p}_\mathpzc{l}}^{*} 
\partial_{\boldsymbol{\q}_1} \ldots \partial_{\boldsymbol{\q}_\mathpzc{m}} 
d\boldsymbol{\p}_1 \ldots d\boldsymbol{\p}_\mathpzc{l}
d\boldsymbol{\q}_1 \ldots d\boldsymbol{\q}_\mathpzc{m}
\\
=
\int \ud^4 x_1 \ldots \ud^4x_n \ud^4 x \, \mathbb{A}_{{}_{\textrm{int}}}^{(n)}(x_1, \ldots, x_n; x) \, g(x_1) \ldots g(x_n) \, \phi(x),
\end{multline*}
with vector-valued distributional kernels $\kappa_{\mathpzc{l}\mathpzc{m}}$ in the sense of \cite{obataJFA}, with the values in the distributions
over the test nuclear space
\begin{gather*}
\mathscr{E} \ni \phi,
\,\,\,\,\,\,\,\,\,
\textrm{or}
\,\,\,\,\,\,\,\,\,
\mathscr{E}^{\otimes \, (n-p)}\otimes (\oplus_{1}^{d}\mathscr{E})^{\otimes \, p}  \ni g^{\otimes \, (n-p)} \otimes h^{\otimes \, p},
\\
\textrm{or}
\,\,\,\,\,
\mathscr{E}^{\otimes \, n} \ni g^{\otimes \, n} 
\,\,\,\,\,\,\,\,\,
\textrm{or, respectively,}
\,\,\,\,\,\,\,\,\,
\mathscr{E}^{\otimes \, n} \otimes (\oplus_{1}^{d}\mathscr{E}) \ni g^{\otimes \, n} \otimes \phi
\end{gather*}
with 
\[
\mathscr{E} = \mathcal{S}(\mathbb{R}^4;\mathbb{C}). 
\]
Each of the $3$-dim Euclidean integration $d\boldsymbol{\p}_i$ with respect to the spatial momenta $\boldsymbol{\p}_i$ components
$\boldsymbol{\p}_{i1}, \boldsymbol{\p}_{i2}, \boldsymbol{\p}_{i3}$,
also includes here summation over the corresponding discrete spin components $s_i\in(1,2,\ldots)$ hidden under the symbol $\boldsymbol{\p}_i$. 

The class to which the operators $S_n$ and $\mathbb{A}_{{}_{\textrm{int}}}^{(n)}$ belong, expressed in terms of the Hida test space,
depend on the fact if there are massless free fields present in the interaction Lagrange density operator $\mathcal{L}$ or not.
Namely: 
\[
S_n \in
\begin{cases}
\mathscr{L}\big(\mathscr{E}^{\otimes \, n}, \, \mathscr{L}((E),(E))\big), 
& \text{if all fields in $\mathcal{L}$ are massive},\\
\mathscr{L}\big(\mathscr{E}^{\otimes \, n}, \, \mathscr{L}((E),(E)^*)\big), & \text{if there are massless fields in $\mathcal{L}$}.\\
\end{cases}
\]
and the same holds for $S_{n,p}$ with $\mathscr{E}^{\otimes \, n}$ repaced by 
$\mathscr{E}^{\otimes \, n} \otimes (\oplus_{1}^{d}\mathscr{E})^{\otimes \, p}$ 
if both $g$ and $h$, are, respectively, $\mathbb{C}$ and $\mathbb{C}^d$-valued Schwartz test functions. 
But the same distribution valued kernels
of $S_{n,p}$ can be evaluated at the \emph{Grassmann-valued test functions} $h$, 
in the sense of \cite{Berezin}, so that in this case
\[
S_{n,p} \in
\begin{cases}
\mathscr{L}\big((E), (E)\big) \otimes \mathscr{L}(\mathscr{E}^{\otimes \, n} \otimes \mathscr{E}^p, \mathcal{E}^{p \, *}), 
& \text{if all fields in $\mathcal{L}$ are massive},\\
\mathscr{L}\big((E), (E)^*\big) \otimes \mathscr{L}(\mathscr{E}^{\otimes \, n} \otimes \mathscr{E}^p, \mathcal{E}^{p \, *}), 
& \text{if there are massless fields in $\mathcal{L}$},\\
\end{cases}
\]
with $\mathcal{E}^{p \, *}$ being the subspace of grade $p$ of the \emph{abstract Grassmann algebra} $\oplus_p\mathcal{E}^{p \, *}$ 
\emph{with inner product and involution}
in the sense of \cite{Berezin}. $\mathscr{E}^p$ denotes the space of Grassmann-valued test functions $h^p$ of grade $p$ 
due to \cite{Berezin}, and replacing ordinary test functions $h^{\otimes \, p}$.
Here $\mathscr{L}(E_1,E_2)$ denotes the linear space of linear continuous operators $E_1\longrightarrow E_2$ endowed with the natural topology of uniform
convergence on bounded sets. 

In our previous paper \cite{CPQEDWN} we have shown 
that the causality axioms (I)-(V) for $S_{n}$ (and for $S_{n,p}$ with the analogous axioms):
\begin{gather*}
\textrm{(I)} \,\,\,\,\,\,\,\,\,\,\,\,\,\,\,\,\,\,\,\,   
              S_{n}(x_1, \ldots, x_{n}) = S_{k}(x_1, \ldots, x_{k})S_{n-k}(x_{k+1}, \ldots, x_{n}),
\,\,\,\, \textrm{whenever $\{x_{k+1}, \ldots, x_{n} \} \preceq \{x_1, \ldots, x_k\}$},   \\
\textrm{(II)}  \,\,\,\,\,\,\,\,\,\,\,\,\,\,\,\,\,\,\,\,      
U_{b,\Lambda} S_n(x_1, ..,x_n) U_{b, \Lambda}^{+} = S_n(\Lambda^{-1}x_1 - b, .., \Lambda^{-1}x_n - b),\,\,\,\,\,\,\,
\,\,\,\,\,\,\,\,\,\,\,\,\,\,\,\,\,\,\,\,\,\,\,\,\,\,\,\,\,\,\,\,\,\,\,\,\,\,\,\,\,\,\,\,\,\,\,\,\,\,\,\,\,\,\,\,\,\,
\,\,\,\,\,\,\,\,\,\,\,\,\,\,\,\,\,\,\,\,\,\,\,\,\,\,\,\,\,\,\,\,\,\,\,\,\,\,\,\,\, \\
\textrm{(III)}  \,\,\,\,\,\,\,\,\,\,\,\,\,\,\,\,\,\,\,\,        
\overline{S}_{n}(x_1, \ldots, x_{n}) = \eta S_n(x_1, \ldots, x_{n})^{+} \eta,\,\,\,\,\,\,\,\,\,\,\,\,\,\,\,\,\,\,\,\,\,\,\,\,\,\,
\,\,\,\,\,\,\,\,\,\,\,\,\,\,\,\,\,\,\,\,\,\,\,\,\,\,\,\,\,\,\,\,\,\,\,\,\,\,\,\,\,\,\,\,\,\,\,\,\,\,\,\,\,\,\,\,\,\,\,\,\,\,\,\,\,
\,\,\,\,\,\,\,\,\,\,\,\,\,\,\,\,\,\,\,\,\,\,\,\,\,\,\,\,\,\,\,\,\,\,\,\,\,\,\,\,\,\,\,\,\,\,\,\,\,\,\,\,\,\,\,\,\,\,\,\,\, \\
\textrm{(IV)}  \,\,\,\,\,\,\,\,\,\,\,\,\,\,\,\,\,\,\,\,                    
S_{1}(x_1) = i \mathcal{L}(x_1), \,\,\,\,\,\,\,\,\,\,\,\,\,\,\,\,\,\,\,\,\,\,\,\,\,\,\,\,\,\,\,\,\,\,\,\,\,\,\,\,\,\,\,\,\,\,\,\,\,\,\,\,\,\,\,\,\,
\,\,\,\,\,\,\,\,\,\,\,\,\,\,\,\,\,\,\,\,\,\,\,\,\,\,\,\,\,\,\,\,\,\,\,\,\,\,\,\,\,\,\,\,\,\,\,\,\,\,\,\,\,\,\,\,\,\,\,\,\,\,\,\,\,\,\,\,\,\,\,\,\,
\,\,\,\,\,\,\,\,\,\,\,\,\,\,\,\,\,\,\,\,\,\,\,\,\,\,\,\,\,\,\,\,\,\,\,\,\,\,\,\,\,\,\,\,\,\,\,\,\,\,\,\,\,\,\,\,\,\,\,\,\,\,\,\,\,\,\,\,\,\,\,\,\, 
\end{gather*}
\begin{enumerate}
\item[(V)] \,\,\,\,\,\,\,\,\,\,\,\,
The singularity degree of the retarded part of a kernel
should coincide with the singularity degree of this kernel, for the kernels of the of the generalized integral kernel 
and causal operators $D_{(n)}$ which are equal to linear combiations of products of the generalized operators $S_k$,
\,\,\,\,\,\,\,\,\,\,\,\,\,\,\,\,\,\,\,\,\,\,\,\,\,\,\,\,\,\,\,\,\,\,\,\,\,\,\,\,\,\,\,\,\,\,\,\,\,\,\,\,\,\,\,\,\,\,\,\,
\end{enumerate}
and involving the products
\begin{equation}
S_{k}(x_1, \ldots, x_{k})S_{n-k}(x_{k+1}, \ldots, x_{n})
\end{equation}
are indeed meaningful also in case the interaction Lagrangian density operator $\mathcal{L}$ contains massless fields, as e.g. in QED.
We have also  given there the Wick theorem decomposition for such products.
The above-mentioned product operation for $S_n$ and $S_{n,p}$, regarded as finite sums of integral kernel
operators with the Hida operators as the creation-annihilation operators, is well-defined, and given as a limit, in which
the massless free fields are replaced with the corresponding massive fields, and by passing to the zero mass limit.

Using this approach, we give here the solution to the adiabatic limit problem for interacting fields in QED. 
Namely, in case of QED, the limit $g\rightarrow 1$ exists for the higher order contributios
$\psi_{{}_{\textrm{int}}}^{(n)}(g^{\otimes \, n})$ 
and $A_{{}_{\textrm{int}}}^{(n)}(g^{\otimes \, n})$
 to interacting fields $\psi_{{}_{\textrm{int}}}$ and $A_{{}_{\textrm{int}}}$ in QED, and is equal to a finite sum of integral kernel operators
with vector valued kernels in the sense of \cite{obataJFA}, and belongs to
\[
\mathscr{L}\big( \oplus_{1}^{d}\mathscr{E}, \, \mathscr{L}((E),(E)^*)\big), \,\,\,\,\phi \in \oplus_{1}^{d}\mathscr{E}.
\]
Although there does not exist the product operation in the \emph{whole} class $\mathscr{L}\big(\mathscr{E}^{\otimes \, n}, \, \mathscr{L}((E),(E)^*)\big)$
or $\mathscr{L}\big((E), (E)^*\big) \otimes \mathscr{L}(\mathscr{E}^{\otimes \, n} \otimes \mathscr{E}^p, \mathcal{E}^{p \, *})$
of operators, the higher order contributions $S_n$ or $S_{n,p}$ to the scattering operator, which also define the interacting fields,
are of special class, and admit the operation of product defined by the above-mentioned limit operation. 
Moreover, the adiabatic limit for interacting fields exists only if the normalization in the Epstein-Glaser splitting 
into the retarded and advanced part is ``natural'' (e.g.
in which the natural equations of motion for the interacting fields are fulfilled), so the arbitrariness
in the splitting can thus be naturally eliminated by using the Hida operators. 

We will also prove that this limit for interacting fields does not exist if the charged field is massless, irrespectively of the choice of the normalization
in the Epstein-Glaser splitting of the causal distributions in the construction of the scattering operator. This may be regarded as the 
theoretical proof of the experimentally confirmed fact: that all particles with nonzero electric charge are massive. This proof can also 
be extended to the weak hypercharge with the same $U(1)$-type interaction, giving theoretical proof that neutrino is a massive particle.

The adiabatic limit $g\rightarrow 1$ problem for strong interactions on the Minkowski space-time seems to be devoid of any deeper physical sense, 
due to the phenomenon called ``asymptotic freedom'' which allows the perturbative method to be applicable in this case only in the high energy limit
of the scattering processes.  

This paper is organized in the following manner. In Section \ref{OperatorS} we briefly remind the Epstein-Glaser inductive step
construction of $S_n$, based on the Bogoliubov axioms (I)-(V). In Section \ref{OperatorS} we concentrate only on the splitting method,
replacing the standard renormalization technique in computation of the scalar coefficients in $S_n$, giving explicit
computation of the scalar coefficients in $S_2$ for spinor QED, as the splitting method plays important role in the proof of the main
theorem. We do not enter here into any details of the proof that the axioms (I)-(V) make sense
also in case the free fields, their Wick products, and $S_n$, are understood as finite sums of integral kernel operators in the sense
of \cite{obataJFA}, as this problem we have undertaken elsewhere \cite{CPQEDWN}. In Section \ref{IntFields} we remind shortly
the construction of the interacting fields due to \cite{Bogoliubov_Shirkov}. Finally, in  Section \ref{MainTheorem} we give the formulation and proof
of the main theorem, asserting existence of the adiabatic limit of the higher order contributions to interacting fields in QED in the theory (I)-(V)
in which the free fields, and $S_n$, $S_{n,p}$, and higher order contributions to interacting fields are understood 
as the integral kernel operators in the sense of \cite{obataJFA}. In this proof the Hida space $(E)$, pertinent to the construction
of the integral kernel operators (thus our free fields, their Wick products, $S_n$, $S_{n,p}$, $\ldots$) plays essential role, together with the
necessity and sufficiency criteria  \cite{obataJFA} put on the kernels which are about to represent the corresponding well-defined generalized integral kernel 
operators transforming continuously the Hida space into itself or into its strong dual. 
We add also Subsection \ref{Comparison} in which we illustrate the essential
difference between the perturbative QFT theory based on the Wightman distributions and axioms (I)-(V) 
and the perturbative QFT theory based on the integral kernel operators and axioms (I)-(V). We explain in Subsection \ref{Comparison} 
the reason why the application of the integral kernel operators is effective in the investigation of the adiabatic limit whereas the Wightman's
distributions are not.

\section{The scattering operator}\label{OperatorS}

We remind the inductive step invented by Epstein and Glaser, \cite{Epstein-Glaser}, \cite{Scharf}, based solely
on the Bogoliubov's  \cite{Bogoliubov_Shirkov} axioms (I)-(IV) and the preservation of the singularity degree axiom (V),
\cite{Epstein-Glaser}, given in Introduction. We remind this construction for $S_n$. The general case $S_{n,p}$ 
with multicomponent switching on function $(g,h)$ is completely analogous \cite{Epstein-Glaser}.    

Namely, suppose all 
$\{S_k\}_{{}_{k\leq n-1}}$ are already constructed. In order to construct  
$S_n(Z, x_n) = S(Z, x_n)$ \,\,\,\, (here $X, Y, Z$ denote the sets of space-time variables 
such that  $X \cup Y = \{x_1, \ldots, x_n\} = Z$, $X \cap Y = \emptyset$) we construct the following
generalized operators
\[
A'_{(n)}(Z, x_n)  
= \sum \limits_{X\sqcup Y=Z, X\neq \emptyset} \overline{S}(X)S(Y,x_n), \,\,\,
R'_{(n)}(Z, x_n)  
= \sum \limits_{X\sqcup Y=Z, X\neq \emptyset} S(Y,x_n)\overline{S}(X),
\]
where the sums run over all divisions $X\sqcup Y=Z$ of the set $Z$ of variables $\{x_1, \ldots, x_{n-1} \}$
into two disjoint subsets $X$ and $Y$ with $X\neq \emptyset$.
and which, by the inductive assumption, are known. Next we construct the following generalized operators
\[
\begin{split}
A_{(n)}(Z, x_n) = \sum \limits_{X\sqcup Y=Z} \overline{S}(X)S(Y,x_n)
= A'_{(n)}(Z, x_n)  + S(x_1, \ldots, x_n), \\
R_{(n)}(Z, x_n) = \sum \limits_{X\sqcup Y=Z} S(Y,x_n)\overline{S}(X)
= R'_{(n)}(Z, x_n) + S(x_1, \ldots, x_n),
\end{split}
\]
where the sums run over all divisions $X\sqcup Y=Z$ of the set $Z$ of variables $\{x_1, \ldots, x_{n-1} \}$
into two disjoint subsets $X$ and $Y$ including the empty set $X = \emptyset$.
The unprimed generalized operator $R_{(n)}$ has retarded support, meaning that it is supported within the range of each
of the variables $x_1, \ldots, x_{n-1}$ which lie in the future closed light cone emerging from $x_n$, and the 
unprimed generalized operator $A_{(n)}$ has advanced support, meaning that it is supported within the range of each
of the variables $x_1, \ldots, x_{n-1}$ which lie in the past closed light cone emerging from $x_n$, for the proof compare
\cite{Epstein-Glaser} or \cite{Scharf}. The last two statements are consequences of the Bogoliubov causality axioms (I)-(IV),
compare \cite{Epstein-Glaser} or \cite{Scharf}. The same proof can be repeated also in case the free fields, their Wick products
and $S_n$ in (I)-(V), are understood as the integral kernel operators in the sense of \cite{obataJFA}.  
Therefore, 
\[
D_{(n)} = R'_{(n)} - A'_{(n)} = R_{(n)} - A_{(n)} 
\,\,\,\,\,\,\,\,\,\,\,\,\,\,\,\, \textrm{is causally supported},
\]
with
\[
R_{(n)} \,\,\, \textrm{-- being the retarded part of $D_{(n)}$}
\,\,\,\,\,\,\,\,\,\,\,\,\,\,\,\,\,\,\,\,
A_{(n)} \,\,\, \textrm{-- being the advanced part of $D_{(n)}$}.
\]
Next we compute the retarded $R_{(n)}$  and the advanced part $A_{(n)}$ of the given generalized operator $D_{(n)}$ 
independently of the axioms (I)-(V), just observing that each of the numerical tempered distributional coefficient in each of the Wick monomials
in the Wick decomposition of $D_{(n)}$ is a numerical causal distribution, and applying the ordinary theory of distribution
splitting. This splitting is in general not unique, with the number of arbitrary constants in the splitting 
depending on the singularity degree of the splitted distribution. Thus, we arrive at the formula 
\[
\boxed{
 S_n(x_1, \ldots, x_n) =  R_{(n)}(x_1, \ldots, x_n)-R'_{(n)}(x_1, \ldots, x_n) 
}
\]
which is determined up to a finite number of constants, with the number of constants depending on the singularity degree of 
the scalar causal coefficients in the Wick decomposition of $D_{(n)}$.

Let for example
\[
\mathcal{L}(x) = \boldsymbol{{:}} \boldsymbol{\psi}(x)^{+}\gamma_{0} \gamma^\mu \boldsymbol{\psi}(x) A_\mu(x) \boldsymbol{{:}},
\]
as in spinor QED, but our analysis is general, and we can replace $\mathcal{L}$ with any Wick polynomial of free fields
with each of the Wick monomials possibly containing massless free field factors.

In particular for the second order contribution the formula
reduces to
\begin{equation}\label{S2}
S_2(x,y) = R_{(2)}(x,y) - R'_{(2)}(x,y) =  \textrm{ret} \, \big[\mathcal{L}(y)\mathcal{L}(x) -\mathcal{L}(x)\mathcal{L}(y) \big] 
- \mathcal{L}(y)\mathcal{L}(x),
\end{equation}
where $R_{(2)}(x,y)$ is the retarded part of a causally supported
distribution $D_{(2)}(x,y)$. In particular all scalar factors in terms of the Wick decomposition of 
\[
D_{(2)}(x,y)= S(y)\overline{S}(x) - \overline{S}(x)S(y) = -i^2\big[\mathcal{L}(y)\mathcal{L}(x) - \mathcal{L}(x)\mathcal{L}(y)\big]
\]
(those with pairings in the Wick decomposition of
$D_{(2)}(x,y)$) are indeed causal distributions, \emph{i.e.} distributions of one space-time
variable $x-y$, which is supported within the light cone. Therefore the method of splitting of causal distributions
into the retarded and advanced part (due to Epstein-Glaser \cite{Epstein-Glaser} or \cite{Scharf})
can indeed be used for the computation of $S_2$.

Let us note that the general splitting method, to which Epstein and Glaser or Scharf refer in \cite{Epstein-Glaser} or \cite{Scharf},
works not only for the strictly causally supported distributions. In fact it is the local property around $x-y=0$
of a distribution $d(x-y)$ which is to be splitted, which is important for the splitting, and is governed
by a single number $\omega$ pertinent to the distribution $d$, and called \emph{singular order $\omega$ of} $d$ at zero
(or at infinity for the Fourier transformed $\widetilde{d}$). 
Namely, for the plane wave kernels 
\begin{equation}\label{FFkernels}
\kappa^{(1)}_{0,1}, \kappa^{(1)}_{1,0}, \ldots, \kappa^{(q)}_{0,1}, \kappa^{(q)}_{1,0},
\end{equation}
of the free fields all contraction distributions (products of pairings) 
\begin{equation}\label{ProductsOfPairings}
\Big(\kappa^{(1)}_{0,1} \dot{\otimes} \ldots \dot{\otimes} \kappa^{(q)}_{0,1}\Big)
\otimes_q
\Big(\kappa^{(1)}_{0,1} \dot{\otimes} \ldots \dot{\otimes} \kappa^{(q)}_{0,1}\Big)(x,y) = \kappa_{q}(x-y)
\end{equation}
have finite order $\omega$ of singularity at zero and can be splitted as in the cited works. 
Contractions $\otimes_q$ are always understood for the spin momentum variables
in the tensor products of single particle test spaces $E_1$, $E_2$, $\ldots$ and their duals, and expressed
through sum/integration with respect to the $q$ pairs of the contracted spin momentum variables.
In general 
 (\ref{ProductsOfPairings}) is not causally suported. 
This splitting of (\ref{ProductsOfPairings}) is possible because the following differences or sums (depending on the parity of the number $q$) 
of such $\otimes_q$-contractions (\ref{ProductsOfPairings}) (being equal to products of pairings)
\begin{multline}\label{CausalDifferencesOfProductsOfPairings}
\Big(\kappa^{(1)}_{0,1} \dot{\otimes} \ldots \dot{\otimes} \kappa^{(q)}_{0,1}\Big)
\otimes_q
\Big(\kappa^{(1)}_{1,0} \dot{\otimes} \ldots \dot{\otimes} \kappa^{(q)}_{1,0}\Big)(x,y)
-(-1)^q
\Big(\kappa^{(1)}_{1,0} \dot{\otimes} \ldots \dot{\otimes} \kappa^{(q)}_{1,0}\Big)
\otimes_q
\Big(\kappa^{(1)}_{0,1} \dot{\otimes} \ldots \dot{\otimes} \kappa^{(q)}_{0,1}\Big)(y,x)
\\
= \kappa_q(x-y) -(-1)^q \kappa_q(y-x)
\end{multline}
have always causal support, to which the Epstein-Glaser splitting can be applied. The condition
for the convergence of the retarded part of the difference or sum (\ref{CausalDifferencesOfProductsOfPairings}) 
of the contractions (\ref{ProductsOfPairings}) implies convergence for the retarded part of  (\ref{ProductsOfPairings}).
The only difference is that now the retarded part of each $\kappa_q(x-y)$ and $\kappa_q(y-x)$ in (\ref{CausalDifferencesOfProductsOfPairings})
taken separately, is not Lorentz invariant with the time-like unit versor $v$ of the reference frame in the theta function
$\theta(v\cdot x)$ giving the support $\textrm{supp} \theta$ of the retarded part. Only in the sum/diffrene of the retarded 
parts of $\kappa_q(x-y)$ and $\kappa_q(y-x)$ in (\ref{CausalDifferencesOfProductsOfPairings}) the dependence on $v$ drops out, with the retarded part
of the whole distribution (\ref{CausalDifferencesOfProductsOfPairings}) being Lorentz invariant.

After \cite{Scharf} we briefly report a practical method for calculation which, 
as far as we know, goes back to \cite{Scharf}. Namely, the Fourier transform of the scalar $\otimes_q$-contraction
$\kappa_q$ can be computed explicitly quite easily, on using the completeness relations of the plane wave kernels. 
Below in this Subsection, as an example, we give Fourier transforms $\widetilde{\kappa_q}$ of all basic scalar $\kappa_q$, $q=1,2,3$, $\otimes_q$-contractions 
for QED. It turns out that for $q>1$ (the most interesting case)  $\widetilde{\kappa_q}$ are regular, \emph{i.e.} function-like,
distributions, analytic in $p$ everywhere, except the the finite set of characteristic submanifolds of the type $p\cdot p = const.$, or $p_0=0$,
where they have $\theta(p\cdot p - const.)$, $\theta(p_0)$-type finite jump. Let $\omega$ be equal to the order of singularity at zero of $\kappa_q$. Next  we observe
that the subspace of all $\phi \in \mathscr{E}$, whose all derivatives vanish up to $\omega$, can be written
as the image of a continuous idempotent operator $\Omega'$, (\ref{Omega'}), acting on $\mathscr{E} = \mathcal{S}(\mathbb{R}^4)$,
and  $\textrm{ret} \, \kappa_q(x-y)$ can be understood as defined on $\chi(x,y)=(\Omega'\phi)(x-y)\varphi(y)$ by the multiplication by $\theta(x-y)$,
now for $\phi,\varphi$ ranging over general elements in $\mathscr{E}$. By the axiom (V) this $\textrm{ret} \, \kappa_q$ is defined 
up to the finite linear combination of terms $C_\alpha \delta^{(\alpha)}(x-y)$ with arbitrary constants $C_\alpha$ and for $|\alpha|\leq \omega$.
Indeed, let, for each multi-index $\alpha$, such that $0 \leq |\alpha| \leq \omega$,
$\omega_{{}_{o \,\, \alpha}} \in \mathscr{E}$ on $\mathbb{R}^{4}$ be such
functions that\footnote{Such functions $\omega_{{}_{o \,\, \alpha}} \in \mathscr{E}$, $0 \leq |\alpha| \leq \omega$ do exist.
Indeed, let
\[
f_{{}_{\alpha}}(x) =
{\textstyle\frac{x^\alpha}{\alpha!}}, \,\,\, x \in \mathbb{R}^{4}, 0 \leq |\alpha| \leq \omega,
\]
\[
\alpha ! = \prod \limits_{\mu=0}^{3} \alpha_{\mu} !,
\,\,\, x^\alpha = \prod \limits_{\mu=0}^{3} (x_{\mu})^{\alpha_{\mu}}, \,\,\, \mu = 0,1,2,3.
\]
It is well-known fact that there exists $w \in \mathscr{E} = \mathcal{S}(\mathbb{R}^4)$,
which is equal $1$ on some neighborhood of zero.
Then we can put
\[
\omega_{{}_{o \,\, \alpha}} \overset{\textrm{df}}{=} f_{{}_{\alpha}}.w.
\]}
\[
D^\beta \omega_{{}_{o \,\, \alpha}} (0) = \delta^{\beta}_{\alpha}, \,\,\,\, 0 \leq |\alpha|, |\beta| \leq \omega.
\]
Then, for any $\phi \in \mathscr{E}$ we put
\begin{equation}\label{Omega'}
\Omega' \phi = \phi - \sum \limits_{0 \leq |\alpha| \leq \omega} D^\alpha \phi(0) \,\, \omega_{{}_{o \,\, \alpha}}
\,\,\,
=
 \phi - \sum \limits_{0 \leq |\alpha| \leq \omega} D^\alpha \phi(0) \,\, {\textstyle\frac{x^\alpha}{\alpha!}}.w
\end{equation}
Now, the evaluation $\langle \textrm{ret} \, \kappa_q, \chi \rangle$, with $\chi(x,y)=(\Omega'\phi)(x-y)\varphi(y)$ and $\phi, \varphi \in \mathscr{E}$, becomes equal 
\[
\widetilde{\varphi}(0) \,\, \big\langle\kappa_q, \theta \Omega'\phi \big\rangle = 
\big\langle \widetilde{\kappa_q}, \widetilde{\theta} \ast \widetilde{\Omega'\phi} \big\rangle, 
\] 
and is expected to be a well-defined continuous functional of $\widetilde{\phi},\widetilde{\varphi} \in \mathcal{S}(\mathbb{R}^4)$,
or equivalently
\[
\big\langle \widetilde{\kappa_q}, \widetilde{\theta} \ast \widetilde{\Omega'\phi} \big\rangle, 
\]
should give a well-defined functional of $\phi \in \mathcal{S}(\mathbb{R}^4)$. It defines
$\langle \textrm{ret} \, \kappa_q, \chi \rangle$ for all $\chi(x,y)=\phi(x-y)\varphi(y)$ with
$\phi \in \textrm{Im} \, \Omega'$, and because $\textrm{Ker} \, \Omega' \neq\{0\}$ for $q > 1$, its definition
is not unique and, by axiom (V), can be extended by addition of any functional which is zero 
on $\chi(x-y)\varphi(y) = \phi(x-y)\varphi(y)$ with
$\phi \in \textrm{Im} \, \Omega'$. Equivalently the retarded part
\begin{gather}
\textrm{ret} \, \kappa_q \overset{\textrm{df}}{=} \kappa_q \circ (\theta \Omega'),
\label{Def(retkappaq)}
\\
\big\langle \textrm{ret} \, \kappa_q, \phi \big\rangle
= \big\langle \widetilde{\textrm{ret} \, \kappa_q}, \widetilde{\phi} \big\rangle
=\big\langle\kappa_q, \theta \Omega'\phi \big\rangle  
=\big\langle \widetilde{\kappa_q}, \widetilde{\theta} \ast \widetilde{\Omega'\phi} \big\rangle
\label{FT(retkappaq)}
\end{gather}
of $\kappa_q$ is not unique, and we can add to it
\begin{equation}\label{freedom}
\sum\limits_{|\alpha|=0}^{\omega} C_\alpha \delta^{\alpha},
\end{equation} 
which is most general on $\textrm{Ker} \, \Omega'$ and which is zero on $\textrm{Im} \, \Omega'$. 
In order to show (\ref{Def(retkappaq)}) is a well-defined tempered distribution, we proceed after \cite{Scharf},
using the explicit formula for the function $\widetilde{\kappa_q}$ in (\ref{FT(retkappaq)}). 

Since
\[
\widetilde{x^\alpha w}(p) = (iD_{p})^\alpha \widetilde{w}(p),
\]
and
\[
D^\alpha\phi(0) = (-1)^\alpha \big\langle D^\alpha \delta, \phi \big\rangle 
= (-1)^\alpha \big\langle \widetilde{D^\alpha\delta}, \widetilde{\phi} \big\rangle
=
 (2\pi)^{-4/2} \big\langle (ip)^\alpha, \widetilde{\phi}  \big\rangle,
\]
immediately from (\ref{FT(retkappaq)}) we obtain
\begin{multline*}
\big\langle \widetilde{\textrm{ret} \, \kappa_q}, \widetilde{\phi} \big\rangle
= \big\langle \widetilde{\kappa_q}, \widetilde{\theta \Omega'\phi}\big\rangle
= 
(2\pi)^{-4/2} \Bigg\langle \widetilde{\kappa_q},  \,\,\,\, \widetilde{\theta}
\,\,
\ast \,\, \Big[ \widetilde{\phi} - \sum\limits_{|\alpha|=0}^{\omega} {\textstyle\frac{1}{\alpha!}} (iD_{p})^\alpha \widetilde{w}
(2\pi)^{-4/2} \big\langle (ip')^\alpha, \widetilde{\phi} \big\rangle \Big] \Bigg\rangle
\\
=
(2\pi)^{-4/2} \Big\langle \widetilde{\theta} \ast \widetilde{\kappa_q}, \,\,\,
\widetilde{\phi} - \sum\limits_{|\alpha|=0}^{\omega} {\textstyle\frac{1}{\alpha!}} (iD_{p})^\alpha \widetilde{w}
(2\pi)^{-4/2} \big\langle (ip')^\alpha, \widetilde{\phi} \,\, \Big\rangle
\end{multline*}
were the convolution $\widetilde{\theta} \ast \widetilde{\kappa_q}$ is defined only on the subtracted $\widetilde{\phi}$,
\emph{i.e.} on Fourier transforms of test functions with all derivatives vanishing  at zero up to order $\omega$. Interchanging the integration
variables $p'$ and $p$ in the subtracted terms we get
\[
\big\langle \widetilde{\textrm{ret} \, \kappa_q}, \widetilde{\phi} \big\rangle 
= (2\pi)^{-4/2} \int d^4 k \, \widetilde{\theta}(k) \,\, \Big\langle \widetilde{\kappa_q}(p-k)
- (2\pi)^{-4/2} \sum\limits_{|\alpha|=0}^{\omega} {\textstyle\frac{(-1)^\alpha}{\alpha!}}p^\alpha 
\int d^4 p' \,  \widetilde{\kappa_q}(p'-k) D_{p'}^{\alpha}\widetilde{w}, \, \widetilde{\phi} \Big\rangle.
\]
By partial integration in the $p'$ variable we arrive at the formula
\begin{equation}\label{FT(retkappaq)1}
\widetilde{\textrm{ret} \, \kappa_q}(p) = {\textstyle\frac{1}{(2\pi)^2}} \int d^4k \, \widetilde{\theta}(k)
\Bigg[
\widetilde{\kappa_q}(p-k) - {\textstyle\frac{1}{(2\pi)^2}} \sum\limits_{|\alpha|=0}^{\omega}
{\textstyle\frac{p^\alpha}{\alpha!}}\int d^4 p' \, D^\alpha \widetilde{\kappa_q}(p'-k)\widetilde{w}(p')  
\Bigg],
\end{equation}
which, in general should be understood in the distributional sense (converging when
integrated in $p$ variable with a test function of $p$). But in practical computations we are interesting
in the regularity domain, outside the characteristic submanifold $p\cdot p = const.$, where $\widetilde{\textrm{ret} \, \kappa_q}$
is represented by ordinary function of $p$ and where 
this integral should be convergent in the ordinary sense. 
Then we assume that there exists a point $p''$ around which (\ref{Def(retkappaq)}) is regular, and moreover
possess all derivatives in the usual sense (should not be mixed with distributional) at $p''$, 
up to order $\omega$ equal to the singularity degree. This is not entirely arbitrary assumption, because
if the formula (\ref{Def(retkappaq)}) makes any sense at all as a functional of $\phi$, it must have support
in the half space (or forward cone in case we replace $\kappa_q$ with the causal $\kappa_q - (-1)^q\check{\kappa_q}$),
so its Fourier transform should be a boundary value of an analytic function, regular all over $\mathbb{R}^4$,
but compare the comment we give below on this topic.
Next, we subtract  the first 
Taylor terms of $\widetilde{\textrm{ret} \, \kappa_q}$ up to order $\omega$ around the regularity point $p''$ 
\begin{equation}\label{FT(retkappa)p''}
\big[\widetilde{\textrm{ret} \, \kappa_q}\big]_{{}_{{}_{p''}}}(p) = \widetilde{\textrm{ret} \, \kappa_q}(p)  
- \sum\limits_{|\beta|=0}^{\omega}
{\textstyle\frac{(p-p'')^\beta}{\beta!}} D^\beta \widetilde{\textrm{ret} \, \kappa_q}(p'') 
\end{equation}
and get another possible retarded part $\big[\widetilde{\textrm{ret} \, \kappa_q}\big]_{{}_{{}_{p''}}}$, 
Fourier transformed, of $\kappa_q$, as the added terms have the form (\ref{freedom}). 
By construction (\ref{FT(retkappa)p''}) is so normalized, that all its derivatives vanish up to order $\omega$
at $p''$. This point is called \emph{normalization point}.
Next we compute $D^\beta \widetilde{\textrm{ret} \, \kappa_q}$ from the formula (\ref{FT(retkappaq)1}), and substitute
into the formula (\ref{FT(retkappa)p''}). Using
\[
\sum\limits_{\beta\leq \alpha} {\textstyle\frac{(p-p'')^\beta}{\beta!}}
 {\textstyle\frac{{p''}^{\alpha-\beta}}{(\alpha-\beta)!}} 
= 
{\textstyle\frac{1}{\alpha!}}
\sum\limits_{\beta\leq \alpha} {\alpha \choose \beta} (p-p'')^\beta {p''}^{\alpha-\beta}
= {\textstyle\frac{p^\alpha}{\alpha!}},
\]
we see that all terms with the auxiliary function $\widetilde{w}$
drop out and we obtain for $\big[\widetilde{\textrm{ret} \, \kappa_q}\big]_{{}_{{}_{p''}}}$ the formula
\begin{equation}\label{FT(retkappaq)2}
\big[\widetilde{\textrm{ret} \, \kappa_q}\big]_{{}_{{}_{p''}}}(p) = {\textstyle\frac{1}{(2\pi)^2}} \int d^4k \widetilde{\theta}(k)
\Bigg[
\widetilde{\kappa_q}(p-k) - \sum\limits_{|\beta|=0}^{\omega}
{\textstyle\frac{(p-p'')^\beta}{\beta!}} D^\beta \widetilde{\kappa_q}(p''-k) 
\Bigg],
\end{equation}
by construction so normalized that all derivatives of $\big[\widetilde{\textrm{ret} \, \kappa_q}\big]_{{}_{{}_{p''}}}$ 
vanish at $p''$ up to order $\omega$, compare \cite{Scharf}, p. 179. 
We need to rewrite the last integral (\ref{FT(retkappaq)2}) with $\widetilde{\kappa_q}$ taken at the same point in the two terms
in (\ref{FT(retkappaq)2}), in order to combine the two terms into a single term, as these terms taken separately
are divergent. At this point the case in which we have causal $\kappa_q-(-1)^q\check{\kappa_q}$ and Lorentz convariant, instead of
$\kappa_q$, is simpler, because $\kappa_q-(-1)^q\check{\kappa_q}$ has causal support and is Lorentz convariant. 
In this causal case we can use the special $\theta(x) = \theta(x_0) = \theta(v\cdot x)$ with $v=(1,0,0,0)$, and insert
\begin{equation}\label{FTtheta}
\widetilde{\theta}(k_0,\boldsymbol{k}) = 2\pi {\textstyle\frac{i}{k_0+i\epsilon}} \,
\delta(\boldsymbol{k}),
\end{equation}
and use Lorentz frame in which $p=(p_0,0,0,0)$ and then using integration by parts we remove the differentiainion
operation in momentum variables  off $\widetilde{\kappa_q}$ in the second term in (\ref{FT(retkappaq)2}).  
Next we use invariance and analytic continuation to extend 
$\widetilde{\textrm{ret} \, \kappa_q}$ all over $p$.
Because  $\kappa_q$ is not causal, nor Lorentz invariant, we need to consider separately the case
$p=(p_0,0,0,0)$ inside the positive and negative cone and $p=(0,0,0,p_3)$ outside the cone in the momentum space
as well as $\theta(x) = \theta(v\cdot x)$ with a more general timelike unit $v$
in
\[
\widetilde{\theta}(k_0,\boldsymbol{k}) = 2\pi {\textstyle\frac{i}{k_0 +i\epsilon v_0}} \,
\delta(\boldsymbol{k}- k_0{\textstyle\frac{\boldsymbol{v}}{v_0}} ),
\]
in order to reconstruct $\widetilde{\textrm{ret} \, \kappa_q}$, now depending on $v$, as $\widetilde{\kappa_q}$
is not causal. 

Still proceeding after \cite{Scharf}, let us concentrate now on the computation of the retarded part the causal $d= \kappa_q-(-1)^q\check{\kappa_q}$,which
is simpler and moreover, on such QFT in which the normalization point $p''$ can be put equal zero. 
This is e.g. the case for QED's with massive charged fields. 
We reconstruct $\widetilde{\textrm{ret} \, d}(p)$ first for $p$ in the cone $V^+$.  
In this case computations simplify, as we can put $p''=0$, and moreover
we use Lorentz covariance of the causal $d$, and choose a Lorentz frame in which $p=(p_0,0,0,0)$.
Moreover, because $d$ is causal its retarded part is independent of the choice of the time like
unit versor $v$ in the theta function $\theta(v\cdot x)$, so that we can choose $v=(1,0,0,0)$, thus assuming that it is
always parallel  to $p\in V^+$, and thus varies with $p$. After these choices, the last formula
(\ref{FT(retkappaq)2}) for the retarded part, applied to $d$ instead of $\kappa_q$, and with substituted
Fourier transform (\ref{FTtheta}), reads
\[
\widetilde{\textrm{ret} \, d}(p_0,0,0,0) 
= {\textstyle\frac{i}{2\pi}} \int dk_0 
{\textstyle\frac{1}{k_0+i\epsilon}}
\Bigg[
\widetilde{d}(p_0-k_0,0,0,0) - \sum\limits_{a=0}^{\omega}
{\textstyle\frac{(p_0)^a}{a!}}(-1)^a D_{k_0}^{a} \widetilde{d}(q_0-k_0,0,0,0)\big|_{{}_{q_0=0}} 
\Bigg].
\]
Integrating the subtracted terms by parts and then using the intergration variable $k'_{0}= k_0-p_0$ in the first term, we get
\[
\widetilde{\textrm{ret} \, d}(p_0) = {\textstyle\frac{i}{2\pi}} \int dk'_{0} 
\,
\Bigg[
{\textstyle\frac{1}{p_0 + k'_{0}+i\epsilon}}
 - \sum\limits_{a=0}^{\omega}
{\textstyle\frac{(p_0)^a}{a!}} 
{\textstyle\frac{\partial^a}{\partial {k'}_{0}^{a}}}
{\textstyle\frac{1}{k'_{0}+i\epsilon}}
\Bigg] \,
\widetilde{d}(-k'_{0}),
\]
with
\begin{equation}
{\textstyle\frac{1}{p_0 + k'_{0}+i\epsilon}}
 - \sum\limits_{a=0}^{\omega}
{\textstyle\frac{(p_0)^a}{a!}} 
{\textstyle\frac{\partial^a}{\partial {k'}_{0}^{a}}}
{\textstyle\frac{1}{k'_{0}+i\epsilon}} 
=
\Big({\textstyle\frac{-p_0}{k'_{0}+i\epsilon}}\Big)^{\omega+1}
{\textstyle\frac{1}{p_0 +k'_{0}+i\epsilon}}.
\end{equation}
To recover the formula for arbitrary $p\in V^+$ we introduce the variable $t=k_0/p_0$ and obtain
\begin{equation}\label{FT(retd)}
\widetilde{\textrm{ret} \, d}(p) = {\textstyle\frac{i}{2\pi}} \int\limits_{-\infty}^{+\infty} d t
\,
{\textstyle\frac{\widetilde{d}(tp),}{(t-i\epsilon)^{\omega+1}(1-t+i\epsilon)}},
\end{equation}
or
\begin{equation}\label{FT(retCausalkappaq)}
\mathscr{F} \Big(\textrm{ret} \, \big[\kappa_q-(-1)^q\check{\kappa_q}\big]\Big)(p) = {\textstyle\frac{i}{2\pi}} \int\limits_{-\infty}^{+\infty} d t
\,
{\textstyle\frac{\mathscr{F}\big(\kappa_q-(-1)^q\check{\kappa_q}\big)(tp),}{(t-i\epsilon)^{\omega+1}(1-t+i\epsilon)}},
\,\,\,\, p \in V^+,
\end{equation}
and with  normalization point $p''=0$.
To compute the retarded part outside the cone $V^+$ we have several possibilities, e.g. we can use the analytic 
continuation method, or choose the normalization point $p''$ outside the cone and repeat the computation.

Let us give the basic distributions for spinor QED with the standard realizations of the e.m. potential and Dirac free fields
(constructed with the Hida operators as the creation-annihilation operators).
These are the retarded and advanced parts of the $\otimes_q$-contractions, $q=1,2,3$, which appear in the Wick decomposition
of the product
\[
\mathcal{L}(x_1)\mathcal{L}(x_2),
\]
where $\mathcal{L}(x)$ is the interaction Lagrangian density generalized operator for spinor QED.

In QED we have the free fields $\mathbb{A} = A,\boldsymbol{\psi}$ equal to the free e.m. potential field $A$ or the free
Dirac bispinor field $\boldsymbol{\psi}$, regarded as sums of two integral kernel operators
\[
\mathbb{A}(\phi) = \psi^{(-)}(\phi) + \psi^{(+)}(\phi) = \Xi(\kappa_{0,1}(\phi)) +
\Xi(\kappa_{1,0}(\phi))
\]
with the integral kernels $\kappa_{l,m}$ represented by ordinary functions:
\[
\begin{split}
\kappa_{0,1}(\nu, \boldsymbol{\p}; \mu, x) =
{\textstyle\frac{g_{\nu \mu}}{(2\pi)^{3/2}\sqrt{2p^0(\boldsymbol{\p})}}}
e^{-ip\cdot x}, \,\,\,\,\,\,
p = (|p_0(\boldsymbol{\p})|, \boldsymbol{\p}), \, p\cdot p=0, \\
\kappa_{1,0}(\nu, \boldsymbol{\p}; \mu, x) =
{\textstyle\frac{g_{\nu \mu}}{(2\pi)^{3/2}\sqrt{2p^0(\boldsymbol{\p})}}}
e^{ip\cdot x},
\,\,\,\,\,\,
p \cdot p = 0,
\end{split}
\]
for the free e.m.potential field $\mathbb{A}=A$ (in the Gupta-Bleuler gauge) and
\[
\kappa_{0,1}(s, \boldsymbol{\p}; a,x) = \left\{ \begin{array}{ll}
(2\pi)^{-3/2}u_{s}^{a}(\boldsymbol{\p})e^{-ip\cdot x}, \,\,\, \textrm{$p = (|p_0(\boldsymbol{\p})|, \boldsymbol{\p}), \, p \cdot p = m^2$} & \textrm{if $s=1,2$}
\\
0 & \textrm{if $s=3,4$}
\end{array} \right.,
\]
\[
\kappa_{1,0}(s, \boldsymbol{\p}; a,x) = \left\{ \begin{array}{ll}
0 & \textrm{if $s=1,2$}
\\
(2\pi)^{-3/2} v_{s-2}^{a}(\boldsymbol{\p})e^{ip\cdot x}, \,\,\, \textrm{$p \cdot p = m^2$} & \textrm{if $s=3,4$}
\end{array} \right.
\]
for the free Dirac spinor field $\mathbb{A}=\boldsymbol{\psi}$, and
which are in fact the respective plane wave solutions of d'Alembert and of Dirac equation, which span the corresponding generalized eigen-solution sub spaces. 
Here $g_{\nu\mu}$ are the components of the space-time Minkowski metric tensor, and 
\[
u_s(\boldsymbol{\p}) =  \frac{1}{\sqrt{2}} \sqrt{\frac{E(\boldsymbol{\p}) + m}{2 E(\boldsymbol{\p})}}
\left( \begin{array}{c}   \chi_s + \frac{\boldsymbol{\p} \cdot \boldsymbol{\sigma}}{E(\boldsymbol{\p}) + m} \chi_s
\\                                           
              \chi_s - \frac{\boldsymbol{\p} \cdot \boldsymbol{\sigma}}{E(\boldsymbol{\p}) + m} \chi_s                         \end{array}\right),
\,\,\,\,\,\,\,\,
v_s(\boldsymbol{\p}) =  \frac{1}{\sqrt{2}} \sqrt{\frac{E(\boldsymbol{\p}) + m}{2 E(\boldsymbol{\p})}}
\left( \begin{array}{c}   \chi_s + \frac{\boldsymbol{\p} \cdot \boldsymbol{\sigma}}{E(\boldsymbol{\p}) + m} \chi_s
\\                                           
              -\big(\chi_s - \frac{\boldsymbol{\p} \cdot \boldsymbol{\sigma}}{E(\boldsymbol{\p}) + m}\chi_s \big)                          \end{array}\right), 
\]
where
\[
\chi_1 = \left( \begin{array}{c} 1  \\
                                                  0 \end{array}\right), \,\,\,\,\,
\chi_2 = \left( \begin{array}{c} 0  \\
                                                  1 \end{array}\right),
\]
are the Fourier transforms of the complete system of the free Dirac equation in the chiral represenation in which 
\[
\gamma^0 = \left( \begin{array}{cc}   0 &  \bold{1}_2  \\
                                           
                                                   \bold{1}_2              & 0 \end{array}\right), \,\,\,\,
\gamma^k = \left( \begin{array}{cc}   0 &  -\sigma_k  \\
                                           
                                                   \sigma_k             & 0 \end{array}\right),
\]
with the Pauli matrices
\[
\boldsymbol{\sigma} = (\sigma_1,\sigma_2,\sigma_3) = 
\left( \,\, \left( \begin{array}{cc} 0 & 1 \\

1 & 0 \end{array}\right), 
\,\,\,\,\,
\left( \begin{array}{cc} 0 & -i \\

i & 0 \end{array}\right),
\,\,\,
\left( \begin{array}{cc} 1 & 0 \\

0 & -1 \end{array}\right)
\,\,
\right).
\]
The single particle momentum test space $E_1$ of the Dirac field is equal $\mathcal{S}(\mathbb{R}^3; \mathbb{C}^4)$ and the sigle particle
momentum test space $E_2$ of the free e.m. potential field is equal to $\mathcal{S}^{0}(\mathbb{R}^3;\mathbb{C})$ -- the closed subspace
of  $\mathcal{S}(\mathbb{R}^3; \mathbb{C}^4)$ consisting of all those functions whose all derivaives vanish at zero.

The $\otimes_1$-contractions (pairings) are the following (with the plane wave kernels
$\kappa_{0,1}, \kappa_{1,0}$, $\kappa_{1,0}, \ldots$ defining the respecive free fields):

\begin{multline*}
\sum\limits_{a,b}\int \big[\boldsymbol{\psi}^{(-) \, a}(x), \boldsymbol{\psi}^{\sharp (+) \, b}(x)]_{{}_{+}} \, \phi_a(x)\varphi_b(y) \ud^4x \ud^4y
\\
= \sum\limits_{a,b,s}\int \kappa_{1,0}(s, \boldsymbol{\p};a,x)  \kappa_{0,1}^{\sharp}(s, \boldsymbol{\p};b,y)
\, \phi_a(x)\varphi_b(y) \, \ud^3\boldsymbol{\p} \, \ud^4x \ud^4y 
\\
= \kappa_{1,0}(\phi) \otimes_1 \kappa_{0,1}^{\sharp}(\varphi) = \big\langle \kappa_{1,0}(\phi), \kappa_{0,1}^{\sharp}(\varphi) \big\rangle
= -i S^{(+)}(\phi\otimes \varphi),
\end{multline*}
\[
\kappa_{0,1}(\phi) \otimes_1 \kappa_{1,0}^{\sharp}(\varphi) = \big\langle \kappa_{0,1}(\phi), \kappa_{1,0}^{\sharp}(\varphi) \big\rangle
= -i S^{(-)}(\phi\otimes \varphi),
\]
\[
\kappa_{0,1}^{\sharp}(\varphi) \otimes_1 \kappa_{1,0}(\phi) = \big\langle \kappa_{0,1}^{\sharp}(\varphi), \kappa_{1,0}(\phi) \big\rangle
= -i S^{(+)}(\phi\otimes \varphi),
\]
Similarly we have the contraction formula for the pairing
\[
\quad\underbracket{
A_{\mu}(x)
A_{\nu}}(y) = i g_{\mu\nu} D_{0}^{(+)}(x-y) 
\]
and $ig_{\mu\nu}D_{0}^{(-)}(x-y)$.

We have the following $\otimes_2$-contractions $\kappa_2$:
\begin{multline*}
\int \textrm{tr}\big[\gamma^\mu S^{(-)}(x-y) \gamma^\nu S^{(+)}(y-x)] \, \phi_\mu(x)\varphi_\nu(y) \ud^4x \ud^4y
\\
= 
\sum_{\substack{a,b,c,d \\ s,s'}} 
\int
\kappa^{\sharp}_{0,1}(s, \boldsymbol{\p}; a,x)\gamma^{\mu}_{ab}\kappa_{0,1}(s', \boldsymbol{\p}';b,x)
\kappa^{\sharp}_{1,0}(s', \boldsymbol{\p}';c,y)\gamma^{\nu}_{cd}\kappa_{1,0}(s, \boldsymbol{\p}; d,y) 
\, \times
\\
\times
\, \phi_\mu(x)\varphi_\nu(y) \ud^3\boldsymbol{\p} \ud^3\boldsymbol{\p}' \ud^4x \ud^4y
\end{multline*}
\begin{multline*}
=
\big(\kappa^{\sharp}_{0,1} \dot{\otimes}\gamma^\mu \kappa_{0,1}\big)(\phi_\mu) \otimes_2
(\kappa^{\sharp}_{1,0} \dot{\otimes} \gamma^\nu \kappa_{1,0}\big)(\varphi_\nu)
\\
=
\sum_{\substack{a,b,c,d}} \int
\gamma^{\mu}_{ab} \gamma^{\nu}_{cd}
\quad\underbracket{
\boldsymbol{\psi}^{\sharp}_{a}(x)
\boldsymbol{\psi}_{d}}(y)
\quad\underbracket{
\boldsymbol{\psi}_{b}(x)
\boldsymbol{\psi}^{\sharp}_{c}}(y)
\, \phi_\mu(x)\varphi_\nu(y) \ud^4x \ud^4y
\end{multline*}
or 
\[
\sum\limits_{a,b,c,d}
\gamma^{\mu}_{ab} \gamma^{\nu}_{cd}
\quad\underbracket{
\boldsymbol{\psi}^{\sharp}_{a}(x)
\boldsymbol{\psi}_{d}}(y)
\quad\underbracket{
\boldsymbol{\psi}_{b}(x)
\boldsymbol{\psi}^{\sharp}_{c}}(y)
=
\big(\kappa^{\sharp}_{0,1} \dot{\otimes}\gamma^\mu \kappa_{0,1}\big) \otimes_2
(\kappa^{\sharp}_{1,0} \dot{\otimes} \gamma^\nu \kappa_{1,0}\big)(x,\mu, y,\nu) = \kappa^{\mu\nu}_{2}(x-y)
\]
Similarly
\begin{multline*}
\sum\limits_{a,d,\mu}\int \big[\gamma^\mu S^{(-)}\gamma_\mu\big]^{ad}(x-y)D_{0}^{(+)}(y-x) \phi_a(x)\varphi_d(y) \ud^4x \ud^4y
\\
= 
\sum_{\substack{a,b,c,d \\ \mu, \nu, s,s'}} 
\int
g_{\mu\nu}
\gamma^{\mu}_{ab}\kappa_{0,1}(s, \boldsymbol{\p}; b,x) \kappa_{1,0}(s', \boldsymbol{\p}';\mu,x) \kappa^{\sharp}_{1,0}(s, \boldsymbol{\p};c,y)\gamma^{\nu}_{cd}
\kappa_{0,1}(s', \boldsymbol{\p}';\nu,y)
\, \times
\\
\times
\, \phi_a(x)\varphi_d(y) \ud^3\boldsymbol{\p} \ud^3\boldsymbol{\p}' \ud^4x \ud^4y
\end{multline*}
\begin{multline*}
=
\sum\limits_{\mu}
\big(\gamma^\mu\kappa_{0,1} \dot{\otimes} \kappa_{1,0 \,\, \mu}\big)(\phi) \otimes_2
(\kappa^{\sharp}_{1,0} \dot{\otimes} \gamma_\mu \kappa_{0,1 \,\, \nu}\big)(\varphi)
\\
=
\sum_{\substack{a,b,c,d \\ \mu, \nu}} \int
\gamma^{\mu}_{ab} \gamma^{\nu}_{cd}
\quad\underbracket{
\boldsymbol{\psi}_{b}(x)
\boldsymbol{\psi}^{\sharp}_{c}}(y)
\quad\underbracket{
A_{\mu}(x)
A_{\nu}}(y)
\, \phi_a(x)\varphi_d(y) \ud^4x \ud^4y
\end{multline*}
or
\[
\sum_{\substack{b,c \\ \mu, \nu}}
\gamma^{\mu}_{ab} \gamma^{\nu}_{cd}
\quad\underbracket{
\boldsymbol{\psi}_{b}(x)
\boldsymbol{\psi}^{\sharp}_{c}}(y)
\quad\underbracket{
A_{\mu}(x)
A_{\nu}}(y)
=
\sum\limits_{\mu\nu}
\big(\gamma^\mu\kappa_{0,1} \dot{\otimes} \kappa_{1,0 \,\, \mu}\big) \otimes_2
(\kappa^{\sharp}_{1,0} \dot{\otimes} \gamma_\mu \kappa_{0,1 \,\, \nu}\big)(a,x,d,y) = \kappa_{2}^{ad}(x-y)
\]

Finally we have one $\otimes_3$-contraction:
\begin{multline*}
\sum_{\substack{a,b,c,d \\ \mu, \nu}} 
\gamma^{\mu}_{ab} \gamma^{\nu}_{cd}
\quad\underbracket{
\boldsymbol{\psi}^{\sharp}_{a}(x)
\boldsymbol{\psi}_{d}}(y)
\quad\underbracket{
\boldsymbol{\psi}_{b}(x)
\boldsymbol{\psi}^{\sharp}_{c}}(y)
\quad\underbracket{
A_{\mu}(y)
A_{\nu}}(x)
\\
= \textrm{tr}\big[\gamma^\mu S^{(-)}(x-y) \gamma_\mu S^{(+)}(y-x)]D_{0}^{(+)}(y-x)
\\
=
\sum\limits_{\mu\nu}
\big(\kappa^{\sharp}_{0,1} \dot{\otimes}\gamma^\mu \kappa_{0,1} \dot{\otimes} \kappa_{0,1 \,\,\mu}\big) \otimes_3
(\kappa^{\sharp}_{1,0} \dot{\otimes} \gamma_\mu \kappa_{1,0} \dot{\otimes} \kappa_{1,0 \,\,\nu}\big)(x, y) = \kappa_3(x-y)
\end{multline*}

Using the contraction formula in explicit form and the completeness relations for the plane wave kernels $\kappa_{0,1}$, $\kappa_{1,0}$ 
of the free fields we can compute the Fourier transforms 
(compare \cite{Scharf}) of these contraction distributions in explicit form:
\[
\begin{split}
\widetilde{\kappa_{2}^{\mu\nu}}(p) =- {\textstyle\frac{(2\pi)^{-3}}{3}} \big({\textstyle\frac{p^\mu p^\nu}{p^2}} - g^{\mu\nu}\big)
(p^2+2m^2)\sqrt{1-{\textstyle\frac{4m^2}{p^2}}}\theta(p^2-4m^2)\theta(-p_0),
\\
\widetilde{\kappa_{2}}(p) = {\textstyle\frac{1}{2^5\pi^3}} \theta(p^2-m^2)\theta(-p_0)\big(1- {\textstyle\frac{m^2}{p^2}}\big)
\gamma^\mu \big[ m + {\textstyle\frac{\slashed{p}}{2}}\big(1+{\textstyle\frac{m^2}{p^2}}\big) \big]  \gamma_\mu,
\\
\widetilde{\kappa_{3}}(p) = i(2\pi)^{-5} \theta(p^2-4m^2)\theta(-p_0)\Big[ 
\big({\textstyle\frac{p^4}{24}} + {\textstyle\frac{m^2}{12}}p^2 + m^4\big)\sqrt{1-{\textstyle\frac{4m^2}{p^2}}}
\\
+ {\textstyle\frac{m^4}{p^2}}(4m^2-3p^2) \textrm{ln} \big(\sqrt{{\textstyle\frac{p^2}{4m^2}}} + \sqrt{{\textstyle\frac{p^2}{4m^2}}-1}\big)
\Big].
\end{split}
\]

For example in order to get the vacuum polarization term 
\[
-i \Pi^{\mu\nu}(x_1-x_2) \, {:}A_\mu(x_1)A_\nu(x_2){:},
\]
in $S_2$ we apply the Wick theorem  to the operators
\[
\mathcal{L}(x_2)\mathcal{L}(x_1) - \mathcal{L}(x_1)\mathcal{L}(x_2)
\,\,\,\,\,
\textrm{and}
\,\,\,\,\,
\mathcal{L}(x_2)\mathcal{L}(x_1)
\]
and collect all terms proportional to
${:}A_\mu(x_1)A_\nu(x_2){:}$. Next, we compute the scalar coefficient $d^{\mu\nu}$ $= \kappa^{\mu\nu}_{2} - \check{\kappa}^{\mu\nu}_{2}$
\begin{gather*}
d^{\mu\nu}(x_1-x_2) = e^2 \textrm{Tr} \, \big[\gamma^\mu S^{(+)}(x_1-x_2)\gamma^\nu S^{(-)}(x_2-x_1) \big]
-e^2 \textrm{Tr} \, \big[\gamma^\mu S^{(+)}(x_2-x_1)\gamma^\nu S^{(-)}(x_1-x_2)\big],
\\
\widetilde{d^{\mu\nu}}(p)= \big({\textstyle\frac{p^\mu p^\nu}{p^2}} - g^{\mu\nu}\big)f(p),
\,\,\,\, f(p) = g(p)-g(-p),
\,\,\,\, g(p) = \theta(p^2-4m^2) \, (p^2+2m^2) \sqrt{1-{\textstyle\frac{4m^2}{p^2}}},
\end{gather*}
multiplying ${:}A_\mu(x_1)A_\nu(x_2){:}$ in the Wick decomposition of 
\[
\mathcal{L}(x_2)\mathcal{L}(x_1) - \mathcal{L}(x_1)\mathcal{L}(x_2),
\]
and from (\ref{S2}) finally get
\[
\Pi^{\mu\nu}(x_1-x_2) = \textrm{ret} \, d^{\mu\nu}(x_1-x_2) -e^2 \textrm{Tr} \, \big[\gamma^\mu S^{(+)}(x_2-x_1)\gamma^\nu S^{(-)}(x_1-x_2)\big].
\]
Here we compute $\textrm{ret} \, d^{\mu\nu}$, or its Fourier transform $\widetilde{\textrm{ret} \, d^{\mu\nu}}$, insertig the above 
$\widetilde{d^{\mu\nu}}$ for $\widetilde{d}$ into the formula (\ref{FT(retd)}). In this way we obtain
\begin{equation}\label{Example:Pi}
\widetilde{{\Pi}_{\mu \nu}}(p) = 
(2\pi)^{-4} \big({\textstyle\frac{p_\mu p_\nu}{p^2}} - g_{\mu\nu}\big) \widetilde{\Pi}(p),
\,\,\,\,\,\,
\widetilde{\Pi}(p) =
{\textstyle\frac{1}{3}} p^4
\int\limits_{4m^2}^{\infty} {\textstyle\frac{s+2m^2}{s^2(p^2-s+i0)}}\sqrt{1-{\textstyle\frac{4m^2}{s}}} ds.
\end{equation}

Analogously we compute the scalar coefficients at the remaining Wick monomials in $S_2$ using the formulas for the Fourier transforms of
the $q$-contractions $\otimes_q$, $\widetilde{d}=\widetilde{\kappa_q} - (-1)^q \check{\widetilde{\kappa_q}}$, the splitting formula (\ref{FT(retd)}) and (\ref{S2}).

Of course dispersion integral (\ref{FT(retd)}) has also its higher dimensional
analogue for translationally invariant causal distributions $d$ in more space-time variables, which is derived in the analogous way as
(\ref{FT(retd)}) and which is used in the construction of $S_n$ with $n>2$.

\section{Interacting fields}\label{IntFields}

Using the causal perturbative 
method (compare \cite{Epstein-Glaser}, \cite{Scharf}) we construct inductively 
the scattering operator functional
\begin{multline*}
S(g\mathcal{L} + h\mathbb{A}) 
=
\boldsymbol{1}+
\sum\limits_{n=0}^{+\infty} {\textstyle\frac{1}{n!}} \sum\limits_{p=0}^{n}
S_{n-p,p}\big(g^{\otimes \, (n-p)} \otimes h^{\otimes \, p}\big)
\\
= \boldsymbol{1}+
\sum \limits_{n,m=0}^{\infty} {\textstyle\frac{1}{(n+m)!}} 
\int \ud^4 x_1 \cdots \ud^4 x_n y_1 \cdots \ud^4 y_m S(x_1, \ldots, x_n,y_1, \ldots, y_m) \,
g(x_1) \cdots g(x_n) h(y_1) \cdots h(y_m), \,\,\, n+m\neq 0
\end{multline*}
\begin{multline*}
S^{-1}(g\mathcal{L}+h\mathbb{A}) 
=
\boldsymbol{1}+
\sum\limits_{n=0}^{+\infty} {\textstyle\frac{1}{n!}} \sum\limits_{p=0}^{n}
\overline{S_{n-p,p}}\big(g^{\otimes \, (n-p)} \otimes h^{\otimes \, p}\big)
\\
= \boldsymbol{1} + 
\sum \limits_{n,m =0}^{\infty} {\textstyle\frac{1}{(n+m)!}} 
\int \ud^4 x_1 \cdots \ud^4 x_n\ud^4 y_1 \cdots \ud^4 y_m \overline{S}(x_1, \ldots, x_n,y_1, \ldots, y_m) \,
g(x_1) \cdots g(x_n)h(y_1) \cdots h(y_m), \,\,\, n+m\neq 0
\end{multline*}
on using the interaction Lagrange density $g\mathcal{L} + h\mathbb{A}$ with two-component
$(g,h)$ ``intensity-of-interaction'' function $(g,h)$ serving as a tool for implementing the causality condition
\[
S\Big((g_1+g_2)\mathcal{L} + (h_1+h_2)\mathbb{A}\Big)
= S\big(g_2\mathcal{L} + h_2\mathbb{A}\big)S\big(g_1\mathcal{L} + h_1\mathbb{A}\big)
\,\,\,\,\,\,\,\,
\textrm{\tiny whenever $\textrm{supp} \, (g_1,h_1) \preceq \textrm{supp} \, (g_2,h_2)$},
\]
with $h$ and $h^{\otimes \, p}$ eventually replaced with Grassmann-valued test function $h$ of grade $1$ and its Grassmann $p$-product 
$h^p$ of grade $p$ in the sense of \cite{Berezin} (compare our previous paper and \cite{Bogoliubov_Shirkov}, 
\cite{Epstein-Glaser}, \cite{DKS1}, \cite{Scharf}, \cite{DutFred}).
The kernel $S(x_1, \ldots, x_n,y) = S_{n,1}(x_1, \ldots, x_n,y)$ of the $n+1$-order term
\[ 
{\textstyle\frac{1}{(n+1)!}} 
\int \ud^4 x_1 \cdots \ud^4 x_n\ud^4 y S(x_1, \ldots, x_n,y) \,
g(x_1) \cdots g(x_n)h(y) 
\]
contribution to $S(g\mathcal{L} + h\mathbb{A})$ into which $h$ enters linearly,
we denote shortly by $S(Z,y)$, using the abbreviated notation $Z$ of Epstein-Glaser 
for the set $\{x_1, \ldots, x_n\}$ of space-time variables. The kernels of the $n$-th order contributions 
\[
{\textstyle\frac{1}{(n)!}} 
\int \ud^4 x_1 \cdots \ud^4 x_n S(x_1, \ldots, x_n) \,
g(x_1) \cdots g(x_n)
\,\,\,
\textrm{and} \,\,\,
{\textstyle\frac{1}{(n)!}} 
\int \ud^4 x_1 \cdots \ud^4 x_n \overline{S}(x_1, \ldots, x_n) \,
g(x_1) \cdots g(x_n)
\]
to $S(g\mathcal{L})$, and respectively to $S^{-1}(g\mathcal{L})$,
we denote simply by $S(x_1, \ldots, x_n) = S(Z)$ and $\overline{S}(x_1, \ldots, x_n)
= \overline{S}(Z)$.

The variational derivative at $h=0$ in definition of interacting fields is understood,
at least initially, only formally at the present stage of the theory and
means nothing else but taking the sum 
\[
\int \Sigma(g,x)h(x) dx
\]
of all terms which are linear in $h$ in the formal expansion for 
$S(g\mathcal{L} +h\mathbb{A})$ and putting the kernel $\Sigma(g,x)$ of this term as the derivative. 
A rigorous definition of the higher order contributions to $S(g,h) = S(g\mathcal{L}+h\mathbb{A})$,
with the creation-annihilation operators understood as the Hida operators, including the case of the Grassmann-valued
test functions $h$, we have treated in our previous paper \cite{CPQEDWN}. 
For a rigorous treatement of the variational derivatives with 
respect to Grassmann-valued test functions $h$, evaluated at $h\neq 0$, where we have to distinguish the left-hand-side
and right-hand-side derivatives, compare \cite{Berezin}.

Note here that $\mathbb{A}$
need not be equal to any one of the elementary free fields of the theory, but it can be equal to any 
local field expressed as a Wick polynomial of free fields. We apply this
construction to the physical observables $\mathbb{A}$ -- the Noether conserved currents, which are Hermitian symmetric
(equal to their Hermitian adjoints $(\, \cdot \, )^+$, \emph{i.e.} the linear transpositions preceded and followed by the complex conjugation operation,
eventually with the additional application of the Gupta-Bleuler operator in case of the theory with gauge fields. 

Performing the formal functional variation we can easily see that
$\mathbb{A}^{(n)}(x_1, \ldots, x_n,x) = \mathbb{A}^{(n)}(Z,x)$
is eqaual
\begin{multline*}
\mathbb{A}^{(n)}(Z,x) =  {\textstyle\frac{1}{i}} \sum \limits_{X \sqcup Y=Z}
\overline{S}(X)S(Y,x),
\\ 
\textrm{\tiny sum over all partitions $X \sqcup Y$ of $Z$ including $X = \emptyset$}
\\
= {\textstyle\frac{1}{i}} \sum' \limits_{X \sqcup Y=Z}
\overline{S}(X)S(Y,x) + S(Z,x),
\\
\textrm{\tiny sum over all partitions $X \sqcup Y$ of $Z$ with $X \neq \emptyset$},
\end{multline*}
compare \cite{DKS1} and \cite{Scharf}.
This is precisely the kernel $A_{(n+1)}(Z,x)$ of Epstein-Glaser, computed for the scattering 
operator $S(g,h) = S(g\mathcal{L}+h\mathbb{A})$, with the support properties 
summarized in \cite{Epstein-Glaser}. In particular, it follows
that $\mathbb{A}^{(n)}(Z,x)$ is equal to the advanced part of the kernel 
$D_{(n+1)}(Z,x)$ in the inductive constrution of the $n+1$-order contribution (defined above)
to the scattering operator $S(g\mathcal{L}+h\mathbb{A})$ , multiplied
by $-i$. More precisely let us introduce after Epstein and Glaser, besides
\[
A'_{(n+1)}(Z,x) = \sum' \limits_{X \sqcup Y=Z}
\overline{S}(X)S(Y,x)
\]
the kernel
\[
R'_{(n+1)}(Z,x) = \sum' \limits_{X \sqcup Y=Z} S(Y,x) \overline{S}(X)
\]
(note that the primed sums run over those partitions which do not include the 
empty set $X=\emptyset$).
Introducing further, after Epstein and Glaser, the kernel
\[
D_{(n+1)}(Z,x) = R'_{(n+1)}(Z,x) - A'_{(n+1)}(Z,x)
\]
we see that
\[
\mathbb{A}^{(n)}(Z,x) = {\textstyle\frac{1}{i}} A_{(n+1)}(Z,x)
\]
where
\[
A_{(n+1)}(Z,x) = \textrm{advanced part} \big[D_{(n+1)}(Z,x)  \big]
\]
in the decomposition 
\[
D_{(n+1)}(Z,x) = R_{(n+1)}(Z,x)-A_{(n+1)}(Z,x) 
\]
of $D_{(n+1)}$ into the advanced $A_{(n+1)}$ and retarded part $R_{(n+1)}(Z,x)$.

In particular for $\mathbb{A}(x) = A^\mu(x)$ equal to the $\mu$-th component 
of electromagnetic potential field we have the following formula
\[
A^{\mu \, (1)}(x_1,x) = {\textstyle\frac{1}{i}} \textrm{advanced part} \big[D_{(2)}(x_1,x)  \big]
\]
for the kernel of the first order contribution to the interacting
field $A_{{}_{\textrm{int}}}^{\mu}(x)$,
where ($\eta$ is the Gupta-Bleuler operator in the total Fock space of the free
$A$ and $\boldsymbol{\psi}$ fields acting trivially in the factor Fock space
of the field $\boldsymbol{\psi}$)
\begin{multline*}
D_{(2)}(x_1,x) = \eta \big(i\mathcal{L}(x_1)\big)^{+} \eta \,\, iA^\mu(x) 
\,\,\,
-
\,\,\,
iA^\mu(x)   \,\,\, \eta \big(i\mathcal{L}(x_1)\big)^{+} \eta 
\\
=
eA^\mu(x) \,  \boldsymbol{{:}} \psi^\sharp(x_1) \gamma_\nu A^\nu(x_1) \psi(x_1) \boldsymbol{{:}}
\,\,\,\,
- 
\,\,\,\,
e \boldsymbol{{:}} \psi^\sharp(x_1) \gamma_\nu  A^\nu(x_1) \psi(x_1) \boldsymbol{{:}} \,  A^\mu(x) 
\\
=
-e[A^\nu(x_1), A^\mu(x) ] \,
\boldsymbol{{:}} \psi^\sharp(x_1) \gamma_\nu \psi(x_1) \boldsymbol{{:}} 
\,\,\,\,\,\,
=
-ie D_{0}(x_1-x) \,
\boldsymbol{{:}} \psi^\sharp(x_1) \gamma_\mu \psi(x_1) \boldsymbol{{:}}. 
\end{multline*}

For $\mathbb{A}(x) = \boldsymbol{\psi}^a(x)$ equal to the $a$-th component 
of Dirac's spinor field we have the following formula
\[
\boldsymbol{\psi}^{a \, (1)}(x_1,x) = {\textstyle\frac{1}{i}} \textrm{advanced part} \big[D_{(2)}(x_1,x)  \big]
\]
for the kernel of the first order contribution to the interacting
field $\boldsymbol{\psi}_{{}_{\textrm{int}}}^{a}(x)$, with
\[
D_{(2)}(x_1,x) = \eta \big(i\mathcal{L}(x_1)\big)^{+} \eta \,\, i\boldsymbol{\psi}^a(x) 
\,\,\,
-
\,\,\,
i\boldsymbol{\psi}^a(x)  \,\,\, \eta \big(i\mathcal{L}(x_1)\big)^{+} \eta 
=
ie S(x-x_1)\gamma_\nu \boldsymbol{\psi}(x_1)A^\nu(x_1).
\]

The computation is reduced to the computation of $S(Y,x) = S_{k,1}(Y,x)$ and $S(X)$ $= S_k(X)$ $=S_{k,0}(X)$, $k\leq n$, and then $D_{(n+1)}(Z,x)$, 
with the only nontrivial step lying in the splitting into advanced and retarded part, 
which can essentially be borrowed from \cite{Scharf} 
(in our previous paper \cite{CPQEDSIS} we have given
a method which reduces this problem to the computation of splittings of scalar distributions equal to the 
scalar coefficients which enter the Wick decomposition of $D_{n}$, together with the 
explicit examples of computations of the splittings in case of QED), 
and  can therefore be omitted. We give only the explicit formulas for the contributions of the first and second order to interacting fields 
(with summation convention and Dirac adjoined bispinor 
$\boldsymbol{\psi}^\sharp(x) = \boldsymbol{\psi}(x)^\sharp = \boldsymbol{\psi}(x)^+\gamma_0$, 
compare \cite{DKS1}, \cite{Scharf}, \cite{DutFred} where $\boldsymbol{\psi}^\sharp(x)$
is denoted by $\overline{\boldsymbol{\psi}}(x)$, and with the 
symbol $\{x_1 \leftrightarrow x_2 \}$ denoting expression immediately preceding it with the variables
$x_1$ and $x_2$ interchanged) 
\[
\boldsymbol{\psi}_{{}_{\textrm{int}}}^{a}(g, x) =
\boldsymbol{\psi}^{a}(x) + \sum \limits_{n=1}^{\infty} \frac{1}{n!}
\int \limits_{\mathbb{R}^{4n}} \ud^4x_1 \cdots \ud^4 x_n \boldsymbol{\psi}^{a \, (n)}(x_1, \ldots, x_n; x)
g(x_1) \cdots g(x_n), 
\]
with
\[
\boldsymbol{\psi}^{a \, (1)}(x_1; x) = 
-e S_{{}_{\textrm{ret}}}^{aa_1}(x-x_1) \gamma^{\nu_1 \, a_1a_2} \boldsymbol{\psi}^{a_2}(x_1)A_{\nu_1}(x_1), 
\]
\begin{multline*}
\boldsymbol{\psi}^{a \, (2)}(x_1, x_2; x) = 
e^2 \Bigg\{ S_{{}_{\textrm{ret}}}^{aa_1}(x-x_1) \gamma^{\nu_1 \, a_1a_2}S_{{}_{\textrm{ret}}}^{a_2a_3}(x_1-x_2)
\gamma^{\nu_2 \, a_3a_4} \,  {:}\boldsymbol{\psi}^{a_4}(x_2)  A_{\nu_1}(x_1)A_{\nu_2}(x_2) {:} \\ 
- S_{{}_{\textrm{ret}}}^{aa_1}(x-x_1) \gamma^{\nu_1 \, a_1a_2} \,
{:} \boldsymbol{\psi}^{a_2}(x_1) 
\boldsymbol{\psi}^{\sharp \, a_3}(x_2) \gamma_{\nu_1}^{a_3a_4} \boldsymbol{\psi}^{a_4}(x_2){:} \,
D^{{}^{\textrm{ret}}}_{0}(x_1-x_2) \\
+S_{{}_{\textrm{ret}}}^{aa_1}(x-x_1) \Sigma_{{}_{\textrm{ret}}}^{a_1a_2}(x_1-x_2)\boldsymbol{\psi}^{a_2}(x_2)
\Bigg\} \,\,\, +  \,\,\, \Bigg\{ x_1 \longleftrightarrow x_2 \Bigg\},
\end{multline*}
\[
\textrm{e. t. c.}
\]
and  
\[
{A_{{}_{\textrm{int}}}}_{\mu}(g, x) =
A_{\mu}(x) + \sum \limits_{n=1}^{\infty} \frac{1}{n!}
\int \limits_{\mathbb{R}^{4n}} \ud^4x_1 \cdots \ud^4 x_n A_{\mu}^{\, (n)}(x_1, \ldots, x_n; x)
g(x_1) \cdots g(x_n),
\]
with
\[
A_{\mu}^{\, (1)}(x_1;x) = -e D^{{}^{\textrm{av}}}_{0}(x_1-x) \,
{:}\boldsymbol{\psi}^{\sharp \, a_1}(x_1) \gamma_{\mu}^{a_1a_2} \boldsymbol{\psi}^{a_2}(x_1){:},
\]
\begin{multline*}
A_{\mu}^{\, (2)}(x_1, x_2; x) = 
e^2 \Bigg\{ 
{:}\boldsymbol{\psi}^{\sharp \, a_1}(x_1) 
\Big( 
 \gamma_{\mu}^{a_1a_2} S_{{}_{\textrm{ret}}}^{a_2a_3}(x_1-x_2) \gamma^{\nu_1 \, a_3a_4}
D^{{}^{\textrm{av}}}_{0}(x_1-x) A_{\nu_1}(x_2) \\
+ \gamma^{\nu_1 \, a_1a_2}S_{{}_{\textrm{av}}}^{a_2a_3}(x_1-x_2) \gamma_{\mu}^{a_3a_4}
D^{{}^{\textrm{av}}}_{0}(x_2-x)A_{\nu_1}(x_1)
\Big)  \boldsymbol{\psi}^{a_4}(x_2){:} \\
+ D^{{}^{\textrm{av}}}_{0}(x_1-x) {\Pi^{{}^{\textrm{av}}}}_{\mu}^{\nu_1}(x_2-x_1)A_{\nu_1}(x_2)
\Bigg\} \,\,\, + \,\,\, \Bigg\{ x_1 \longleftrightarrow x_2 \Bigg\}
\end{multline*}
\[
\textrm{e. t. c.}
\]
where $g$ is the intensity-of-interaction function over space-time which is assumed to be an element of
the ordinary Schwartz space $\mathcal{S}(\mathbb{R}^4; \mathbb{C})$, and which plays a technical role in 
realizing the causality condition in the form we have learned from Bogoliubov and Shirkov
\cite{Bogoliubov_Shirkov}, compare also \cite{Epstein-Glaser}, \cite{DKS1}, \cite{Scharf}, \cite{DutFred}.
This intensity function $g$ modifies the interaction into unphysical in the regions
which lie outside the domain on which $g$ is constant and equal to $1$. It is therefore important
problem to pass to a ``limit'' case of physical interaction with $g=1$ everywhere over the space-time.  

We should note that the only higher order sub-contributions of $\mathbb{A}_{{}_{\textrm{int}}}(x_1, \ldots, x_n,x)$ 
to interacting fields $\mathbb{A}_{{}_{\textrm{int}}}$,  
which have nonzero contribution in the limit $g \rightarrow 1$,  have 
the form of repeated convolution 
of Wick products $W$ of free fields with the 
retarded and advanced parts of the products of pairings which enter the Wick product decomposition 
of the product $\mathcal{L}(x)\mathcal{L}(y)$, when they are $\ud^4x_1, \ldots \ud^4x_n$-integrated. 
Explicit form of these products of parings $D_0(x), \Pi^{\mu}_{\nu}(x), \ldots$ 
and their retarded and advanced parts $D_{0}^{{}^{\textrm{ret}}}(x)$,$D_{0}^{{}^{\textrm{av}}}(x)$, 
${\Pi^{{}^{\textrm{av}}}}_{\mu}^{\nu}(x)$, ${\Pi^{{}^{\textrm{ret}}}}_{\mu}^{\nu}(x), \ldots$,
will be given in the next Section and in our previous paper \cite{CPQEDSIS}. 
We will use the fact that sub-contributions to  
$\mathbb{A}_{{}_{\textrm{int}}}(x_1, \ldots, x_n,x)$
are naturally divided into two classes:
\begin{enumerate}
\item[1)]
those which correspond to ``connected Feynman amputed graphs '' 
which, when $\ud^4x_1, \ldots \ud^4x_n$-integrated, have the form of repeated convolutions
of Wick products $W$ of free fields with the 
retarded and advanced parts of the products of pairings, 
\item[2)]
and those which
correspond to ``disconnected amputed graphs'', and their total sum is identically zero when $\ud^4x_1, \ldots \ud^4x_n$ integrated
with the switching-off function $g^{\otimes \, n}(x_1, \ldots, x_n)$. 
\end{enumerate}

\section{Interacting fields in the adiabatic limit $g\rightarrow 1$. The main theorem}\label{MainTheorem}

In this Section we give the proof of the following
\begin{twr}
Each higher order contribution $\psi_{{}_{\textrm{int}}}^{(n)}$, $A_{{}_{\textrm{int}}}^{(n)}$ in the causal perturbative series 
for the interacting fields $\psi_{{}_{\textrm{int}}}$  in the
Bogoliubov's causal QED's (e.g. spinor QED), with the creation-annihilitation operators
of the free fields understood as the Hida operators, becomes equal to a finite sum of integral kernel
operators with vector-valued kernels in the sense of Obata. 
The values of the kernels are continuous functionals
of $g^{\otimes \, n}$ or, respectively, of $g^{\otimes \, n}\otimes \phi$, where $g$ is the intensity of interaction function $g\in \mathscr{E}$ and $\phi$
is the space-time test function belonging to $\oplus_{1}^{d}\mathscr{E}$. 
The higher order contributions to interacting fields in spinor QED with massive charged field make sense of finite sums of generalized
integral kernel operators with vector-valued kernels even for $g\rightarrow 1$ (in the adiabatic limit) 
in the sense of Obata, if and only if the normalization in the Epstein-Glaser splitting in the construction of the scattering
operator is ``natural''. In this case
the values of the kernels are continuous functionals of $\phi$.
\end{twr}   

Here in QED we have $d=4$ with the free fields $\mathbb{A} = A,\boldsymbol{\psi}$ equal to the free e.m. potential field $A$ or the free
Dirac bispinor field $\boldsymbol{\psi}$, regarded as sums of two integral kernel operators.

Let us denote the kernels $\kappa_{0,1},\kappa_{1,0}$ of
of the free fields $\boldsymbol{\psi}$ and $A$, respectively, by $\kappa_{0,1}^{{}^{1}},\kappa_{1,0}^{{}^{1}}$ and  
$\kappa_{0,1}^{{}^{2}},\kappa_{1,0}^{{}^{2}}$.  


Below we are using the closed subspace $\mathcal{S}^{0}$ of the Schwartz space which consists of the functions
whose all derivatives vanish at zero, and the inverse Fourier transform image $\mathcal{S}^{00}$ of $\mathcal{S}^{0}$.

Recall, that the kernels of Wick products
of free fields, always have the product form, or  pointed product taken at the common space-time point,
depending if the free fields are taken at various independent space-time points or at the cammon space-time point,
or evenually contain both:
\[
\kappa_{l_1m_1}^{{}^{n_1}} \otimes \ldots \otimes \kappa_{l_Mm_M}^{{}^{n_M}},
\,\,\,
\textrm{or}
\,\,\,
\kappa_{l_1m_1}^{{}^{n_1}} \dot{\otimes} \ldots \dot{\otimes} \kappa_{l_Mm_M}^{{}^{n_M}},
\,\,\,
\textrm{or eventually both}
\,\,\,
\kappa_{l_1m_1}^{{}^{n_1}} \otimes \ldots \dot{\otimes} \kappa_{l_Mm_M}^{{}^{n_M}}.
\] 
Recall, that the product of pairings are equal to scalar contractions of the kernels of the Wick products
composed of the interaction Lagrangian operator. It can be seen that all contractions are equal to products
of the basic products of pairings-- the contractions which are present in the Wick product decomposition
formula of the product $\mathcal{L}(x)\mathcal{L}(y)$.  
First, we give a proof of two auxiliary Lemmas, concerned with the kernels which have the form of repeated convolutions 
of the kernels of Wick products of free fields with tempered distributions $d_1, d_2, \dots$
equal to the retarded or advanced parts of the products of pairings. We assume that singularity degree is preserved
during the splitting into retarded and advanced part and the normalization
in the splitting is ``natural''. Definition of the ``natural'' choice of the splitting we are giving in the subsequent 
Subsection.

Namely, we have the following Lemma which allows us to operate with convolutions
of integral kernel operators with tempered distributions $d \in \mathcal{S}(\mathbb{R}^4; \mathbb{C})^*$:

\begin{lem}\label{S*Xi} 
Let $d \in \mathcal{S}(\mathbb{R}^4; \mathbb{C})^*$, and let 
\[
\kappa_{l,m} \in \mathscr{L} \big(\mathscr{E} , \,\, \big(E_{i_1} \otimes \cdots \otimes E_{i_{l+m}} \big)^* \, \big)
\cong \mathscr{L} \big(E_{i_1} \otimes \cdots \otimes E_{i_{l+m}}, \,\, \mathscr{E}^* \big)
\]
with the kernel
\[
\kappa_{l,m} = \Big(\kappa_{l_1,m_1}^{{}^{n_1}} \Big) \overline{\dot{\otimes}} \cdots \overline{\dot{\otimes}} \,\, 
\Big(\kappa_{l_M,m_M}^{{}^{n_M}} \Big)
\]
corresponding to the Wick product (at the same space-time point $x$) 
\[
\Xi_{l,m}(\kappa_{lm}(x)) =
\boldsymbol{}: \Xi_{l_1,m_1}\Big(\kappa_{l_1,m_1}^{{}^{n_1}}(x)\Big) \cdots 
\Xi_{l_M,m_M}\Big(\kappa_{l_M,m_M}^{{}^{n_M}}(x)\big) \boldsymbol{:} 
\]
of the integral kernel operators 
\[
\Xi_{l_k,m_k}\Big(\kappa_{l_k,m_k}^{{}^{n_k}}(x)\Big). 
\]
Let the integral kernel $d \ast \kappa_{l,m}$  be equal
\begin{multline*}
\langle d \ast \kappa_{l,m}(\xi_{i_1} \otimes \cdots \otimes \xi_{i_{l+m}}), \phi \rangle
=  \int \limits_{\mathbb{R}^4} d \ast \kappa_{lm}(\xi_1, \ldots, \xi_{l+m})(x) \, \phi(x) \, \ud^4 x \\
\int \limits_{\mathbb{R}^4 \times \mathbb{R}^4} d(x-y)\kappa_{l,m}(w_{i_1}, 
\ldots, w_{i_{l+m}}; y) \, \xi_{i_1}(w_{i_1}), 
\ldots, \xi_{i_{l+m}}(w_{i_{l+m}}) \, \phi(x) \,
\ud w_{i_1} \cdots \ud w_{i_{l+m}}
\ud^4 y \ud^4 x, \\
\,\,\,\, \xi_{i_k} \in E_{i_k}, \, \phi \in \mathscr{E} = \mathcal{S}(\mathbb{R}^4; \mathbb{}C) 
\,\, \textrm{or} \,\, \mathscr{E} = \mathcal{S}^{00}(\mathbb{R}^4; \mathbb{C}). 
\end{multline*}
Then
\begin{enumerate}
\item[1)]
if the convolution
\[
d_n \ast d_{n-1} \ast \ldots \ast d_1 \ast \kappa_{l,m}
\] 
exists, then it is continuous, \emph{i.e.}
\[
d_n \ast d_{n-1} \ast \ldots \ast d_1 \ast \kappa_{l,m} \in \mathscr{L} \big(E_{i_1} \otimes \cdots \otimes E_{i_{l+m}}, \,\, \mathscr{E}^* \big)
\]
provided 
\[
\kappa_{l,m} = \Big(\kappa_{l_1,m_1}^{{}^{n_1}} \Big) \dot{\otimes} \cdots \dot{\otimes} \,\, \Big(\kappa_{l_M,m_M}^{{}^{n_M}} \Big), \,\, l+m =M,
\]
and each $d_i$ is equal to the product of pairings or to the retarded or advanced part of the causal combinations of products
of pairings and $M>1$, which we encounter as the higher order contributions to interacting fields in spinor QED.

\item[2)]
Let moreover, in case $M=1$, $\kappa_{l_1,m_1}^{{}^{n_1}} = \kappa_{0,1},\kappa_{1,0}$ be equal to the kernel of a free field
with a mass $m_{i_1}$. If further among the distributions $d_n, d_{n-1}, \ldots, d_1$ there are no
(retarded or advanced parts of the) commutation functions of a free field of mass $m_2 = m_{i_1}$, then the convolutions 
\[
d_{n} \ast \ldots \ast d_1 \ast \kappa_{0,1}(\xi), d_{n} \ast \ldots \ast d_1 \ast \kappa_{1,0}(\xi) , \,\,\,\, \xi \in E,
\]
are well-defined and
\[
d_{n} \ast \ldots \ast d_1 \ast \kappa_{0,1}, \, d_{n} \ast \ldots \ast d_1 \ast \kappa_{1,0} \in \mathscr{L} \big(E_{i_1}, \,\, \mathscr{E}^* \big).
\]

\item[3)]
If $\kappa_{l_1,m_1}^{{}^{n_1}} = \kappa_{0,1},\kappa_{1,0}$ is the kernel of a free field
with the mass not equal to the mass of the free field whose commutation function (or its retarded or 
advanced part) is equal $d$ then the convolutions
\[
d \ast \kappa_{0,1}, d \ast \kappa_{1,0} ,  \in \mathscr{L} \big(E_{i_1}^*, \,\, \mathscr{E}^* \big) 
 = \mathscr{L} \big( \mathscr{E}, E_{i_1} \big)
\subset \mathscr{L} \big(E_{i_1}, \mathscr{E}^* \big).
\]
are well-defined.
\item[4)]
If $\kappa_{l_1,m_1}^{{}^{n_1}} = \kappa_{0,1},\kappa_{1,0}$ is the kernel of a free field
with the mass equal to the mass of the free field whose commutation function (or its retarded or 
advanced part) is equal $d$ then the convolutions
\[
d \ast \kappa_{0,1}, d \ast \kappa_{1,0},  
\]
are not well-defined.

\end{enumerate}
\end{lem}

\qedsymbol \,
The map 
\begin{multline*}
E_1 \otimes \ldots \otimes E_{{}_{M}} \ni
\xi_1 \otimes \ldots \xi_{{}{M}} \longmapsto \kappa_{l,m}(\xi_1 \otimes \ldots \xi_{{}{M}}) 
\\
= \Big(\kappa_{l_1,m_1}^{{}^{n_1}} \Big) \dot{\otimes} \cdots \dot{\otimes} \,\, \Big(\kappa_{l_M,m_M}^{{}^{n_M}} \Big)
(\xi_1 \otimes \ldots \xi_{{}{M}})
\\
= \Big(\kappa_{l_1,m_1}^{{}^{n_1}} \Big)(\xi_1) \,\, \cdots  \,\, \Big(\kappa_{l_M,m_M}^{{}^{n_M}} \Big)(\xi_{{}{M}})
\in \mathcal{S}(\mathbb{R}^4)^*
\end{multline*}
is continuous, which easily follows from this product form and from the explicit formula for $\kappa_{0,1}^{{}^{i}},\kappa_{1,0}^{{}^{i}}$,
$i=1,2$. Because, for fixed tempered distributions
$d_1, \ldots, d_n$, the map
\[
E_{i_1} \otimes \cdots \otimes E_{i_{l+m}} \ni \xi \longmapsto d_n \ast \ldots \ast d_1 \ast \kappa_{l,m}(\xi) \in \mathcal{S}(\mathbb{R}^4)^*
\]
is continuous with respect to the ordinary strong dual topology on $\mathcal{S}(\mathbb{R}^4)^*$
whenever the convolution
\[
d_n \ast \ldots \ast d_1 \ast \kappa_{l,m}(\xi)
\] 
is well-defined (Banach-Steinhaus theorem), 
then we need only prove the very existence of the said convolutions. But the asserted existence (eventually non-existence) 
follows immediatelly from the explicit formulas
for these convolutions, for the plane wave kernels
\[
\kappa_{l_k,m_k}^{{}^{n_k}} 
\]
of the free fields (or their derivatives) and for the advanced and retarded parts of the causal combinations $d_1, \ldots, d_n$
of the products of pairings of the free fields 
and with the assumed``natural'' splitting  into retarded and advanced parts
(below in the next Subsection we analyse the existence problem together with the explicit
formulas for these convolutions). 

Indeed, let us show that continuity follows from the very existence of the convolution.

It is sufficient to consider the case $\mathscr{E} = \mathscr{E}_{1} = \mathcal{S}(\mathbb{R}^4; \mathbb{C})$,
because $\mathscr{E}_{1}^{*}$ is continuously embedded into 
$\mathscr{E}_{2}^{*} = \mathcal{S}^{00}(\mathbb{R}^4; \mathbb{C})^{*}$.

Because the Schwartz' algebra $\mathcal{O}'_{C}(\mathbb{R}^4; \mathbb{C})$ of convolutors
of $\mathcal{S}(\mathbb{R}^4; \mathbb{C})^*$
(for definition of $\mathcal{O}'_{C}$ compare e.g. \cite{Schwartz}),
is dense in $\mathcal{S}(\mathbb{R}^4; \mathbb{C})^*$ in the strong dual topology, then for $\epsilon >0$
we can find $d_{i \, \epsilon} \in \mathcal{O}'_{C}$ such that 
\[
\underset{\epsilon \rightarrow 0}{\textrm{lim}}d_{i \, \epsilon} = d_i, \,\,\, i= 1, \ldots, n,
\]
in the strong topology of the dual space $\mathcal{S}(\mathbb{R}^4; \mathbb{C})^*$ of tempered distributions.
Let 
\[
d_\epsilon = d_{n \, \epsilon} \ast \ldots \ast d_{1 \, \epsilon}.
\]
Let $\xi$ be any element of
\[
E_{i_1} \otimes \cdots \otimes E_{i_{l+m}}.
\]
For $\epsilon >0$ we define the following linear operator $\Lambda_\epsilon$
\[
\Lambda_\epsilon(\xi) \overset{\textrm{df}}{=}
d_\epsilon \ast \kappa_{l,m}(\xi), \,\,\,
\xi \in E_{i_1} \otimes \cdots \otimes E_{i_{l+m}},
\]
on 
\[
E_{i_1} \otimes \cdots \otimes E_{i_{l+m}}.
\]
Because $d_\epsilon \in \mathcal{O}'_{C}$, $\epsilon>0$, and because 
\[
\kappa_{l,m} \in \mathscr{L} \big(E_{i_1} \otimes \cdots \otimes E_{i_{l+m}}, \,\, \mathscr{E}^* \big),
\]
then for each $\epsilon >0$ the operator 
\[
\Lambda_\epsilon: E_{i_1} \otimes \cdots \otimes E_{i_{l+m}} \longrightarrow
\mathscr{E}^*
\]
is continuous, i.e. 
\[
\Lambda_\epsilon \in \mathscr{L} \big(E_{i_1} \otimes \cdots \otimes E_{i_{l+m}}, \,\, \mathscr{E}^* \big).
\]

Let $\xi \in E_{i_1} \otimes \cdots \otimes E_{i_{l+m}}$.
By the existence assumption, for each $\xi \in E_{i_1} \otimes \cdots \otimes E_{i_{l+m}}$
\[
\underset{\epsilon \rightarrow 0}{\textrm{lim}} \Lambda_\epsilon(\xi) = 
\underset{\epsilon \rightarrow 0}{\textrm{lim}} S_\epsilon \ast \kappa_{l,m}(\xi) \,\,\,
\]
in strong dual topology of $\mathscr{E}^*$
exists 
\[
\underset{\epsilon \rightarrow 0}{\textrm{lim}} \Lambda_\epsilon(\xi) = S \ast \kappa_{l,m}(\xi),
\]
where the notation $d \ast \kappa_{l,m}(\xi)$,  used for this limit, is rather symbolic
(here we do not pretend to give any deeper justification for this notation).

Because  $E_{i_1} \otimes \cdots \otimes E_{i_{l+m}}$ is a complete Fr\'echet space then by the Banach-Steinhaus
theorem (e.g. Thm. 2.8 of \cite{Rudin}) it follows that
$d \ast \kappa_{l,m}$ is a continuous linear operator 
$E_{i_1} \otimes \cdots \otimes E_{i_{l+m}} \rightarrow \mathscr{E}^{*}$, i.e.
\[
d \ast \kappa_{l,m} \in \mathscr{L} \big(E_{i_1} \otimes \cdots \otimes E_{i_{l+m}}, \,\, \mathscr{E}^* \big).
\]
If $\mathscr{E} = \mathcal{S}^{00}(\mathbb{R}^4; \mathbb{C})$ then $d$ can be extended over to an element of
$\mathcal{S}^{00}(\mathbb{R}^4; \mathbb{C})^*$ (Hahn-Banach theorem), and the above proof can be repeated, 
because the algebra of convolutors of $\mathcal{S}^{00}(\mathbb{R}^4; \mathbb{C})^*$ is dense in 
$\mathcal{S}^{00}(\mathbb{R}^4; \mathbb{C})^*$ and contains $\mathcal{O}_{C}(\mathbb{R}^4; \mathbb{C})$
This completes the proof of continuity. 

\qed

\begin{rem*}
We should emphasize here the fact that the space $E_2$ is equal
\[
\mathcal{S}^{0}(\mathbb{R}^3; \mathbb{C}^4) \neq \mathcal{S}(\mathbb{R}^3; \mathbb{C}^4)
\]
intervenes here non trivially. For the wrong space $\mathcal{S}(\mathbb{R}^3; \mathbb{C}^4)$
used for $E_2$ this Lemma would be false, concernng the existence part, as we will see in the next Subsection,
where the existnce is investigated. 
But this Lemma
is important for the construction of higher order contributions to interacting
fields understood as well-defined integral kernel operators with vector-valued kernels.
Analogue situation we encounter for any other zero mass field for which the corresponding space
$E_2$ must be equal $\mathcal{S}^{0}(\mathbb{R}^3; \mathbb{C}^r)$.
\end{rem*}

From Lemma \ref{S*Xi} and Theorems 3.6 and 3.9 of \cite{obataJFA} (or their generalization to the Fermi case 
and general Fock space) it follows the following

\begin{lem}\label{corS*Xi}
\item[1)]
Let
$d_i$ be equal to the product of pairings or to the retarded or advanced part of the causal combinations of products
of pairings and $M>1$, 
which we encounter as the kernels of the higher order contributions to interacting fields in spinor QED
and with the ``natural'' splitting of the causal distributions in the computation of the scattering operator.
Let us asume that the convolution 
\[
d_n \ast d_{n-1} \ast \ldots \ast d_1 \ast \kappa_{l,m}
\] 
exists.
Then the operator
\begin{multline*}
d_n \ast \ldots \ast d_1 \ast \Xi_{l,m}(\kappa_{l,m})(x) 
\\
= 
\int \limits_{[\mathbb{R}^4]^{\times n}} d_n(x-y_n) d_{n-1}(y_{n}-y_{n-1}) \ldots d_1(y_2-y_1) \Xi_{l,m}\big(\kappa_{l,m}(y_1)\big) \, \ud^4 y_1 \ldots \ud^4y_n \\
= \Xi_{l,m}\Bigg(\int \limits_{[\mathbb{R}^4]^{\times n}} d_n(x-y_n) d_{n}(y_{n-1}-y_{n-2}) \ldots d_1(y_2-y_1) \kappa_{l,m}(y_1) \, \ud^4 y_1 \ldots \ud^4y_n \,  \Bigg)
\\
= \Xi_{l,m}\big( d_n \ast \ldots \ast d_1 \ast \kappa_{lm}(x) \big) 
\end{multline*}
defines integral kernel operator 
\[
\Xi_{l,m}\big( d_n \ast \ldots \ast d_1 \ast \kappa_{lm} \big) 
\in \mathscr{L}\big((\boldsymbol{E}) \otimes \mathscr{E}, \, (\boldsymbol{E})^*\big) \cong
\mathscr{L}\big(\mathscr{E}, \, \mathscr{L}((\boldsymbol{E}), (\boldsymbol{E})^*)\big)
\]
with the vector-valued kernel
\[
d_n \ast \ldots \ast d_1 \ast \kappa_{lm} \in \mathscr{L}\big(\mathscr{E}, \, \big(E_{i_1} \otimes \cdots \otimes E_{i_{l+m}} \big)^* \, \big)
\cong \mathscr{L} \big(E_{i_1} \otimes \cdots \otimes E_{i_{l+m}}, \,\, \mathscr{E}^* \big).
\]
The contributions to interacting fields
whose kernels do not fulfil this condition cancel out and are identically zero for each intensity of interaction function
$g\in \mathcal{S}(\mathbb{R}^4)$, before passing to the limit $g\rightarrow 1$.

\item[2)]
Let moreover, for the higher order contributions, in case $M=1$, $\kappa_{l_1,m_1}^{{}^{n_1}} = \kappa_{0,1},\kappa_{1,0}$ 
be equal to the kernel of a free field
with a mass $m_{i_1}$. If further among the distributions $d_n, d_{n-1}, \ldots, d_1$ there are no
(retarded or advanced parts of the) commutation functions of a free field of mass $m_2 = m_{i_1}$, then
\begin{multline*}
d_n \ast \ldots \ast d_1 \ast \Xi_{0,1}(\kappa_{0,1})(x) 
\\
= 
\int \limits_{[\mathbb{R}^4]^{\times n}} d_n(x-y_n) d_{n-1}(y_{n}-y_{n-1}) \ldots d_1(y_2-y_1) \Xi_{0,1}\big(\kappa_{0,1}(y)\big) \, \ud^4 y_1 \ldots \ud^4 y_n 
\\
= \Xi_{0,1}\Bigg(\int \limits_{[\mathbb{R}^4]^{\times n}} d_n(x-y_n) d_{n-1}(y_{n}-y_{n-1}) \ldots d_1(y_2-y_1)\kappa_{0,1}(y) \, \ud^4 y_1 \ldots \ud^4 y_n \,  \Bigg)
\\
= \Xi_{0,1}\big( d_n \ast \ldots \ast d_1 \ast \kappa_{0,1}(x) \big) 
\end{multline*}
defines integral kernel operator 
\[
\Xi_{0,1}\big( d_n \ast \ldots \ast d_1 \ast \kappa_{lm} \big) 
\in  \mathscr{L}\big((\boldsymbol{E}) \otimes \mathscr{E}, \, (\boldsymbol{E})^*\big) \cong
\mathscr{L}\big(\mathscr{E}, \, \mathscr{L}((\boldsymbol{E}), (\boldsymbol{E})^*)\big)
\]
with the vector-valued kernel
\[
d_n \ast \ldots \ast d_1 \ast \kappa_{0,1} \in \mathscr{L} \big(E_{i_1}, \,\, \mathscr{E}^* \big);
\]
and similarly for the kernel $\kappa_{1,0}$.
The contributions to interacting fields
whose kernels do not fulfil this condition are identically zero,  if we use the ``natural'' splitting of the causal distributions
in the construction of the scattering operator.

\item[3)] 
If  moreover, in case $M=1$, $\kappa_{l_1,m_1}^{{}^{n_1}} = \kappa_{0,1},\kappa_{1,0}$ is the kernel of a free field
with the mass not equal to the mass of the free field whose commutation function (or its retarded or 
advanced part) is equal $d$ then the integral kernel operators
\[
d \ast \Xi_{0,1}(\kappa_{0,1}) = \Xi_{0,1}\big( d \ast \kappa_{0,1} \big), \, 
d \ast \Xi_{1,0}(\kappa_{1,0}) = \Xi_{0,1}\big( d \ast \kappa_{1,0} \big),
\]
are well-defined and
\begin{multline*}
d \ast \Xi_{0,1}(\kappa_{0,1}) = \Xi_{0,1}\big( d \ast \kappa_{0,1} \big), \, 
d \ast \Xi_{1,0}(\kappa_{1,0}) = \Xi_{0,1}\big( d \ast \kappa_{1,0} \big) 
\\
\in \mathscr{L}\big(\mathscr{E}, E_{i_1}\big) = \mathscr{L}\big(E_{i_1}^{*}, \mathscr{E}^* \, \big)
\subset \mathscr{L} \big(E_{i_1}, \,\, \mathscr{E}^* \big).
\end{multline*}

\item[4)]
If $\kappa_{l_1,m_1}^{{}^{n_1}} = \kappa_{0,1},\kappa_{1,0}$ is the kernel of a free field
with the mass equal to the mass of the free field whose commutation function (or its retarded or 
advanced part) is equal $d$ then the integral kernel operators
\[
d \ast \Xi_{0,1}(\kappa_{0,1}) = \Xi_{0,1}\big( d \ast \kappa_{0,1} \big), \, 
d \ast \Xi_{1,0}(\kappa_{1,0}) = \Xi_{0,1}\big( d \ast \kappa_{1,0} \big), 
\]
are not well-defined.
\end{lem}

\subsection{Proof of the main theorem}\label{Proof}

The proof follows by induction and the repeated application of the 
fundamental Lemma \ref{S*Xi} and Lemma \ref{corS*Xi}.

In spinor QED we do not encounter, among the higher order contributions to interacting fields in the adiabatic limit $g=1$, the integral
kernel operators of the form counted as case 4) of Lemmas \ref{S*Xi} and \ref{corS*Xi}. There are however higher order
terms equal to repeated convolutions of the commutation and/or retarded/advanced parts of the causal combinations of the products of pairings
with just one free field operator (case $M=1$ of Lemmas \ref{S*Xi} and \ref{corS*Xi}) which however do not respect the condition
of case 2) of these Lemmas. All contributions of this form to $\boldsymbol{\psi}_{{}_{\textrm{int}}}$ and $A_{{}_{\textrm{int}}}$,
have even  $2k$-th order and have the general form
\begin{equation}\label{LoopTermsINpsi(2k)andA(2k)}
\ldots \underbrace{\big(S_{{}_{\textrm{ret}, \textrm{av}}} \ast \Sigma_{{}_{\textrm{ret}, \textrm{av}}} \ast\big)}_\textrm{$k$ terms} \ldots  \ast \boldsymbol{\psi}, \,\,\, \textrm{and respectively} \,\,\,
\ldots \underbrace{\big(D_{0}^{{}^{\textrm{av}, \textrm{ret}}} \ast \Pi^{{}^{\textrm{av}, \textrm{ret}} \, \nu_k}_{\mu_k} \ast \big)}_\textrm{$k$ terms} \ldots \ast A.
\end{equation}
Here 
\begin{gather*}
\Pi^{\textrm{ret} \, \mu \nu}(x) = C_{2}^{\textrm{ret} \, \mu \nu}(x), \,\,\, \Pi^{\textrm{av} \, \mu \nu}(x) = - C_{2}^{\textrm{ret} \mu \nu}(-x),
\\
C_2(x-y) = \big(\kappa_{0,1}^{1 \, \sharp}  \dot{\otimes}  \gamma^\mu \kappa_{0,1}^{1}\big)
\otimes_2 
\big(\kappa_{1,0}^{1 \, \sharp} \dot{\otimes} \gamma_{\nu}\,\, \kappa_{1,0}^{1}\big)(x,y),
\\
\Sigma^{\textrm{ret}}(x) = K_{2}^{\textrm{ret}}(x), \,\,\, \Sigma^{\textrm{av}}(x) = K_{2}^{\textrm{ret}}(-x),
\\
K_{2}^{ab}(x-y) = \big(\gamma^\mu \,\, \kappa_{0,1}^{1 \, \sharp}  \dot{\otimes} \kappa_{0,1 \,\, \mu}^{2}\big)
\otimes_2 
\big(\gamma^\nu \,\, \kappa_{1,0}^{1 \, \sharp} \dot{\otimes} \kappa_{1,0 \,\, \nu}^{2}\big)(a,x,b,y),
\\
\Upsilon^{\textrm{ret}}(x) = C_{3}^{\textrm{ret}}(x), \,\,\, \Upsilon^{\textrm{av}}(x) = - C_{3}^{\textrm{ret}}(-x),
\\
C_3(x-y) = \big(\kappa_{0,1}^{1 \, \sharp}  \dot{\otimes}  \gamma^\mu \kappa_{0,1}^{1} \dot{\otimes} \kappa_{0,1 \,\, \mu}^{2}\big)
\otimes_3 
\big(\kappa_{1,0}^{1 \, \sharp} \dot{\otimes} \gamma_{\nu} \,\, \kappa_{1,0}^{1} \dot{\otimes} \kappa_{1,0 \,\, \nu}^{2}\big).
\end{gather*}
Here $\otimes_2,\otimes_3$ denote contractions with respect to spin-momentum variables and $\kappa_{1,0 \,\, \mu}^{2}(x) = \kappa_{1,0}^{2}(\mu, x)$,
$\kappa_{0,1 \,\, \mu}^{2}(x) = \kappa_{0,1}^{2}(\mu, x)$ denote the positive and negative frequency kernels of the free
e.m. potential field $A_\mu(x)$, $\kappa_{1,0}^{1}(a,x), \kappa_{1,0}^{1}(a,x)$ the positive and negative frequency kernels of 
the free spinor free field $\boldsymbol{\psi}^{a}(x)$, $\kappa^{1 \, \sharp}_{1,0}(a,x), \kappa^{1 \, \sharp}_{1,0}(a,x)$ 
the positive and negative frequency kernels of 
the Dirac-conjugated spinor free field $\boldsymbol{\psi}^{\sharp \, a}(x)$ and finally 
\[
\gamma^{\mu} \,\, \kappa_{1,0}^{1}(a,x) = \sum\limits_{b} \big[\gamma^{\mu} \big]^{ab} \,\, \kappa_{1,0}^{1}(b,x).
\]
Summations are performed with respect to the repeated Lorentz indices.

For example the third term in the second order contributions to interacting fields $\boldsymbol{\psi}_{{}_{\textrm{int}}}$ and $A_{{}_{\textrm{int}}}$ have, in each case, the form
\begin{equation}\label{LoopTermsINpsi(2)andA(2)}
S_{{}_{\textrm{ret}}} \ast \Sigma_{{}_{\textrm{ret}}} \ast \boldsymbol{\psi}, \,\,\, \textrm{and respectively} \,\,\,
D_{0}^{{}^{\textrm{av}}} \ast \Pi^{{}^{\textrm{av}} \, \nu}_{\mu} \ast A_\nu.
\end{equation}
But here in both these convolution operators we have the retarded/advanced part of the commutation function, $D_0$ or, respectively, $S$,
of the same field with which it is convoluted. However, for the natural choice of the normalization in the causal Epstein-Glaser splitting
of the causal combinations of products of pairing distributions\footnote{Recall that in general the splitting is not unique and contains freedom with the number of the
arbitrary constants depending on the singularity degree at zero of the causal distribution which is to be splitted.} all these
terms, in particular (\ref{LoopTermsINpsi(2)andA(2)}), are well-defined. In fact both terms (\ref{LoopTermsINpsi(2)andA(2)}),
and \emph{a fortiori} all terms (\ref{LoopTermsINpsi(2k)andA(2k)}),
become equal to zero for the standard normalization. This normalization in the choice of the splitting is not arbitrary
and in particular follows if we put the condition that the natural equations of motion for the interacting fields $\boldsymbol{\psi}_{{}_{\textrm{int}}}(g;x)$,
$A_{{}_{\textrm{int}}}(g;x)$ are fulfilled, when regarded perturbatively term by term in the expansion into integral kernel operators
regarded as a series of integral kernel operators in functional powers of the intensity of interaction function $g$,
compare \cite{DKS1}. Here we are giving another justification for the choice of the natural normalization in the splitting procedure:
for any other normalization in the splitting the terms of the type under case 2) of Lemmas \ref{S*Xi} and \ref{corS*Xi} in which the condition put on the mass
in 2) is violated, e.g. (\ref{LoopTermsINpsi(2)andA(2)}) and (\ref{LoopTermsINpsi(2k)andA(2k)}), are no longer well-defined
and the convolutions in such terms become meaningless as generalized integral kernel operators with $g=1$
for the other non-natural normalizations in the splitting. Strictly speaking the existence of (\ref{LoopTermsINpsi(2)andA(2)}),
forcing existence of (\ref{LoopTermsINpsi(2k)andA(2k)}) determines the splitting for the vacuum polarization distribution $\Pi, \Pi^{\textrm{av}, \textrm{av}}$
and self-energy distribution $\Sigma, \Sigma^{\textrm{av}, \textrm{av}}$. In order to fix the splitting in the computation of the  
vacuum graph distribution $\Upsilon$, and its retarded and advanced part, we need to impose the condition of the existence of the adiabatic limit
\[
\underset{g\rightarrow 1}{\textrm{lim}} \big\langle \Phi_0, S_n(g) \Phi_0 \big\rangle
\]
for the  vacuum $\Phi_0$ in the full Fock space of all free fields underlying QED (the free spinor and e.m. potential fields)
and for each $n$-th order contribution $S_n$ to the scattering operator,
compare \cite{Scharf}.

In fact the higher order sub-contributions are of two kinds. To the first kind belong all those sub-contributions, which make sense as
integral kernel operators simply for $g=1$. They correspond to the connected Feynman (amputed) graphs.
To the second kind of $n$-th order sub-contributions
\[
F(x_1, \ldots, x_n, x) \,\,\,
\textrm{in}
\,\,\,
A_{{}_{\textrm{int}}}^{(n)}(x_1, \ldots, x_n, x),
\,\,\,
\boldsymbol{\psi}_{{}_{\textrm{int}}}^{(n)}(x_1, \ldots, x_n, x),
\]
belong the sub-contributions for which we cannot
simply put $g=1$ in
\[
\int F(x_1, \ldots, x_n, x) g(x_1) \ldots g(x_n) \, d^4 x_1 \ldots d^4 x_n,
\]
with the space-time variables $X = \{x_1, \ldots, x_n, x \}$ divided into several disjoint subsets, $X_1, \ldots, X_k$
each separate subset of space-time variables entering as variables of the kernel $\kappa_i(X_i)$ of the corresponding separate factor $F_i(X_i)$ of
the kernels
\[
\kappa(X) = \kappa_1(X_1) \ldots \kappa_k(X_k)
\]
of the contribution
\[
F(X) = F_1(X_1) \ldots F_k(X_k).
\]
The kernels of $F(X)$ are equal to the sums of product of the kernels of the factors $F_1(X_1), \ldots, F_k(X_k)$.
The kernels $\kappa_i(X_i)$ of the factor $F_i(X_i)$, with $X_i = \{x_{i1}, \ldots, x_{ij_n}, x \}$,
containing the space-time variable $x$, when
$d^4x_{i1} \ldots d^4x_{ij_i} = dX_i$-integrated with respect to the corresponding set of space-time variables $\{x_{i1}, \ldots, x_{ij_n} \}$, except $x$,
becomes equal to a convolution
\[
\Bigg(\int \kappa_i(X_i) \, dX_i\Bigg)(x) = S_{i1} \ast \ldots \ast S_{ij_{i}}(x)
\]
with all $S_{i1, \ldots}$ being equal to some tempered distributions (products of advanced or retarded parts of the basic distributions
in differences of the corresponding space-time variables), possibly with the last one $S_{ij_i}$
being a product of the kernels of the free fields in some space-time variables $x_{j_i}$.
Such contributions,
\[
F_1(X_1) \ldots F_{k}(X_{k})
\]
when  $d^4 x_1 \ldots d^4x_n$-integrated with $g=1$,
have kernels equal to the sum of the integrals
\[
\int \kappa_1(X_1) \ldots \kappa_{k}(X_{k}) dX_1 \ldots dX_k
\]
which are not convergent. They correspond to disconnected Feynman (amputed)
graphs. An example of the $3$-rd order sub-contribution
to $A_{{}_{\textrm{int}} \, \mu}$ of the second kind (corresponding to a disconnected graph) would be obtained for $g=1$ by integrating
\begin{equation}\label{TypicalDivergentContribution}
\Upsilon^{\textrm{av}}(x_1-x_3) \,\,  {:}\boldsymbol{\psi}^\sharp \gamma_\mu \boldsymbol{\psi} {:} (x_2) \,\, D_{0}^{\textrm{av}}(x_2-x)
-
\,\,
\Upsilon^{\textrm{av}}(x_3-x_1) \,\, {:}\boldsymbol{\psi}^\sharp \gamma_\mu \boldsymbol{\psi} {:} (x_2) \,\, D_{0}^{\textrm{av}}(x_2-x)
\\
+ \,\,\,\, \ldots
\end{equation}
(where dots represent the sum of similar paired differences for all permutations of the variables $x_2$, $x_2$, $x_3$)
with respect to $d^4x_1 d^4x_2 d^4 x_3$, which, with each term in (\ref{TypicalDivergentContribution})
taken separately,  would of course be divergent (for $g=1$). But for each $g\in \mathcal{S}(\mathbb{R}^4)$
the contribution (\ref{TypicalDivergentContribution}) is identically zero, being equal to the $d^4x_1 d^4x_2 d^4 x_3$-integral
of (\ref{TypicalDivergentContribution}) multiplied by the (symmetric) function
\[
g^{\otimes \, 3}(x_1,x_2, x_3) = g(x_1)g(x_2)g(x_3)
\]
in the space-time variables $x_1,x_2,x_3$, with the pairs of terms in the differences in (\ref{TypicalDivergentContribution})
cancelling out each other.
Because all contributions, with the variables factoring in the similar way into disjoint sets of variables
entering into respective various factors, always enter in pairs cancelling each other, similarly as in (\ref{TypicalDivergentContribution}),
when integrated with  the (symmetric) $g^{\otimes \, n}$, then the contributions of second kind are zero
for each $g \in \mathcal{S}(\mathbb{R}^4)$ and have zero contribution in the limit $g \rightarrow 1$.
Thus all contributions corresponding to disconnected graphs drop out before passing to the limit $g \rightarrow 0$.
However, because of the presence of the sub-contributions of the second kind, we cannot simply state that the interacting fields make
sense (for the ``natural'' splitting) in spinor massive QED simply for $g$ put equal $1$, but instead it is the case for the
limit $g \rightarrow 1$ with $g$ tending to $1$ in the strong dual space $\mathcal{S}(\mathbb{R}^4)^*$.

Below, in this Subsection, we are giving
detailed analysis of all contributions to interacting fields in the adiabatic limit up to order 2,
including (\ref{LoopTermsINpsi(2)andA(2)}), proving that they make sense only
for the ``natural'' splitting, and in this case (\ref{LoopTermsINpsi(2)andA(2)}) are equal to zero, and thus also all
(\ref{LoopTermsINpsi(2k)andA(2k)}) are identically zero. In addition, we will analyze kernels of sub-contributions to interacting fields
in the adiabatic limit with arbitrary
many vacuum polarization loop insertions to interacting fields of odd $n=2k+1$-order (analysis for arbitrary many self-energy loop insertions
is analogous).

Let us give the explicit form of the distributions $d$:
$S_{{}_{\textrm{ret}}}^{a_1a_2}(x_1-x_2)$, $S_{{}_{\textrm{av}}}^{a_1a_2}(x_1-x_2)$,
$g^{\mu \nu}D^{{}^{\textrm{av}}}_{0}(x_1-x_2)$, $g^{\mu \nu}D^{{}^{\textrm{ret}}}_{0}(x_1-x_2)$, $\Sigma_{{}_{\textrm{av}}}^{ab}(x_1-x_2)$, 
$\Sigma_{{}_{\textrm{ret}}}^{ab}(x_1-x_2)$, ${\Pi^{{}^{\textrm{av}}}}_{\mu \nu}(x_1-x_2)$,
${\Pi^{{}^{\textrm{ret}}}}_{\mu \nu}(x_1-x_2)$, $\ldots$
which are present in higher order contributions to interacting fields in spinor QED in the last Theorem,
compare Section \ref{IntFields}. It is important only that they are well-defined
tempered distributions $S$, for which the convolutions of Lemma \ref{S*Xi} exist, and it is important that higher order contributions are
of the form of repeated convolution operation with these tempered distributions (in fact a finite number of such distributions
for each QFT case, here we have spinor QED). 
The said distributions, their retarded and advanced parts are obtained by the Epstein-Glaser splitting into the 
retarded and advanced parts of the causally supported scalar distributions which appear
in the higher order contributions to the scattering
generalized operator under the assumed preservation of singularity degree and the ``natural'' choice of the splitting. 
Explicit computation and explicit form of them
the reader will find in the monograph \cite{Scharf}. 
In our previous paper (\emph{``Causal perturbative QED and white noise. A simplification of the inductive step''}) 
we have given a simple method for the computation of this set of distributions for general causal QFT. 
For the sake of completeness we write here the Fourier transforms of the complete finite 
set of these distributions for the spinor QED and ``natural'' normalization:
\begin{equation}\label{Pi}
\widetilde{{\Pi}_{\mu \nu}}(p) = 
(2\pi)^{-4} \big({\textstyle\frac{p_\mu p_\nu}{p^2}} - g_{\mu\nu}\big) \widetilde{\Pi}(p),
\,\,\,\,\,\,
\widetilde{\Pi}(p) =
{\textstyle\frac{1}{3}} p^4
\int\limits_{4m^2}^{\infty} {\textstyle\frac{s+2m^2}{s^2(p^2-s+i0)}}\sqrt{1-{\textstyle\frac{4m^2}{s}}} ds,
\end{equation}
\begin{equation}\label{Piav}
\widetilde{{\Pi^{{}^{\textrm{av}}}}_{\mu \nu}}(p) = 
(2\pi)^{-4} \big({\textstyle\frac{p_\mu p_\nu}{p^2}} - g_{\mu\nu}\big) \widetilde{\Pi{{}^{\textrm{av}}}}(p),
\,\,\,\,\,\,
\widetilde{\Pi{{}^{\textrm{av}}}}(p) =
{\textstyle\frac{1}{3}} p^4
\int\limits_{4m^2}^{\infty} {\textstyle\frac{s+2m^2}{s^2(p^2-s- \, i \,  p_0 \, 0)}}\sqrt{1-{\textstyle\frac{4m^2}{s}}} ds,
\end{equation}
\begin{equation}\label{Sigma}
\widetilde{\Sigma}(p) = 
(2\pi)^{-4}
\Big\{
\big(1-{\textstyle\frac{m^2}{p^2}}\big)
\big[\textrm{ln} \big|1- {\textstyle\frac{p^2}{m^2}} \big| -i\pi \, \theta(p^2 - m^2)\big]
\, 
\big[ 
m-{\textstyle\frac{\slashed{p}}{4}}\big(1+{\textstyle\frac{m^2}{p^2}}\big)\big]
+ {\textstyle\frac{m^2}{p^2}}{\textstyle\frac{\slashed{p}}{4}}
-{\textstyle\frac{m}{4}}
+{\textstyle\frac{1}{8}}(\slashed{p}-m)
\Big\},
\end{equation}
\begin{multline}\label{Sigmaret}
\widetilde{\Sigma_{{}_{\textrm{ret}}}}(p) = 
(2\pi)^{-4}
\Big\{
\big(1-{\textstyle\frac{m^2}{p^2}}\big) \big[\textrm{ln} \big|1- {\textstyle\frac{p^2}{m^2}} \big| -i\pi \, \textrm{sgn} \, p_0 \,\, \theta(p^2 - m^2)\big]
\, \times 
\\
\times \, 
\big[ 
m - {\textstyle\frac{\slashed{p}}{4}}\big(1+{\textstyle\frac{m^2}{p^2}}\big)
\big]
+ {\textstyle\frac{m^2}{p^2}}{\textstyle\frac{\slashed{p}}{4}}
-{\textstyle\frac{m}{4}}
+{\textstyle\frac{1}{8}}(\slashed{p}-m)
\Big\},
\end{multline}
\begin{multline}\label{Upsilon}
\widetilde{\Upsilon}(p)= \widetilde{\Upsilon''}(p) -\widetilde{\Upsilon'}(p),
\\
\widetilde{\Upsilon''}(p) =
i (2\pi)^{-6} m^4 
\Bigg\{
{\textstyle\frac{5p^4}{48m^4}} + {\textstyle\frac{2p^2}{3m^2}} + 1 
+\big(3- {\textstyle\frac{4m^2}{p^2}}\big) \, \textrm{ln}^2\Big(\sqrt{{\textstyle\frac{-p^2}{4m^2}}} + \sqrt{1-{\textstyle\frac{p^2}{4m^2}}}\Big)
\\
+
\big({\textstyle\frac{p^4}{24m^4}} + {\textstyle\frac{p^2}{12m^2}} + 1\big) \sqrt{1-{\textstyle\frac{4m^2}{p^2}}}
\textrm{ln}{\textstyle\frac{\sqrt{1-{\textstyle\frac{4m^2}{p^2}}}-1}{\sqrt{1-{\textstyle\frac{4m^2}{p^2}}}+1}}
\Bigg\}, 
\end{multline}
\begin{multline*}
\widetilde{\Upsilon'}(p) =
(2\pi)^{-5} \, \theta(p^2 - 4m^2) \, \theta(-p_0) \, \times
\\
\times \,
\Big\{
\big({\textstyle\frac{p^4}{24}} + {\textstyle\frac{m^2}{12}} p^2 + m^4\big) \sqrt{1-{\textstyle\frac{4m^2}{p^2}}}
+ {\textstyle\frac{m^4}{p^2}}(4m^2-3p^2)
\textrm{ln}\Big(\sqrt{{\textstyle\frac{p^2}{4m^2}}} + \sqrt{{\textstyle\frac{p^2}{4m^2}}-1}\Big)
\Big\}. 
\end{multline*}
Here $p^4 = (p \cdot p)^2$, and $p^2 = p\cdot p$ is the Lorentz invariant square with the Feynman slash
notation $\slashed{p} = p_\mu \gamma^\mu$. The formula for $\widetilde{\Upsilon''}$
is valid for time-like $p$ with $p^2>0$, and the general formula follows by analytic continuation. The remaining 
distributions are the standard pairings and commutation functions of the free fields as well as their retarded and advanced parts, 
so there is no necessity to write them explicitly here.

Let us give several explicit formulas for the convolution kernels
\[
d \ast {}^{i}\kappa_{0,1}(\xi), \,\,\,\, d \ast \big[ \kappa_{0,1}^{i} \dot{\otimes} \kappa_{0,1}^{j}(\xi_1 \otimes \xi_2) \big]= 
d \ast \big[ \kappa_{0,1}^{i}(\xi_1) \cdot \kappa_{0,1}^{j}(\xi_2) \big], \,\,\,\, \ldots
\]
where $d$ range over the basic distributions and $\kappa_{0,1}^{i}, \kappa_{1,0}^{i}$ range over the positive and negative energy
plane wave kernels of the free fields in spinor (or various) QED's. Let for example $d$ be equal to the retarded part
\[
S_{\textrm{ret}}(x) = \theta(x)S(x)
=
{\textstyle\frac{1}{(2\pi)^4}}
\int
{\textstyle\frac{m+\slashed{k}}{m^2 - k^2-i\epsilon k_{0}}} e^{-ik\cdot x} \, \ud^4 k
\]
\emph{i.e.} the retarded part of the
(anti)commutation function $S(x)$ for the Dirac conjugated and the Dirac field itself, and let $\kappa_{0,1}^{2}$, $\kappa_{1,0}^{2}$ be the
positive and negative energy plane wave distribution kernels of the free electromagnetic potential field. Then 
\begin{equation}\label{Sret*2kappa01}
S_{\textrm{ret}} \ast \big[\kappa_{0,1}^{2}(\xi)\big](y) = \int S_{\textrm{ret}}(y-x) \,\, \cdot \,\,\, \kappa_{0,1}^{2}(\xi)(\mu, x)  \, \ud^4 x 
=
\int \ud^3 \boldsymbol{\p} {\textstyle\frac{\xi^\mu(\boldsymbol{\p})}{|\boldsymbol{\p}|}}
{\textstyle\frac{m+\slashed{p}}{m^2}} e^{-i|\boldsymbol{\p}|y_0 + i \boldsymbol{\p} \cdot \boldsymbol{\y}},
\end{equation}
\begin{equation}\label{Sret*2kappa10}
S_{\textrm{ret}} \ast \big[\kappa_{1,0}^{2}(\xi)\big](y) = \int S_{\textrm{ret}}(y-x) \,\, \cdot \,\,\, \kappa_{1,0}^{2}(\xi)(\mu, x)  \, \ud^4 x 
=
\int \ud^3 \boldsymbol{\p} {\textstyle\frac{\xi^\mu(\boldsymbol{\p})}{|\boldsymbol{\p}|}}
{\textstyle\frac{m+\slashed{p}}{m^2}} e^{i|\boldsymbol{\p}|y_0 - i \boldsymbol{\p} \cdot \boldsymbol{\y}},
\end{equation}
(up to a constant factor equal to a power of $2\pi$). Similarly, for the negative and positive energy plane wave kernels $\kappa_{0,1}^{1}, \kappa_{1,0}^{1}$ 
defining the free Dirac field and for the negative and positive energy plane wave kernels $\kappa_{0,1}^{2}, \kappa_{1,0}^{2}$
defining the free electromagnetic potential field
and for $\xi_1=\chi, \xi_2=\zeta$, we have
\begin{multline}\label{Sret*[2kappa.1kappa]}
S_{\textrm{ret}} \ast \big[\kappa_{0,1}^{2}(\zeta) \cdot \kappa_{0,1}^{1}(\chi) \big] (y)
= \int S_{\textrm{ret}}(y-x) \,\, \big[ \kappa_{0,1}^{2}(\zeta) \cdot \kappa_{0,1}^{1}(\chi) \big] (x)  \, \ud^4 x
\\
=
\sum\limits_{s}
\int \ud^3 \boldsymbol{\p} \ud^3\boldsymbol{\p} 
{\textstyle\frac{\zeta^\mu(\boldsymbol{\p}')\chi_s(\boldsymbol{\p}) (m+\slashed{p} +\slashed{p'})u_s(\boldsymbol{\p})}{|\boldsymbol{\p}'|(\langle \boldsymbol{\p}' | \boldsymbol{\p} \rangle -|\boldsymbol{\p}'|p_0(\boldsymbol{\p}) )}} 
e^{-i(|\boldsymbol{\p}'|+p_0(\boldsymbol{\p}))y_0 + i (\boldsymbol{\p}'+ \boldsymbol{\p}) \cdot \boldsymbol{\y}}
\end{multline}
\[
p = (p_0(\boldsymbol{\p}), \boldsymbol{\p}), \,\,\, p' = (|\boldsymbol{\p}'|, \boldsymbol{\p}')
\]
with the analogous expressions for the various combinations of the positive and negative energy kernels,
with the corresponding signs in front of the momenta in the numerator of the integrand and eventually with the
Fourier transforms of the positive energy solutions $u_s$ replaced with the Fourier transforms of the negative energy solutions
$v_s$ of the free Dirac equation. Similarly, for the retarded part 
\[
D_{0}^{\textrm{ret}}(x) = \theta(x)D_{0}(x)
=
-{\textstyle\frac{1}{(2\pi)^4}}
\int
{\textstyle\frac{1}{k^2+i\epsilon k_{0}}} e^{-ik\cdot x} \, \ud^4 k
\]
of the massless scalar commutation Pauli-Jordan function, we have the followong convolution kernels (up to a power of $2\pi$)
\begin{equation}\label{D0ret*1kappa01}
D_{0}^{\textrm{ret}} \ast \big[\kappa_{0,1}^{1}(\xi)\big](y) = \int D_{0}^{\textrm{ret}}(y-x) \,\, \cdot\,\,\, \kappa_{0,1}^{1}(\xi)(x)  \, \ud^4 x
 =
-
\sum\limits_{s}\int \ud^3 \boldsymbol{\p} \xi_s(\boldsymbol{\p})
{\textstyle\frac{u_s(\boldsymbol{\p})}{m^2}} e^{-ip_0(\boldsymbol{\p})y_0 + i \boldsymbol{\p} \cdot \boldsymbol{\y}},
\end{equation}
\begin{equation}\label{D0ret*1kappa10}
D_{0}^{\textrm{ret}} \ast \big[{}^{1}\kappa_{1,0}(\xi)\big](y) = \int D_{0}^{\textrm{ret}}(y-x) \,\, \cdot\,\,\, {}^{1}\kappa_{1,0}(\xi)(x)  \, \ud^4 x
 = 
-
\sum\limits_{s}\int \ud^3 \boldsymbol{\p} \xi_s(\boldsymbol{\p})
{\textstyle\frac{u_s(\boldsymbol{\p})}{m^2}} e^{ip_0(\boldsymbol{\p})y_0 - i \boldsymbol{\p} \cdot \boldsymbol{\y}}.
\end{equation}
Similarly
\begin{multline}\label{Sigmaret*1kappa01}
\Sigma_{\textrm{ret}} \ast \big[{}^{1}\kappa_{0,1}(\xi)\big](y) = \int\Sigma_{\textrm{ret}}(y-x) \,\, \cdot\,\,\, {}^{1}\kappa_{0,1}(\xi)(x)  \, \ud^4 x
\\
 =
\sum\limits_{s}\int \ud^3 \boldsymbol{\p} \xi_s(\boldsymbol{\p}) 
\widetilde{\Sigma_{\textrm{ret}}}(p_0(\boldsymbol{\p}),\boldsymbol{\p}) u_s(\boldsymbol{\p})
e^{-ip_0(\boldsymbol{\p})y_0 + i \boldsymbol{\p} \cdot \boldsymbol{\y}},
\end{multline}
\begin{multline}\label{Sigmaret*1kappa10}
\Sigma_{\textrm{ret}} \ast \big[{}^{1}\kappa_{1,0}(\xi)\big](y) = \int\Sigma_{\textrm{ret}}(y-x) \,\, \cdot\,\,\, {}^{1}\kappa_{1,0}(\xi)(x)  \, \ud^4 x
\\
 =
\sum\limits_{s}\int \ud^3 \boldsymbol{\p} \xi_s(\boldsymbol{\p}) 
\widetilde{\Sigma_{\textrm{ret}}}(-p_0(\boldsymbol{\p}),-\boldsymbol{\p}) v_s(\boldsymbol{\p}) 
e^{ip_0(\boldsymbol{\p})y_0 - i \boldsymbol{\p} \cdot \boldsymbol{\y}}.
\end{multline}
\begin{equation}\label{Piav*2kappa01}
\Pi^{\textrm{av}}_{\mu\nu} \ast \big[{}^{2}\kappa_{0,1}(\xi^\nu)\big](y) 
= \int \Pi^{\textrm{av}}_{\mu\nu}(y-x) \,\, \cdot \,\,\, {}^{2}\kappa_{0,1}(\xi^\nu)(\mu, x)  \, \ud^4 x 
=
\int \ud^3 \boldsymbol{\p} {\textstyle\frac{\xi^\nu(\boldsymbol{\p})}{|\boldsymbol{\p}|}}
\widetilde{\Pi^{\textrm{av}}_{\mu\nu}}(|\boldsymbol{\p}|, \boldsymbol{\p}) e^{-i|\boldsymbol{\p}|y_0 + i \boldsymbol{\p} \cdot \boldsymbol{\y}},
\end{equation}
\begin{equation}\label{Piav*2kappa10}
\Pi^{\textrm{av}}_{\mu\nu} \ast \big[{}^{2}\kappa_{1,0}(\xi^\nu)\big](y) = \int \Pi^{\textrm{av}}_{\mu\nu}(y-x) 
\,\, \cdot \,\,\, {}^{2}\kappa_{1,0}(\xi^\nu)(\mu, x)  \, \ud^4 x 
=
\int \ud^3 \boldsymbol{\p} {\textstyle\frac{\xi^\nu(\boldsymbol{\p})}{|\boldsymbol{\p}|}}
\widetilde{\Pi^{\textrm{av}}_{\mu\nu}}(-|\boldsymbol{\p}|, -\boldsymbol{\p}) e^{i|\boldsymbol{\p}|y_0 - i \boldsymbol{\p} \cdot \boldsymbol{\y}}.
\end{equation}
In all these formulas $p_0(\boldsymbol{\p}) = \sqrt{|\boldsymbol{\p}|^2 + m^2}$ with $m$ equal to the mass of the charged field (here the spinor Dirac
field).

Now we note that all the convolution kernels (\ref{Sret*2kappa01})-(\ref{Piav*2kappa10}) are smooth \emph{very slowly increasing} functions
of the space-time variable
$y$, whose derivatives are all bounded, in particular they all belong to the Horv{\'a}th's predual
$\mathcal{O}_C(\mathbb{R}^4)$ of the Schwartz convolutor algebra
$\mathcal{O}'_{C}(\mathbb{R}^4)$ of \emph{rapidly decreasing distributions}.
For definition of $\mathcal{O}_C$ and $\mathcal{O}'_C$,
compare \cite{Horvath}. This easily follows from the explicit expressions
(\ref{Sret*2kappa01})-(\ref{D0ret*1kappa10}) and explicit expressions for the plane wave kernels $\kappa_{0,1}^{i},\kappa_{1,0}^{i}$ 
of the free fields
. Indeed, this follows for (\ref{Sret*2kappa01}) and (\ref{Sret*2kappa10}),
because $\xi \in \mathcal{S}^{0}(\mathbb{R}^3)$ there and the functions $u_s$ and $v_s$ are bounded.
The case (\ref{D0ret*1kappa01}) and (\ref{D0ret*1kappa10}) follows because $\xi \in \mathcal{S}(\mathbb{R}^3)$ there.
The case (\ref{Sigmaret*1kappa01}) and (\ref{Sigmaret*1kappa10}) follows because $\xi \in \mathcal{S}(\mathbb{R}^3)$ there
and $\widetilde{\Sigma_{\textrm{ret}}}(\pm p_0(\boldsymbol{\p}),\pm \boldsymbol{\p})$ is locally integrable function of
$\boldsymbol{\p}$ which grows at most polynomially at infinity and $u_s$
and $v_s$ are bounded functions of $\boldsymbol{\p}$.
The case (\ref{Piav*2kappa01}) and (\ref{Piav*2kappa10}) follows because $\xi \in \mathcal{S}^{0}(\mathbb{R}^3)$ there
and $\widetilde{\Pi^{\textrm{av}}_{\mu\nu}}(\pm|\boldsymbol{\p}|, \pm\boldsymbol{\p})$ is locally integrable function of
$\boldsymbol{\p}$ which grows at most polynomially at infinity.
Finally the case (\ref{Sret*[2kappa.1kappa]}) follows because $\chi \in \mathcal{S}(\mathbb{R}^3)$ and
$\zeta \in \mathcal{S}^{0}(\mathbb{R}^3)$ there, the matrix elements of $(m\pm\slashed{p} \pm\slashed{p'})u_s(\boldsymbol{\p})$,
$(m\pm\slashed{p} \pm\slashed{p'})v_s(\boldsymbol{\p})$ grows not faster than polynomially with $\boldsymbol{\p}$, $\boldsymbol{\p}'$
going to infinity and by using the following elementary estimation
\begin{equation}\label{EstimationForm>0}
\Big|{\textstyle\frac{1}{|\boldsymbol{\p}'|(\langle \boldsymbol{\p}' | \boldsymbol{\p} \rangle -|\boldsymbol{\p}'|p_0(\boldsymbol{\p}) )}}  \Big| <
{\textstyle\frac{4(m+|\boldsymbol{\p}|)}{m^2|\boldsymbol{\p}'|^2}},
\end{equation}
of course applicable only if the charged field is massive with the mass $m \neq 0$.
Therefore, the integrand in (\ref{Sret*[2kappa.1kappa]}) is absolutely integrable, and remains to be absolutely integrable for the
expression analogous to (\ref{Sret*[2kappa.1kappa]}) representing any derivative in $y$ of (\ref{Sret*[2kappa.1kappa]}), similarly as
for (\ref{Sret*2kappa01})-(\ref{D0ret*1kappa10}) and (\ref{Sigmaret*1kappa01})-(\ref{Piav*2kappa10})
.

For example, using the convolution kernels (\ref{Sret*2kappa01})-(\ref{Piav*2kappa10}) we prove that the higher order contributions
(here we restrict ourselves to the first and second order) contributions
to interacting fields are well-defined finite sums of generalized integral kernel operators with vector-vaued kernels in the sense of Obata
\cite{obataJFA}, (particular case of Lemmas \ref{S*Xi} and \ref{corS*Xi}). Indeed, it follows from the existence of the convolutions
with their continuity following immediately from the fact that
\[
\big[ \kappa_{0,1}^{i} \dot{\otimes} \,\,  \kappa_{0,1}^{j}(\xi_i \otimes \xi_j) \big]=
\kappa_{0,1}^{i}(\xi_i) \cdot \kappa_{0,1}^{j}(\xi_j), \ldots
\in \mathcal{O}_C(\mathbb{R}^4) \subset \mathcal{O}_M(\mathbb{R}^4)
\]
(dots stand for this expression with all possible combinations of the positive
$\kappa_{1,0}^{i}$ and negative $\kappa_{0,1}^{i}$ energy plane wave kernels and arbitrary number of factors), 
which depend continuously
on $\xi_i \otimes \xi_j$, $\xi_i \otimes \xi_j \otimes \ldots$, $\ldots$,
and from the following product formula
(for various combinations of the positive $\kappa_{1,0}^{i}$ and negative
$\kappa_{0,1}^{i}$ energy plane wave kernels)
\[
\big[\kappa_{0,1}^{i} \dot{\otimes} \,\, \kappa_{0,1}^{j} \dot{\otimes} \ldots  \dot{\otimes} \,\, {}^{n}\kappa_{0,1}\big] (\xi_i \otimes \xi_j \otimes
\ldots \otimes \xi_n) =
\kappa_{0,1}^{i}(\xi_i) \cdot \kappa_{0,1}^{j}(\xi_j) \cdot \ldots \cdot {}^{n}\kappa_{0,1}(\xi_n),
\]
and the explicit formula for
\[
\kappa_{0,1}^{i}(\xi_i), \kappa_{1,0}^{i}(\xi_i).
\]
Recall that $\xi_i$ in $\kappa_{0,1}^{i}(\xi_i)$ belongs to $\mathcal{S}(\mathbb{R}^3)$ or
$\mathcal{S}^{0}(\mathbb{R}^3)$, respectively, for the massive (Dirac) field ($i=1$) or massless photon field ($i=2$).
The second, and even more important fact, is that the corresponding convolutions, e.g. (\ref{Sret*[2kappa.1kappa]}) exist,
which, using for example existence of  (\ref{Sret*[2kappa.1kappa]}), we  prove that the first
order contribution to the interacting Dirac field in the adiabatic limit $g=1$ is a finite sum
of well-defined integral kernel operators with vector-valued kernels in the sense
of \cite{obataJFA} explained in this and the previous Subsection.
Also, for $\xi_1=\zeta, \xi_2=\chi$ in $E_1 = \mathcal{S}(\mathbb{R}^3)$,
\begin{multline}\label{D0av*[1kappa.1kappa]}
D_{0}^{\textrm{av}} \ast \big[ {}^{1}\kappa_{0,1}^{\sharp}(\zeta) \cdot {}^{1}\kappa_{0,1}(\chi) \big] (y)
= \int D_{0}^{\textrm{av}}(y-x) \,\, \big[ {}^{1}\kappa_{0,1}^{\sharp}(\zeta) \cdot {}^{1}\kappa_{0,1}(\chi) \big] (\mu, x)  \, \ud^4 x
\\
=
\sum\limits_{s,s'}
\int \ud^3 \boldsymbol{\p} \ud^3\boldsymbol{\p}'
{\textstyle\frac{\zeta_{s}(\boldsymbol{\p})v_{s}(\boldsymbol{\p})^{+}\chi_s(\boldsymbol{\p}') u_{s'}(\boldsymbol{\p}')}{((p_0(\boldsymbol{\p})+p_0(\boldsymbol{\p}'))^2
-|\boldsymbol{\p}+\boldsymbol{\p}'|^2 -i\epsilon (p_0(\boldsymbol{\p})+p_0(\boldsymbol{\p}'))}}
e^{-i(p_0(\boldsymbol{\p})+p_0(\boldsymbol{\p}'))y_0 + i (\boldsymbol{\p}+ \boldsymbol{\p}') \cdot \boldsymbol{\y}}
\end{multline}
with the analogous expressions for the various combinations of the positive and negative energy kernels,
with the corresponding signs in front of the momenta  $\pm (p_0(\boldsymbol{\p}),\boldsymbol{\p})$, and  
$\pm (p_0(\boldsymbol{\p}'), \boldsymbol{\p}')$ of the integrand and eventually with the
Fourier transforms of the positive energy solutions $u_s$ replaced with the Fourier transforms of the negative energy solutions
$v_s$ of the free Dirac equation. Therefore, the convolution (\ref{D0av*[1kappa.1kappa]}), in the limit $\epsilon \rightarrow 0^+$,
even with various combinations of the signs in front of the momenta, is a well-defined distribution,
because the function of $(\boldsymbol{\p}, \boldsymbol{\p}')$ in the numerator
belongs to $E_{1}^{\otimes \, 2} = \mathcal{S}(\mathbb{R}^3)^{\otimes \, 2}$. Therefore the first order contribution
to the interacting potential field in the adiabatic limit $g=1$ is equal to a finite sum of integral kernel operators.

As we have already proved (Lemma \ref{S*Xi}), continuity of the convolution kernels follows automatically from the  
very existence. But, as an example, let us prove it independently for the kernels
(\ref{Sret*[2kappa.1kappa]}) defining the first order contribution to the interacting Dirac field given in Section \ref{IntFields}.
In fact, we should insert additional gamma matrix $\gamma^{\mu}$ into (\ref{Sret*[2kappa.1kappa]}) in order to obtain
actually the kernels of the operators which give this first order contribution, which of course does not matter for the continuity
question.   
To this end let $W(B, \varepsilon)$ be a strong zero-neighborhood in $\mathscr{E}^* = \mathcal{S}(\mathbb{R}^4)^*$,
determined by a bounded set $B$ in $\mathscr{E} = \mathcal{S}(\mathbb{R}^4)$ and $\varepsilon >0$.
We will construct zero-neighborhoods $V^0$ and $V$, respectively, in $E_2 = \mathcal{S}^{0}(\mathbb{R}^3)$
and in $E_1 = \mathcal{S}(\mathbb{R}^3)$ such that
\begin{equation}\label{kappapsi(1)(V0,V)inW(B,e)}
S_{\textrm{ret}} \ast \big[ \gamma^{\nu'} \, \kappa_{0,1}^{1}(\chi) \cdot \kappa_{0,1}^{2}(\zeta_{\nu'}) \big] \subset W(B,\varepsilon)
\end{equation}
for all $\zeta \in V^0$ and $\chi \in V$. We have the following estimation for $\phi$ ranging over $B$:
\begin{multline*}
\Big| \big\langle S_{\textrm{ret}} \ast \big[ \gamma^{\nu'} \, \kappa_{0,1}^{1}(\chi) \cdot \kappa_{0,1}^{2}(\zeta_{\nu'}) \big], \phi \big\rangle \Big|
\\
\leq
\sum\limits_{s, \nu'}
\int \ud^3 \boldsymbol{\p} \ud^3\boldsymbol{\p}
{\textstyle\frac{|(m+\slashed{p} +\slashed{p'})\gamma^{\nu'}u_s(\boldsymbol{\p})\widetilde{\phi}(-|\boldsymbol{\p}'|-p_0(\boldsymbol{\p}), -\boldsymbol{\p}'- \boldsymbol{\p})|}{|\boldsymbol{\p}'|(\langle \boldsymbol{\p}' | \boldsymbol{\p} \rangle -|\boldsymbol{\p}'|p_0(\boldsymbol{\p}) )}}
|\zeta_{\nu'}(\boldsymbol{\p}')| \, |\chi_s(\boldsymbol{\p})|
\end{multline*}

\begin{multline*}
\leq
{\textstyle\frac{4}{m^2}}
\sum\limits_{s, \nu'}
\int \ud^3 \boldsymbol{\p} \ud^3\boldsymbol{\p}
|(m+\slashed{p} +\slashed{p'})\gamma^{\nu'}u_s(\boldsymbol{\p})\widetilde{\phi}(-|\boldsymbol{\p}'|-p_0(\boldsymbol{\p}), -\boldsymbol{\p}'- \boldsymbol{\p})|
\,\, \times
\\
\times \,\,
\, (m+|\boldsymbol{\p}|) |\chi_s(\boldsymbol{\p})|
{\textstyle\frac{|\zeta_{\nu'}(\boldsymbol{\p}')|}{|\boldsymbol{\p}'|}}
\,\,
\leq
\,\,
\sum\limits_{s, \nu'} c^{s\nu'}
\int \ud^3 \boldsymbol{\p} (m+|\boldsymbol{\p}|) \, |\chi_s(\boldsymbol{\p})| \,
\int \ud^3\boldsymbol{\p}' {\textstyle\frac{|\zeta_{\nu'}(\boldsymbol{\p}')|}{|\boldsymbol{\p}'|}}
\end{multline*}
\begin{multline*}
=
\sum\limits_{s, \nu'} c^{s\nu'}
\int \ud^3 \boldsymbol{\p}
(1+|\boldsymbol{\p}|^2)^3
\, |\chi_s(\boldsymbol{\p})| \,
{\textstyle\frac{(m+|\boldsymbol{\p}|)}{(1+|\boldsymbol{\p}|^2)^3}}
\int \ud^3\boldsymbol{\p}' {\textstyle\frac{|\zeta_{\nu'}(\boldsymbol{\p}')|}{|\boldsymbol{\p}'|}} (1+|\boldsymbol{\p}'|^2)^2
{\textstyle\frac{1}{(1+|\boldsymbol{\p}'|^2)^2}}
\\
\leq
\sum\limits_{s, \nu'} c^{s\nu'} \Big\| {\textstyle\frac{1}{(1+|\boldsymbol{\p}'|^2)^2}} \Big\|_{1}
\Big\| {\textstyle\frac{(m+|\boldsymbol{\p}|)}{(1+|\boldsymbol{\p}|^2)^3}} \Big\|_{1}
p_0(\zeta_{\nu'}) \, p_1(\chi_s) = \sum\limits_{s, \nu'} c^{s\nu'} c_1 c_2 \,\,  p_0(\zeta_{\nu'})p_1(\chi_s)
\end{multline*}
where in the second inequality we have used the elementary estimation (\ref{EstimationForm>0}).
Here the functions $u_s$ and $\widetilde{\phi}$ are contracted with respect to the spinor index:
$u_s\widetilde{\phi} = \Sigma_a u_{s}^{a}\widetilde{\phi}_a$ and
\[
c^{s\nu'} = \underset{\boldsymbol{\p}',\boldsymbol{\p}' \in \mathbb{R}^3, \phi \in B}{\textrm{sup}}
{\textstyle\frac{4}{m^2}}
|(m+\slashed{p} +\slashed{p'})\gamma^{\nu'}u_s(\boldsymbol{\p})\widetilde{\phi}(-|\boldsymbol{\p}'|-p_0(\boldsymbol{\p}), -\boldsymbol{\p}'- \boldsymbol{\p})|
\]
\[
p_0(\zeta_{\nu'}) = \underset{\boldsymbol{\p}' \in \mathbb{R}^3}{\textrm{sup}}
{\textstyle\frac{|\zeta_{\nu'}(\boldsymbol{\p}')|}{|\boldsymbol{\p}'|}} (1+|\boldsymbol{\p}'|^2)^2
\]
and
\[
p_1(\chi_s) = \underset{\boldsymbol{\p} \in \mathbb{R}^3}{\textrm{sup}}
(1+|\boldsymbol{\p}|^2)^3
\, |\chi_s(\boldsymbol{\p})|
\]
\[
c_1 = \Big\| {\textstyle\frac{1}{(1+|\boldsymbol{\p}'|^2)^2}} \Big\|_{1}, \,\,\,
c_2 = \Big\| {\textstyle\frac{(m+|\boldsymbol{\p}|)}{(1+|\boldsymbol{\p}|^2)^3}} \Big\|_{1},
\]
Therefore, we can put
\[
V^0 = \{\zeta: p_0(\zeta_{\nu'}) < {\textstyle\frac{\sqrt{\varepsilon}}{C}}  \},
\,\,\, V = \{\chi: p_0(\chi_{s}) < {\textstyle\frac{\sqrt{\varepsilon}}{C}}  \},
\,\,\, C = \sum\limits_{s, \nu'} c^{s\nu'} c_1c_2
\]
because $p_0$ is a continuous seminorm on $\mathcal{S}^{0}(\mathbb{R}^3)$,
and it is a well-known fact that
$p_1$  is a continuous seminorm on $\mathcal{S}(\mathbb{R}^3)$.
Because
\[
\Big| \big\langle S_{\textrm{ret}} \ast \big[ \gamma^{\nu'} \, \kappa_{0,1}^{1}(\chi) \cdot \kappa_{0,1}^{2}(\zeta_{\nu'}) \big], \phi \big\rangle \Big|
< \varepsilon
\]
for all $\zeta \in V^0$, $\chi \in V$ and $\phi \in B$, then (\ref{kappapsi(1)(V0,V)inW(B,e)}) follows and the kernel(s)
of the first order contribution to the interacting Dirac field are continuous and belong to
\[
\mathscr{L}(E_1\otimes E_2; \mathscr{E}^*) = \mathscr{L}\big(\mathcal{S}^{0}(\mathbb{R}^3) \otimes \mathcal{S}(\mathbb{R}^3); \mathcal{S}(\mathbb{R}^4)^* \big).
\]

Because we have mass $m>0$ of the charged field (here Dirac field) in the denominator in the above estimations,
coming of course from the elementary estimation (\ref{EstimationForm>0}), then one could perhaps be inclined to think
that the first order contribution to the interacting (Dirac) field is a finite sum of well-defined integral kernel
operators only if the charged field is massive. This conclusion would be false, however, and the mass $m$ in the denominator
is only an artifact of the estimation method, based on (\ref{EstimationForm>0}), which is applicable only if
$m\neq 0$. In fact the first order contributions to interacting fields make sense as sums of generalized integral kernel operators
for $g=1$ also for spinor QED with the massless Dirac field. 

However, for the second and higher order contributions to interacting fields for spinor QED with massless
Dirac field, situation is different. In this massless case the higher order contributions to interacting fields
are meaningless even as the sums of generalized integral kernel operators with vector-valued kernels
in the sense of \cite{obataJFA}.

Let us look at the second order contributions given in Section \ref{IntFields}. In particular, having given
the convolution kernels (\ref{Sret*[2kappa.1kappa]}), with all possible combinations of positive and negative energy plane wave kernels
\[
\begin{split}
S_{\textrm{ret}} \ast \big[ \kappa_{0,1}^{2}(\zeta) \cdot \kappa_{0,1}^{1}(\chi) \big], \,\,
S_{\textrm{ret}} \ast \big[ \kappa_{0,1}^{2}(\zeta) \cdot \kappa_{1,0}^{1}(\chi) \big], \,\,
\\
S_{\textrm{ret}} \ast \big[ \kappa_{1,0}^{2}(\zeta) \cdot \kappa_{0,1}^{1}(\chi) \big], \,\,
S_{\textrm{ret}} \ast \big[ \kappa_{1,0}^{2}(\zeta) \cdot \kappa_{1,0}^{1}(\chi) \big],
\end{split}
\]
we can easily see then, that all convolutions 
\[
\begin{split}
S_{\textrm{ret}} \ast \big[ \kappa_{0,1}^{2}(\zeta) \cdot \kappa_{0,1}^{1}(\chi) \big] \cdot \kappa_{0,1}^{2}(\xi), \,\,
S_{\textrm{ret}} \ast \big[ \kappa_{0,1}^{2}(\zeta) \cdot \kappa_{1,0}^{1}(\chi) \big] \cdot \kappa_{0,1}^{2}(\xi), \,\,
\\
S_{\textrm{ret}} \ast \big[ \kappa_{1,0}^{2}(\zeta) \cdot \kappa_{0,1}^{1}(\chi) \big] \cdot \kappa_{0,1}^{2}(\xi), \,\,
S_{\textrm{ret}} \ast \big[ \kappa_{1,0}^{2}(\zeta) \cdot \kappa_{1,0}^{1}(\chi) \big] \cdot \kappa_{0,1}^{2}(\xi), \,\,
\\
S_{\textrm{ret}} \ast \big[ \kappa_{0,1}^{2}(\zeta) \cdot \kappa_{0,1}^{1}(\chi) \big] \cdot \kappa_{1,0}^{2}(\xi), \,\,
S_{\textrm{ret}} \ast \big[ \kappa_{0,1}^{2}(\zeta) \cdot \kappa_{1,0}^{1}(\chi) \big] \cdot \kappa_{1,0}^{2}(\xi), \,\,
\\
S_{\textrm{ret}} \ast \big[ \kappa_{1,0}^{2}(\zeta) \cdot \kappa_{0,1}^{1}(\chi) \big] \cdot \kappa_{1,0}^{2}(\xi), \,\,
S_{\textrm{ret}} \ast \big[ \kappa_{1,0}^{2}(\zeta) \cdot \kappa_{1,0}^{1}(\chi) \big] \cdot \kappa_{1,0}^{2}(\xi), \,\,
\end{split}
\]
are well-defined. In particular
\begin{multline*}
\Big(S_{\textrm{ret}} \ast \big[ \kappa_{0,1}^{2}(\zeta) \cdot \kappa_{0,1}^{1}(\chi) \big] \cdot \kappa_{0,1}^{2}(\xi)\Big)(y)
\\
=
\sum\limits_{s,s'}
\int \ud^3 \boldsymbol{\p} \ud^3\boldsymbol{\p}' \ud^3\boldsymbol{\p}''
{\textstyle\frac{[m+\slashed{p}+\slashed{p'}+\slashed{p''}]\zeta^\mu(\boldsymbol{\p}')\chi_s(\boldsymbol{\p}) (m+\slashed{p} +\slashed{p'})u_s(\boldsymbol{\p})\xi^\mu(\boldsymbol{\p}'')}{[m^2 - (p+p'+p'')\cdot (p+p'+p'') -i\epsilon(p_0+p'_{0}+p'''_{0})]|\boldsymbol{\p}'|(\langle \boldsymbol{\p}' | \boldsymbol{\p} \rangle -|\boldsymbol{\p}'|p_0(\boldsymbol{\p}) )|\boldsymbol{\p}''|}} \,\, \times
\\
\times \,\,
e^{-i(p+p'+p'')\cdot y}
\end{multline*}
where
\[
p=(p_0,\boldsymbol{\p}) = (p_0(\boldsymbol{\p}), \boldsymbol{\p}), \,\,\, p' = (p'_{0},\boldsymbol{\p}') = (p_0(\boldsymbol{\p}'), \boldsymbol{\p}'), 
\,\,\, p'' = (p''_{0},\boldsymbol{\p}'') = (|\boldsymbol{\p}''|, \boldsymbol{\p}''),
\]
is a well-defined distribution in $y$-variable, because 
the numerator of the integrand, regarded as a function of 
$\boldsymbol{\p} \times \boldsymbol{\p}' \times \boldsymbol{\p}''$,
belongs to 
\[
E_1 \otimes E_2 \otimes E_2 = \mathcal{S}(\mathbb{R}^3) \otimes \mathcal{S}^0(\mathbb{R}^3)  \otimes \mathcal{S}^{0}(\mathbb{R}^3).
\] 
Recall that here $\zeta \in E_2 = \mathcal{S}^{0}(\mathbb{R}^3)$, $\chi \in E_1 = \mathcal{S}(\mathbb{R}^3)$ and $\xi \in E_2 = \mathcal{S}^{0}(\mathbb{R}^3)$. 
Therefore, by the very existence of these convolutions and the continuity proof of Lemma \ref{S*Xi} and Lemma \ref{corS*Xi}
it follows that the first term 
\begin{multline}\label{firstInpsi(2)}
{\textstyle\frac{e^2}{2}} \int\limits_{[\mathbb{R}^4]^{\times 2}} \ud^4 x_1 \ud^4 x_2  \Bigg[ 
 \Big\{S_{{}_{\textrm{ret}}}^{aa_1}(x-x_1) \gamma^{\nu_1 \, a_1a_2}S_{{}_{\textrm{ret}}}^{a_2a_3}(x_1-x_2)
\gamma^{\nu_2 \, a_3a_4} \, {:}\boldsymbol{\psi}^{a_4}(x_2) A_{\nu_1}(x_1)A_{\nu_2}(x_2){:}
\Big\} 
\\
+ \Big\{x_1 \longleftrightarrow x_2 \Big\} 
\Bigg] 
\end{multline}
in the second order contribution $\boldsymbol{\psi}^{(2)}(g=1,x)$ of Section \ref{IntFields}
defines a finite sum of integral kernel operators with vector-valued kernels in the sense of \cite{obataJFA}.

Similarly, because 
the convolution kernels (\ref{D0ret*1kappa01}) and (\ref{D0ret*1kappa10})
\[
D_{0}^{\textrm{ret}} \ast \big[\kappa_{0,1}^{1}(\xi)\big], \,\,
D_{0}^{\textrm{ret}} \ast \big[\kappa_{1,0}^{1}(\xi)\big]
\]
both are smooth functions with each derivative being bounded, then, as is easily seen
\begin{equation}\label{kappaInsecondInpsi(2)}
\begin{split}
D_{0}^{\textrm{ret}} \ast \big[\kappa_{0,1}^{1}(\xi)\big] \cdot \kappa_{0,1}^{1 \, \sharp}(\chi) \cdot \kappa_{0,1}^{1}(\zeta), 
D_{0}^{\textrm{ret}} \ast \big[\kappa_{0,1}^{1}(\xi)\big] \cdot \kappa_{1,0}^{1 \, \sharp}(\chi) \cdot \kappa_{0,1}^{1}(\zeta), \,\,
\\
D_{0}^{\textrm{ret}} \ast \big[\kappa_{1,0}^{1}(\xi)\big] \cdot \kappa_{0,1}^{1 \, \sharp}(\chi) \cdot \kappa_{0,1}^{1}(\zeta), 
D_{0}^{\textrm{ret}} \ast \big[\kappa_{1,0}^{1}(\xi)\big] \cdot \kappa_{1,0}^{1 \, \sharp}(\chi) \cdot \kappa_{0,1}^{1}(\zeta), \,\,
\\
D_{0}^{\textrm{ret}} \ast \big[\kappa_{0,1}^{1}(\xi)\big] \cdot \kappa_{0,1}^{1 \, \sharp}(\chi) \cdot \kappa_{1,0}^{1}(\zeta), 
D_{0}^{\textrm{ret}} \ast \big[\kappa_{0,1}^{1}(\xi)\big] \cdot \kappa_{1,0}^{1 \, \sharp}(\chi) \cdot \kappa_{1,0}^{1}(\zeta), \,\,
\\
D_{0}^{\textrm{ret}} \ast \big[\kappa_{1,0}^{1}(\xi)\big] \cdot \kappa_{0,1}^{1 \, \sharp}(\chi) \cdot \kappa_{1,0}^{1}(\zeta), \,\,
D_{0}^{\textrm{ret}} \ast \big[\kappa_{1,0}^{1}(\xi)\big] \cdot \kappa_{1,0}^{1 \, \sharp}(\chi) \cdot \kappa_{1,0}^{1}(\zeta), \,\,
\end{split}
\end{equation}
all are smooth with each derivative bounded, so allthemore define well-defined distributions in space-time variable.
Therefore by the very existence of these convolutions and the continuity proof of Lemma \ref{S*Xi} and Lemma \ref{corS*Xi}
\begin{multline}\label{secondInpsi(2)}
-{\textstyle\frac{e^2}{2}} \int\limits_{[\mathbb{R}^4]^{\times 2}} \ud^4 x_1 \ud^4 x_2  \Bigg[ 
\Big\{
S_{{}_{\textrm{ret}}}^{aa_1}(x-x_1) \gamma^{\nu_1 \, a_1a_2} \,
{:}\boldsymbol{\psi}^{a_2}(x_1)
\boldsymbol{\psi}^{\sharp \, a_3}(x_2) \gamma_{\nu_1}^{a_3a_4} \boldsymbol{\psi}^{a_4}(x_2){:} \,
D^{{}^{\textrm{ret}}}_{0}(x_1-x_2)
\Big\} 
\\
+ \Big\{x_1 \longleftrightarrow x_2 \Big\} 
\Bigg] 
\end{multline}
in the second order contribution $\boldsymbol{\psi}^{(2)}(g=1,x)$ of Section \ref{IntFields}
defines a finite sum of integral kernel operators with vector-valued kernels in the sense of \cite{obataJFA}. 

In the standard or ``natural'' normalization  in 
the computation of the splitting of the causal distributions, \emph{i.e.} for the standard normalization
of the vacuum polarization distribution $\Pi$ and the self-energy distribution $\Sigma$
it is immediately seen that all the convolution kernels
(\ref{Sigmaret*1kappa01}), (\ref{Sigmaret*1kappa10}) and  
(\ref{Piav*2kappa01}), (\ref{Piav*2kappa10}) are identically zero. Therefore, 
the third term 
\[
{\textstyle\frac{e^2}{2}} \int\limits_{[\mathbb{R}^4]^{\times 2}} \ud^4 x_1 \ud^4 x_2  \Bigg[ 
\Big\{
S_{{}_{\textrm{ret}}}^{aa_1}(x-x_1) \Sigma_{{}_{\textrm{ret}}}^{a_1a_2}(x_1-x_2)\boldsymbol{\psi}^{a_2}(x_2)
\Big\} 
+ \Big\{x_1 \longleftrightarrow x_2 \Big\} 
\Bigg] 
\]
drops out and becomes identically zero, similarly as the third term in the second order contribution to the 
interacting electromagnetic potential field.

We should emphasize here that with any other normalization the convolution kernels (\ref{Sigmaret*1kappa01})-(\ref{Piav*2kappa10})
will be non-zero and in this case the third term (both in the second order contribution to the interacting Dirac field and the electromagnetic
potential field) would not be well-defined as a generalized integral kernel operator. 
In particular the kernels (evaluated at $\xi\in E_1$) of the integral kernel operators
associated to the third term in the second order contribution to the interacting Dirac field 
are equal to the following convolutions
\[
S_{\textrm{ret}} \ast \Sigma_{\textrm{ret}} \ast \big[\kappa_{0,1}^{1}(\xi)\big],
\,\,\, S_{\textrm{ret}} \ast \Sigma_{\textrm{ret}} \ast \big[\kappa_{1,0}^{1}(\xi)\big].
\]
In particular evaluation of the first of these kernels at the space-time test function $\phi \in \mathcal{S}(\mathbb{R}^4)$ would be given by the
limit $\epsilon \rightarrow 0^+$ of the valuation integral (here $p = (p_0(\boldsymbol{\p}), \boldsymbol{\p})$)
\begin{multline*}
\Big\langle S_{\textrm{ret}} \ast \Sigma_{\textrm{ret}} \ast \big[\kappa_{0,1}^{1}(\xi)\big], \phi \Big\rangle
\\
=
\underset{\epsilon\rightarrow0^+}{\textrm{lim}}
\sum\limits_{s}\int \ud^3 \boldsymbol{\p} \ud^4 y \xi_s(\boldsymbol{\p}) 
{\textstyle\frac{
(m+\slashed{p})\widetilde{\Sigma_{\textrm{ret}}}(p_0(\boldsymbol{\p}),\boldsymbol{\p}) u_s(\boldsymbol{\p})}{-i\epsilon p_0(\boldsymbol{\p})}}
e^{-ip_0(\boldsymbol{\p})y_0 + i \boldsymbol{\p} \cdot \boldsymbol{\y}} \phi(y) 
\\
=
\underset{\epsilon\rightarrow0^+}{\textrm{lim}}
\sum\limits_{s}\int \ud^3 \boldsymbol{\p} \xi_s(\boldsymbol{\p}) 
{\textstyle\frac{
(m+\slashed{p})\widetilde{\Sigma_{\textrm{ret}}}(p_0(\boldsymbol{\p}),\boldsymbol{\p}) u_s(\boldsymbol{\p})}{-i\epsilon p_0(\boldsymbol{\p})}}
 \widetilde{\phi}(-p_0(\boldsymbol{\p}), -\boldsymbol{\p})
= +\infty \,\, \textrm{or} \,\, -\infty, 
\end{multline*}
which is meaningless as a value of the vector-valued distribution on the test functions. Difficulty comes of course from the fact that both the
Fourier transforms $\widetilde{S_{\textrm{ret}}}$  and $\widetilde{\kappa_{0,1}^{1}(\xi)}$ are concentrated on the
same orbit $\mathscr{O}_{{}_{\pm m,0,0,0}}$ -- the mass shell -- in the momentum space. In the standard or ``natural'' 
normalization difficulty disappears because in this normalization 
\[
\widetilde{\Sigma_{\textrm{ret}}}(p_0(\boldsymbol{\p}), \boldsymbol{\p}) u_s(\boldsymbol{\p}) = 0 \,\,\, 
\textrm{for each} \,\, \boldsymbol{\p} \in \mathbb{R}^3,
\]
and similarly on the negative energy orbit with the Fourier transforms of the fundamental positive energy solutions $u_s$ replaced
with the negative energy solutions $v_s$,
so that the above valuation limit is well-defined in the standard normalization and is equal zero.
 
We have analogous situation with the third term 
\[
D_{0}^{{}^{\textrm{av}}} \ast \Pi^{{}^{\textrm{av}} \, \mu\nu} \ast A_\nu
\]
in the second order contribution to the interacting electromagnetic potential field,
where both $\widetilde{D_0}$ and $\widetilde{{}^{2}\kappa_{0,1}(\xi)}$ are concentrated on the light cone
in the momentum space. Only with the standard normalization in which
\[
\widetilde{\Pi^{\textrm{av}}_{\mu\nu}}(\pm|\boldsymbol{\p}|, \pm \boldsymbol{\p})
= 0 \,\,\, 
\textrm{for each} \,\, \boldsymbol{\p} \in \mathbb{R}^3,
\]
this term is well-defined and in fact becomes identically zero. For other normalizations in the splitting determining
the vacuum polarization distribution $\Pi$ the third term 
\[
D_{0}^{{}^{\textrm{av}}} \ast \Pi^{{}^{\textrm{av}} \, \mu\nu} \ast A_\nu
\]
in the second order contribution to the interacting electromagnetic potential
field would be meaningless as a generalized integral kernel operator in the adiabatic limit $g=1$. 

Thus we have alredy shown that the full second order contribution $\boldsymbol{\psi}^{(2)}(g=1,x)$
to $\boldsymbol{\psi}_{{}_{\textrm{int}}}(g=1,x)$
is equal to a finite sum of well-defined integral kernel operators with vector-valued kernels in the sense of \cite{obataJFA}
 in spinor QED with massive charged (here Dirac) field and ``natural''
splitting in the construction of the scattering operator.

Proof that the full second order contribution $A^{(2)}(g=1)$ to the interacting potential field  
$A_{{}_{\textrm{int}}}(g=1)$ is equal to a finite sum of well-defined integral kernel operators with vector-valued kernels in the sense of \cite{obataJFA}
is identical.

Note that it does not follow from the general properties of the Schwartz algebra of multipliers and convolutors, \cite{Schwartz} 
,
that the convolution kernels (\ref{Sret*2kappa01})-(\ref{D0ret*1kappa10}) are well-defined tempered distributions, because
\[
\kappa(\xi) = {}^{i}\kappa_{0,1}(\xi_i), \kappa_{1,0}^{i}(\xi_i) \notin \mathcal{O}'_C(\mathbb{R}^4)
\]
although these $\kappa(\xi)$ belong to $\mathcal{O}_C(\mathbb{R}^4)$. It follows immediately from the fact that
the Fourier transforms $\widetilde{\kappa_{0,1}^{i}(\xi_i)}$, $\widetilde{\kappa_{1,0}^{i}(\xi_i)}$, 
are concentrated on the orbit of the field defined by the kernels
$\kappa_{0,1}^{i}$, $\kappa_{1,0}^{i}$.
Therefore by the Schwartz's Fourier exchange theorem 
it follows that $\widetilde{\kappa_{0,1}^{i}(\xi_i)}$, $\widetilde{\kappa_{1,0}^{i}(\xi_i)}$,  
cannot belong to $\mathcal{O}'_C(\mathbb{R}^4)$ because
$\widetilde{\kappa_{0,1}^{i}(\xi_i)}$, 
$\widetilde{\kappa_{1,0}^{i}(\xi_i)} \notin \mathcal{O}_M(\mathbb{R}^4)$. Thus, it is not obvious that
the convolutions (\ref{Sret*2kappa01})-(\ref{D0ret*1kappa10}) exist.

However, the situation for the second and higher order contributions to interacting fields for spinor QED with massless
Dirac field, situation is different,
because the standard normalization in the choice of the splitting,
in the computation of $\Pi^{\mu\nu}$, $\Pi^{\textrm{av} \, \mu\nu}$, $\Pi^{\textrm{ret} \, \mu\nu}$,
$\Sigma^{\mu\nu}$, $\Sigma^{\textrm{av} \, \mu\nu}$, $\Sigma^{\textrm{ret} \, \mu\nu}$,
$\Upsilon^{\mu\nu}$, $\Upsilon^{\textrm{av} \, \mu\nu}$, $\Upsilon^{\textrm{ret} \, \mu\nu}$ cannot be retained.
In particular
\[
\widetilde{\Pi^{\textrm{av} \, \mu\nu}}\Big|_{{}_{p\cdot p =0}}(p) \neq 0,
\]
and for no choice of the splitting we can achieve
\[
\widetilde{\Pi^{\textrm{av} \, \mu\nu}}\Big|_{{}_{p\cdot p =0}}(p) = 0.
\]
Recall that for the massless charged  (here spinor) field the products of parings with positive singularity degree
which we are about to split, have their Fourier transforms which have the singularity of the jump-type discontinuity
on the cone surface $p\cdot p =0$ in the momentum space. Therefore, the central splitting with the normalization point
$p'=0$ cannot be used, compare \cite{Scharf} 
.
By the analysis shown above, the convolution kernels
(\ref{Sigmaret*1kappa01}), (\ref{Sigmaret*1kappa10}) and  
(\ref{Piav*2kappa01}), (\ref{Piav*2kappa10}) cease to be identically zero if $m=0$, and, by the above analysis, the convolutions
\[
D_{0}^{{}^{\textrm{av}}} \ast \Pi^{{}^{\textrm{av}} \, \mu\nu} \ast A_\nu,
\]
equal to the third term in the second order contribution to the interacting electromagnetic potential field,
ceases to be well-defined even as the finite sum of generalized integral kernel operators in the sense
of \cite{obataJFA}. Indeed, let us remind the computation of the splitting in case the charged spinor field is massless,
with $m=0$.
In this case the Fourier transform
of the causal combinations of the product of pairings whose retarded part enters into the
vacuum polarization distribution has the jump singularity $\theta(p^2)$ on the cone in momentum space, and the normalization point
in the computation of this retarded part (compare \cite{Scharf}, pp. 178-181) cannot be chosen at zero.
In particular, we cannot naively put $m=0$ in the formula for the vacuum
polarization in the massive spinor QED in order to compute the vacuum polarization for the massless spinor QED.
Therefore, in QED with massless spinor
field we compute the vacuum polarization using the normalization point $p'$ (denoted by $q$ in \cite{Scharf}, pp. 178-181)
shifted from zero. In particular
for the formula valid in the positive energy cone and in the open domain including it, we use the normalization point
$p'$ somewhere within the positive energy cone in the momentum space. The result is the following
\begin{multline}\label{FTPim=0}
\widetilde{\Pi_{\mu \nu}}(p) =
(2\pi)^{-4} \big({\textstyle\frac{p_\mu p_\nu}{p^2}} - g_{\mu\nu}\big) \widetilde{\Pi}(p),
\\
\widetilde{\Pi}(p) =
{\textstyle\frac{2\pi}{3}} p^2 \theta(p^2) \Bigg[
\textrm{sgn}(p_0)\big[(p-p')\cdot p\big]^3
{\textstyle\frac{i}{2\pi}}
\int\limits_{-\infty}^{+\infty} {\textstyle\frac{t^2 \, dt}{(1-t+i0)(tp^2+p\cdot p' +i0)^3}}
+ \theta(-p_0)
\Bigg],
\end{multline}
and $p'^2>0$ so $p'$ cannot be put equal to zero. The absolute value of this function is everywhere
bounded by a polynomial,
but has the jump singularity $\theta(p^2)$ at the cone $p^2=0$. These properties are preserved
by the splitting, as is easily seen by the dispersion formula
\begin{equation}\label{dispesionFm=0}
\widetilde{d^{\textrm{ret}}}(p) =
\big[(p-p')\cdot p\big]^3
{\textstyle\frac{i}{2\pi}}
\int\limits_{-\infty}^{+\infty} {\textstyle\frac{\widetilde{d}(-tp) \, dt}{(1-t+i0)(tp^2+p\cdot p' +i0)^3}}
\end{equation}
in case $m=0$ (and analogously for the advanced part),
for the splitting of $d$ with singularity degree $\omega=2$ (as is the case for $\Pi_{\mu \nu}$);
compare our previous paper (\emph{``Causal perturbative QED and white noise. A simplification of the inductive step''})
or \cite{Scharf}.

Now we give existence proof the kernels of typical contributions to interacting fields which contain arbitrary many
vacuum polarization loop insertions in the adiabatic limit (simply for $g=1$) in both cases with $m\neq 0$
and in case $m=0$ for spinor QED with the charged spinor free field of mass $m$. Analysis of the kernels
of the contributions containing arbitrary many self-energy loop insertions is completely analogous.

Namely, let us consider the following examples of even $n = 2k+1$-order sub-contributions (with $n$ convolutions $\ast$, summation is performed
with respect to the repeated Lorentz indices)
\[
D_{0}^{{}^{\textrm{av}}} \ast \Pi^{{}^{\textrm{av}} \, \mu_k}_{\mu} \ast \ldots \ast D_{0}^{{}^{\textrm{av}}} \ast \Pi^{{}^{\textrm{av}} \, \nu}_{\mu_1} \ast
D_{0}^{{}^{\textrm{av}}}  \ast {:}\boldsymbol{\psi}^\sharp \gamma_\nu \boldsymbol{\psi}{:},
\]
(containing $k$ loop vacuum polarization graph $\Pi$ insertions) to interacting e.m. potential $A_{{}_{\textrm{int}} \, \mu}(g=1)$
and their kernels equal to the $\epsilon \rightarrow 0$ limits of the kernels of the form (with various combinations of the positive and negative
frequency kernels $\kappa_{1,0}^{\sharp}$, $\kappa_{0,1}^{\sharp}$ of the Dirac conjugated field and $\kappa_{1,0}$,$\kappa_{0,1}$ of the Dirac field
and the corresponding signs $\pm$ in front of the four-momenta $p_1,p_2$)
\begin{multline*}
\Big(D_{0 \, \epsilon}^{{}^{\textrm{av}}} \ast \Pi^{{}^{\textrm{av}} \, \mu_k}_{\mu} \ast \ldots \ast D_{0 \, \epsilon}^{{}^{\textrm{av}}} \ast \Pi^{{}^{\textrm{av}} \, \mu_1}_{\nu} \ast D_{0 \, \epsilon}^{{}^{\textrm{av}}}  \ast \big[\kappa_{\mathpzc{l}_1,\mathpzc{m}_1}^{\sharp}(\xi_1) \gamma^\nu \dot{\otimes} \kappa_{\mathpzc{l}_2,\mathpzc{m}_2}(\xi_2)\big]\Big)(x)
\\
=
\sum\limits_{s_1,s_2} \int \ud^3 \boldsymbol{\p}_1\ud^3 \boldsymbol{\p}_2
{\textstyle\frac{\xi_1(s_1, \boldsymbol{\p}_1)\xi_2(s_2, \boldsymbol{\p}_2) u_{s_1}^{\pm}(\boldsymbol{\p}_1)^\sharp \gamma^\nu u_{s_2}^{\mp}(\boldsymbol{\p}_2)}{[(\pm p_1\pm p_2)^2 +i \epsilon \, (\pm p_{10} \pm p_{20})]^{k+1}}} \,\, \times
\\
\times \,\,
\widetilde{\Pi^{{}^{\textrm{av}} \, \mu_k}_{\mu}}(\pm p_1 \pm p_2) \widetilde{\Pi^{{}^{\textrm{av}} \, \mu_{k-1}}_{\mu_k}}(\pm p_1 \pm p_2)
\ldots \widetilde{\Pi^{{}^{\textrm{av}} \, \mu_{1}}_{\nu}}(\pm p_1+ \pm p_2)
e^{i(\pm p_1\pm p_2)\cdot x}
\end{multline*}
\[
p_1, p_2 \in \mathscr{O}_{{}_{m,0,0,0}} = \{p: \,\,\, p\cdot p = m^2, \, p_0>0  \}.
\]
Here $\xi_1, \xi_2 \in E$ of the single particle Hilbert space of the free Dirac field, i.e. $\xi_1, \xi_2 \in \mathcal{S}(\mathbb{R}^3)$,
$u_{s}^{+}(\boldsymbol{\p}) = u_s(\boldsymbol{\p})$, $u_{s}^{-}(\boldsymbol{\p}) = v_s(\boldsymbol{\p})$ are the Fourier transforms of the basic solutions of the free Dirac equation 
, $\gamma^\mu$ are the Dirac gamma matrices, 
and finally
$u_{s}^{\pm}(\boldsymbol{\p})^\sharp$ is the Dirac conjugation of the spinor $u_{s}^{\pm}(\boldsymbol{\p})$. The plus sign stands everywhere
in $\pm p_1$ and in $u_{s_1}^{\pm}(\boldsymbol{\p}_1)^\sharp$ whenewer $(\mathpzc{l}_1,\mathpzc{m}_1) = (1,0)$. The minus
sign stands everywhere
in $\pm p_1$ and in $u_{s_1}^{\pm}(\boldsymbol{\p}_1)^\sharp$ whenewer $(\mathpzc{l}_1,\mathpzc{m}_1) = (0,1)$. Analogously,
the plus sign stands everywhere
in $\pm p_2$ and minus sign in $u_{s_2}^{\pm}(\boldsymbol{\p}_2)$ whenewer $(\mathpzc{l}_2,\mathpzc{m}_2) = (1,0)$. The minus
sign stands everywhere
in $\pm p_2$ and plus sign in $u_{s_2}^{\pm}(\boldsymbol{\p}_2)$ whenewer $(\mathpzc{l}_2,\mathpzc{m}_2) = (0,1)$.

For the ``natural'' normalization in the Epstein-Glaser splitting, and in case $m\neq 0$, the singularity appearing in the limit
\[
{\textstyle\frac{1}{[p^2+\epsilon p_0]^{k+1}}} \overset{\epsilon\rightarrow 0}{\longrightarrow} {\textstyle\frac{1}{(p^{2})^{k+1}}}
-  \textrm{sgn} \, (p_0) \, {\textstyle\frac{i\pi(-1)^{k}}{k!}}\delta^{(k)}(p^2),
\]
is cancelled by the Fourier transform $\widetilde{\Pi^{\textrm{av} \, \mu\nu}}$ of $\Pi^{\textrm{av} \, \mu\nu}$, as $\widetilde{\Pi^{\textrm{av} \, \mu\nu}}
= (\tfrac{p^\mu p^\nu}{p^2} -g^{\mu\nu}) \widetilde{\Pi}(p)$ with a regular $\widetilde{\Pi}$ in the vicinity of the cone $p^2=0$, and equal there to 
$\widetilde{\Pi}(p) = [p^2]^2g_0(p)$ with still regular $g_0$ there.

This is the case for spinor QED with massive charged field. With the ``natural'' splitting this convolution kernels exist as tempered distributions 
in the space-time variable $x$. Namely we have the following
\begin{lem}
The $\epsilon\rightarrow 0$ limit of the kernels
\begin{multline}\label{++}
\Big(D_{0 \, \epsilon }^{{}^{\textrm{av}}} \ast \Pi^{{}^{\textrm{av}} \, \mu_k}_{\mu} \ast \ldots \ast D_{0 \, \epsilon}^{{}^{\textrm{av}}} \ast \Pi^{{}^{\textrm{av}} \, \mu_1}_{\nu} \ast D_{0 \, \epsilon}^{{}^{\textrm{av}}}  \ast \big[\kappa_{1,0}^{\sharp}(\xi_1) \gamma^\nu \dot{\otimes} \kappa_{1,0}(\xi_2)\big]\Big)(x)
\\
=
\sum\limits_{s_1,s_2} \int \ud^3 \boldsymbol{\p}_1\ud^3 \boldsymbol{\p}_2
{\textstyle\frac{\xi_1(s_1, \boldsymbol{\p}_1)\xi_2(s_2, \boldsymbol{\p}_2) u_{s_1}^{+}(\boldsymbol{\p}_1)^\sharp \gamma^\nu u_{s_2}^{-}(\boldsymbol{\p}_2)}{[(p_1+ p_2)^2 +i\epsilon (p_{10} + p_{20})]^{k+1}}} \,\, \times
\\
\times \,\,
\widetilde{\Pi^{{}^{\textrm{av}} \, \mu_k}_{\mu}}(p_1+p_2) \widetilde{\Pi^{{}^{\textrm{av}} \, \mu_{k-1}}_{\mu_k}}(p_1+ p_2)
\ldots \widetilde{\Pi^{{}^{\textrm{av}} \, \mu_{1}}_{\nu}}(p_1+ p_2)
e^{i(p_1+p_2)\cdot x}
\end{multline}
\begin{multline}\label{--}
\Big(D_{0\, \epsilon}^{{}^{\textrm{av}}} \ast \Pi^{{}^{\textrm{av}} \, \mu_k}_{\mu} \ast \ldots \ast D_{0\, \epsilon}^{{}^{\textrm{av}}} \ast \Pi^{{}^{\textrm{av}} \, \mu_1}_{\nu} \ast D_{0 \, \epsilon}^{{}^{\textrm{av}}}  \ast \big[\kappa_{0,1}^{\sharp}(\xi_1) \gamma^\nu \dot{\otimes} \kappa_{0,1}(\xi_2)\big]\Big)(x)
\\
=
\sum\limits_{s_1,s_2} \int \ud^3 \boldsymbol{\p}_1\ud^3 \boldsymbol{\p}_2
{\textstyle\frac{\xi_1(s_1, \boldsymbol{\p}_1)\xi_2(s_2, \boldsymbol{\p}_2) u_{s_1}^{-}(\boldsymbol{\p}_1)^\sharp \gamma_\nu u_{s_2}^{+}(\boldsymbol{\p}_2)}{[(-p_1- p_2)^2 +i\epsilon (-p_{10} - p_{20})]^{k+1}}} \,\, \times
\\
\times \,\,
\widetilde{\Pi^{{}^{\textrm{av}} \, \mu_k}_{\mu}}(-p_1-p_2) \widetilde{\Pi^{{}^{\textrm{av}} \, \mu_{k-1}}_{\mu_k}}(-p_1-p_2)
\ldots \widetilde{\Pi^{{}^{\textrm{av}} \, \mu_{1}}_{\nu}}(-p_1- p_2)
e^{i(-p_1-p_2)\cdot x}
\end{multline}
\begin{multline}\label{+-}
\Big(D_{0\, \epsilon}^{{}^{\textrm{av}}} \ast \Pi^{{}^{\textrm{av}} \, \mu_k}_{\mu} \ast \ldots \ast D_{0\, \epsilon}^{{}^{\textrm{av}}} \ast \Pi^{{}^{\textrm{av}} \, \mu_1}_{\nu} \ast D_{0 \, \epsilon}^{{}^{\textrm{av}}}  \ast \big[\kappa_{1,0}^{\sharp}(\xi_1) \gamma^\nu \dot{\otimes} \kappa_{0,1}(\xi_2)\big]\Big)(x)
\\
=
\sum\limits_{s_1,s_2} \int \ud^3 \boldsymbol{\p}_1\ud^3 \boldsymbol{\p}_2
{\textstyle\frac{\xi_1(s_1, \boldsymbol{\p}_1)\xi_2(s_2, \boldsymbol{\p}_2) u_{s_1}^{+}(\boldsymbol{\p}_1)^\sharp \gamma_\nu u_{s_2}^{+}(\boldsymbol{\p}_2)}{[(p_1- p_2)^2 +i\epsilon (p_{10} - p_{20})]^{k+1}}} \,\, \times
\\
\times \,\,
\widetilde{\Pi^{{}^{\textrm{av}} \, \mu_k}_{\mu}}(p_1-p_2) \widetilde{\Pi^{{}^{\textrm{av}} \, \mu_{k-1}}_{\mu_k}}(p_1- p_2)
\ldots \widetilde{\Pi^{{}^{\textrm{av}} \, \mu_{1}}_{\nu}}(p_1- p_2)
e^{i(p_1-p_2)\cdot x}
\end{multline}
\begin{multline}\label{-+}
\Big(D_{0\, \epsilon}^{{}^{\textrm{av}}} \ast \Pi^{{}^{\textrm{av}} \, \mu_k}_{\mu} \ast \ldots \ast D_{0\, \epsilon}^{{}^{\textrm{av}}} \ast \Pi^{{}^{\textrm{av}} \, \mu_1}_{\nu} \ast D_{0 \, \epsilon}^{{}^{\textrm{av}}}  \ast \big[\kappa_{0,1}^{\sharp}(\xi_1) \gamma^\nu \dot{\otimes} \kappa_{1,0}(\xi_2)\big]\Big)(x)
\\
=
\sum\limits_{s_1,s_2} \int \ud^3 \boldsymbol{\p}_1\ud^3 \boldsymbol{\p}_2
{\textstyle\frac{\xi_1(s_1, \boldsymbol{\p}_1)\xi_2(s_2, \boldsymbol{\p}_2) u_{s_1}^{-}(\boldsymbol{\p}_1)^\sharp \gamma_\nu u_{s_2}^{-}(\boldsymbol{\p}_2)}{[(-p_1+ p_2)^2 +i\epsilon (-p_{10} + p_{20})]^{k+1}}} \,\, \times
\\
\times \,\,
\widetilde{\Pi^{{}^{\textrm{av}} \, \mu_k}_{\mu}}(-p_1+p_2) \widetilde{\Pi^{{}^{\textrm{av}} \, \mu_{k-1}}_{\mu_k}}(-p_1+ p_2)
\ldots \widetilde{\Pi^{{}^{\textrm{av}} \, \mu_{1}}_{\nu}}(-p_1+ p_2)
e^{i(-p_1+p_2)\cdot x}
\end{multline}
\[
p_1, p_2 \in \mathscr{O}_{{}_{m,0,0,0}}
\]
exist as tempered distributions 
in the space-time variable $x$ 
with the ``natural'' splitting applied in the computation of the vacuum polarization distribution $\Pi^{\textrm{av} \, \mu\nu}$ and with massive
spinor field ($m \neq 0$) in causal perturbative spinor QED with the Hida operators as the creation-annihilation operators. 
\label{kernelsg=1withkVPloopsm>0}
\end{lem}

\qedsymbol \,\,\,
We need to show that for any Schwartz test function $\phi$ of $x$ the kernel, with $\epsilon >0$, integrated with $\phi$ 
has a limit $\epsilon \rightarrow 0$. After  this integration we are left with the integrals which in fact not only exist
in the limit  $\epsilon \rightarrow 0$, but in fact are convergent for  $\epsilon$ simply put equal zero. They are, respectively, of the following form
\begin{multline}\label{++int}
\sum\limits_{s_1,s_2} \int \ud^3 \boldsymbol{\p}_1\ud^3 \boldsymbol{\p}_2
{\textstyle\frac{\xi_1(s_1, \boldsymbol{\p}_1)\xi_2(s_2, \boldsymbol{\p}_2) u_{s_1}^{+}(\boldsymbol{\p}_1)^\sharp \gamma^\nu u_{s_2}^{-}(\boldsymbol{\p}_2)}{[(p_1+ p_2)^2]^{k+1}}} \,\, \times
\\
\times \,\,
\widetilde{\Pi^{{}^{\textrm{av}} \, \mu_k}_{\mu}}(p_1+p_2) \widetilde{\Pi^{{}^{\textrm{av}} \, \mu_{k-1}}_{\mu_k}}(p_1+ p_2)
\ldots \widetilde{\Pi^{{}^{\textrm{av}} \, \mu_{1}}_{\nu}}(p_1+ p_2)
\widetilde{\phi}(p_1+p_2)
\end{multline}
\begin{multline}\label{--int}
\sum\limits_{s_1,s_2} \int \ud^3 \boldsymbol{\p}_1\ud^3 \boldsymbol{\p}_2
{\textstyle\frac{\xi_1(s_1, \boldsymbol{\p}_1)\xi_2(s_2, \boldsymbol{\p}_2) u_{s_1}^{-}(\boldsymbol{\p}_1)^\sharp \gamma^\nu u_{s_2}^{+}(\boldsymbol{\p}_2)}{[(-p_1- p_2)^2]^{k+1}}} \,\, \times
\\
\times \,\,
\widetilde{\Pi^{{}^{\textrm{av}} \, \mu_k}_{\mu}}(-p_1-p_2) \widetilde{\Pi^{{}^{\textrm{av}} \, \mu_{k-1}}_{\mu_k}}(-p_1-p_2)
\ldots \widetilde{\Pi^{{}^{\textrm{av}} \, \mu_{1}}_{\nu}}(-p_1- p_2)
\widetilde{\phi}(-p_1-p_2)
\end{multline}
\begin{multline}\label{+-int}
\sum\limits_{s_1,s_2} \int \ud^3 \boldsymbol{\p}_1\ud^3 \boldsymbol{\p}_2
{\textstyle\frac{\xi_1(s_1, \boldsymbol{\p}_1)\xi_2(s_2, \boldsymbol{\p}_2) u_{s_1}^{+}(\boldsymbol{\p}_1)^\sharp \gamma^\nu u_{s_2}^{+}(\boldsymbol{\p}_2)}{[(p_1- p_2)^2]^{k+1}}} \,\, \times
\\
\times \,\,
\widetilde{\Pi^{{}^{\textrm{av}} \, \mu_k}_{\mu}}(p_1-p_2) \widetilde{\Pi^{{}^{\textrm{av}} \, \mu_{k-1}}_{\mu_k}}(p_1- p_2)
\ldots \widetilde{\Pi^{{}^{\textrm{av}} \, \mu_{1}}_{\nu}}(p_1- p_2)
\widetilde{\phi}(p_1-p_2)
\end{multline}
\begin{multline}\label{-+int}
\sum\limits_{s_1,s_2} \int \ud^3 \boldsymbol{\p}_1\ud^3 \boldsymbol{\p}_2
{\textstyle\frac{\xi_1(s_1, \boldsymbol{\p}_1)\xi_2(s_2, \boldsymbol{\p}_2) u_{s_1}^{-}(\boldsymbol{\p}_1)^\sharp \gamma^\nu u_{s_2}^{-}(\boldsymbol{\p}_2)}{[(-p_1+ p_2)^2]^{k+1}}} \,\, \times
\\
\times \,\,
\widetilde{\Pi^{{}^{\textrm{av}} \, \mu_k}_{\mu}}(-p_1+p_2) \widetilde{\Pi^{{}^{\textrm{av}} \, \mu_{k-1}}_{\mu_k}}(-p_1+ p_2)
\ldots \widetilde{\Pi^{{}^{\textrm{av}} \, \mu_{1}}_{\nu}}(-p_1+ p_2)
\widetilde{\phi}(-p_1+p_2)
\end{multline}
with 
\[
p_i(\boldsymbol{\p}_i) = (p_0(\boldsymbol{\p}_i), \boldsymbol{\p}_i) = (\sqrt{|\boldsymbol{\p}_i|^2 + m^2}, \boldsymbol{\p}_i) \in \mathscr{O}_{{}_{m,0,0,0}} = \{p: p^2=m^2, p_0 >0\},
\]
corresponding, respectively, to the valuation of the kernels (\ref{++}), (\ref{--}), (\ref{+-}), (\ref{-+}) at the space-time
test function $\phi \in \mathcal{S}(\mathbb{R}^4)$. 

We will show that in each case the integrand
\[
{\textstyle\frac{\xi_1(s_1, \boldsymbol{\p}_1)\xi_2(s_2, \boldsymbol{\p}_2) u_{s_1}^{\pm}(\boldsymbol{\p}_1)^+\gamma^\nu u_{s_2}^{\mp}(\boldsymbol{\p}_2)}{[(\pm p_1\pm p_2)^2 ]^{k+1}}} \,\, 
\widetilde{\Pi^{{}^{\textrm{av}} \, \mu_k}_{\mu}}(\pm p_1\pm p_2) \widetilde{\Pi^{{}^{\textrm{av}} \, \mu_{k-1}}_{\mu_k}}(\pm p_1\pm p_2)
\ldots \widetilde{\Pi^{{}^{\textrm{av}} \, \mu_{1}}_{\nu}}(\pm p_1\pm p_2)
\widetilde{\phi}(\pm p_1\pm p_2)
\]
with 
\[
p_1, p_2 \in \mathscr{O}_{{}_{m,0,0,0}},
\]
regarded as a function of $(\boldsymbol{\p}_1, \boldsymbol{\p}_2)$, is absolutely integrable.

In this integrand the function
\begin{equation}\label{f}
f(s_1,s_2, \nu, \boldsymbol{\p}_1, \boldsymbol{\p}_2) = 
\xi_1(s_1, \boldsymbol{\p}_1)\xi_2(s_2, \boldsymbol{\p}_2) u_{s_1}^{\pm}(\boldsymbol{\p}_1)^\sharp \gamma^\nu u_{s_2}^{\mp}(\boldsymbol{\p}_2) \,
\widetilde{\phi}\big(\pm p_1(\boldsymbol{\p}_1)\pm p_2(\boldsymbol{\p}_2)\big)
\end{equation}
is a Schwartz function of $(\boldsymbol{\p}_1, \boldsymbol{\p}_2) \in \mathbb{R}^6$, because $u_{s}^{+} = u_{s}$, $u_{s}^{-} = v_s$ 
are smooth and bounded functions of $\boldsymbol{\p}_i$
and $\widetilde{\phi}\big(p_1(\boldsymbol{\p}_1)+p_2(\boldsymbol{\p}_2)\big)$ is a smooth bounded function of 
$(\boldsymbol{\p}_1, \boldsymbol{\p}_2) \in \mathbb{R}^6$ and both $\xi_1, \xi_2$ are Schwartz functions.

Recall that in the ``natural'' splitting we have (\ref{Piav}),
and in the domain $p^2 < 4m^2$ covering the cone $p^2=0$ in the momentum, we have
\[
\widetilde{{\Pi^{{}^{\textrm{av}}}}_{\mu \nu}}(p) = 
(2\pi)^{-4} \big({\textstyle\frac{p_\mu p_\nu}{p^2}} - g_{\mu\nu}\big) \widetilde{\Pi{{}^{\textrm{av}}}}(p),
\,\,\,\,\,\,
\widetilde{\Pi{{}^{\textrm{av}}}}(p) =
{\textstyle\frac{1}{3}} p^4
\int\limits_{4m^2}^{\infty} {\textstyle\frac{s+2m^2}{s^2(p^2-s)}}\sqrt{1-{\textstyle\frac{4m^2}{s}}} ds,
\]
so that in the domain $p^2 < 4m^2$, $\widetilde{\Pi^{\textrm{av} \ \mu \nu}}$ is smooth, equal to the product of the homogeneous of degree zero tensor of second rank 
\begin{equation}\label{ScaleInv}
(2\pi)^{-4} \big({\textstyle\frac{p_\mu p_\nu}{p^2}} - g_{\mu\nu}\big)
\end{equation}
and a smooth function 
\[
\widetilde{\Pi{{}^{\textrm{av}}}}(p) =
{\textstyle\frac{1}{3}} p^4
\int\limits_{4m^2}^{\infty} {\textstyle\frac{s+2m^2}{s^2(p^2-s)}}\sqrt{1-{\textstyle\frac{4m^2}{s}}} ds,
\,\,\,\, p^2 < 4m^2.
\]
We immediately see that $\widetilde{\Pi{{}^{\textrm{av}}}}(p) = [p^2]^2 g_0(p)$ with $g_0$ which is smooth for $p^2<4m^4$,
and it is easily seen that it is also smooth for $p^2>4m^4$. 

We need to show that $g_0$ stays bounded whenever $p^2 \rightarrow {4m^2}^{\pm}$ in the upper and lower case limit. To this end we need 
to compute the integral for $\widetilde{\Pi{{}^{\textrm{av}}}}$ in (\ref{Piav}) explicitly, using the Euler substitution 
\[
{\textstyle\frac{s}{m^2}} = - {\textstyle\frac{(1-\eta)^2}{\eta}}
\]
with the new integration variable $\eta$. This task has beed done in ref. \cite{Scharf}, p. 205, where in particular it was shown that
$g_0$ is a function of $p^2$, stays locally integrable with at most polynomial growth and limits of $g_0(p)$, whenever $p^2 \rightarrow {4m^2}^{\pm}$, stay bounded. 
In particular the product of any
number of factors $\widetilde{\Pi{{}^{\textrm{av}}}}$ stays locally integrable with at most polynomial growth and the same 
is true for the factor $g_0$.

The convergence of (\ref{++int}) and (\ref{--int}) is now trivial because for $p_1,p_2 \in \mathscr{O}_{}{}_{m,0,0,0}$
\[
(p_1+p_2)^2 = (-p_1-p_2)^2 > 2m^2.
\]
Also the convergence of all (\ref{+-int}) and (\ref{-+int}) becomes now trivial for $k>1$ in (\ref{+-int}) and (\ref{-+int}).
Indeed, because
\[
\Big({\textstyle\frac{p_\mu p^\nu}{p^2}} - \delta_{\mu}^{\nu} \Big) 
\Big({\textstyle\frac{p_\nu p^\sigma}{p^2}} - \delta_{\nu}^{\sigma} \Big)
=
(-1)
\Big({\textstyle\frac{p_\mu p^\sigma}{p^2}} - \delta_{\mu}^{\sigma} \Big)
\]
then
\[
\widetilde{\Pi^{{}^{\textrm{av}} \, \mu_k}_{\mu}}(p) \widetilde{\Pi^{{}^{\textrm{av}} \, \mu_{k-1}}_{\mu_k}}(p)
\ldots \widetilde{\Pi^{{}^{\textrm{av}} \, \mu_{1}}_{\nu}}(p)
=
(-1)^{k+1}
\Big({\textstyle\frac{p_\mu p_\nu}{p^2}} - g_{\mu \nu} \Big) 
\big[\widetilde{\Pi^{{}^{\textrm{av}}}}(p)\big]^k.
\]
The integrand of (\ref{+-int}) takes the form
\[
{\textstyle\frac{\xi_1(s_1, \boldsymbol{\p}_1)\xi_2(s_2, \boldsymbol{\p}_2) u_{s_1}^{+}(\boldsymbol{\p}_1)^+\gamma_\nu u_{s_2}^{+}(\boldsymbol{\p}_2)}{[(p_1- p_2)^2 ]^{k+1}}} \,\, 
(-1)^{k+1}
\Big({\textstyle\frac{(p_1-p_2)_\mu (p_1-p_2)_\nu}{(p_1-p_2)^2}} - g_{\mu \nu} \Big) 
\big[\widetilde{\Pi^{{}^{\textrm{av}}}}(p_1 - p_2)\big]^k
\widetilde{\phi}(p_1 - p_2)
\]
and in this case with $k>1$ we can cancel just one $(p_1 -p_2)^2$ in each factor $\widetilde{\Pi^{{}^{\textrm{av}}}}(p_1 -p_2)$ with one $(p_1 -p_2)^2$ in the denominator,
and in two of the factors $\widetilde{\Pi^{{}^{\textrm{av}}}}(p_1 -p_2)$ another one $(p_1 -p_2)^2$ can be cancelled respectively with the last one coming from 
the photon propagator in the denominator and in the denominator in the factor
\begin{equation}\label{ScaleInv(p1-p2)}
\Big({\textstyle\frac{(p_1-p_2)_\mu (p_1-p_2)_\nu}{(p_1-p_2)^2}} - g_{\mu \nu} \Big),
\end{equation}
and still we are left with a locally integrable function and absolutely integrable function (being equal to the product of a Schwartz function and a locally integrable function going to infinity not faster than polynomially). Analysis of the case (\ref{-+int}) with $k>1$ is of course identical.

For the proof of convergence of (\ref{+-int}) and (\ref{-+int}) with $k=1$
 we need to  analyze more carefully the function 
\[
{\textstyle\frac{(p_1 -p_2)_\mu (p_1 -p_2)_\nu}{(p_1 -p_2)^2}}, \,\,\, p_1, p_2 \in \mathscr{O}_{{}_{m,0,0,0}},
\]
as a function of $(\boldsymbol{\p}_1, \boldsymbol{\p}_1) \in \mathbb{R}^6$. In general, for any four-momenta $p$,
the homogeneous of degree zero function (\ref{ScaleInv}) is not locally bounded, streaming, in general, to infinity when
$p^2 \rightarrow 0$. Similarly, the function (\ref{ScaleInv(p1-p2)}), regarded as a function of two unrestricted
four-momenta $(p_1,p_2)$ is not locally bounded when $(p_1-p_2)^2$ approaches zero. But here comes the essential point:
if the four-momenta
$(p_1,p_2)$ are restricted to the positive energy hyperboloid $\{p:p^2=m^2, p_0>0\} = \mathscr{O}_{{}_{m,0,0,0}}$,
\emph{i.e.} $p_i$ are functions, respectively, of the corresponding spatial momenta $\boldsymbol{\p}_i$,
such that
\[
p_i(\boldsymbol{\p}_i) = \big(\sqrt{|\boldsymbol{\p}_i|^2 + m^2}, \boldsymbol{\p}_i \big) \in \mathscr{O}_{{}_{m,0,0,0}},
\]
then (\ref{ScaleInv(p1-p2)}), regarded as a function of the spatial momenta $(\boldsymbol{\p}_1, \boldsymbol{\p}_2) \in \mathbb{R}^6$,
becomes locally integrable of at most polynomial growth in the variables $(\boldsymbol{\p}_1, \boldsymbol{\p}_2) \in \mathbb{R}^6$.
And this is the case only if $m \neq 0$.

In order to show it, we first need to prove the estimation
\begin{equation}\label{estimation1}
0 \leq {\textstyle\frac{|\boldsymbol{\p}_1-\boldsymbol{\p}_2|^2}{-(p_1-p_2)^2}}
\leq {\textstyle\frac{|\boldsymbol{\p}_1|^2+|\boldsymbol{\p}_2|^2 + 2m^2}{2m^2}}
\end{equation}
for $p_1,p_2 \in \mathscr{O}_{{}_{m,0,0,0}}$ and for
\[
{\textstyle\frac{|\boldsymbol{\p}_1-\boldsymbol{\p}_2|^2}{-(p_1-p_2)^2}}
\]
regarded as the function of the spatial momenta
$(\boldsymbol{\p}_1, \boldsymbol{\p}_2)$.
As is easily checked, $(p_1-p_2)^2 \leq 0$ for all $(\boldsymbol{\p}_1, \boldsymbol{\p}_2)$
and $(p_1-p_2)^2 = 0$ if and only if
$\boldsymbol{\p}_1-\boldsymbol{\p}_2 =0$. Thus zeros of $(p_1-p_2)^2$ coincide with the diagonal linear subspace
$\{(\boldsymbol{\p}, \boldsymbol{\p}), \boldsymbol{\p} \in \mathbb{R}^3\}$ in $\mathbb{R}^3 \times \mathbb{R}^3$.
Next we investigate the left and the right hand side functions
\[
{\textstyle\frac{|\boldsymbol{\p}_1-\boldsymbol{\p}_2|^2}{-(p_1-p_2)^2}}
\,\,\,\,
\textrm{and}
\,\,\,\,
{\textstyle\frac{|\boldsymbol{\p}_1|^2+|\boldsymbol{\p}_2|^2 + 2m^2}{2m^2}}
\]
along the straight lines perpendicular to the diagonal (in
 $\mathbb{R}^3 \times \mathbb{R}^3 = \mathbb{R}^6$ regarded as the Euclidean space) and use the fact that the right
hand side function is rotationally invariant in $\mathbb{R}^3 \times \mathbb{R}^3 = \mathbb{R}^6$. Elementary application
of the de l'Hospital rule shows that the limit value of the left hand side function along the straight line perpendicular
to the diagonal at $(\boldsymbol{\p}, \boldsymbol{\p})$, when it reaches the diagonal point $(\boldsymbol{\p}, \boldsymbol{\p})$,
is finite. The limit value when the perpendicular straight line reaches the point $(\boldsymbol{\p}, \boldsymbol{\p})$
depends on the direction $(\boldsymbol{\n}, -\boldsymbol{\n})$ of the perpendicular straight line, and its maximum value
\[
{\textstyle\frac{|\boldsymbol{\p}|^2 +m^2}{m^2}}
\]
is reached when the direction of $\boldsymbol{\n}$ coincides with the direction of $\boldsymbol{\p}$, its minimal value $1$
is reached when the direction $\boldsymbol{\n}$ is perpendicular to $\boldsymbol{\p}$. Comparison of the left and right-hand side functions
along the lines perpendicular to the diagonal, and rotational symmetry of the right-hand side function, shows
the validity of the estimation (\ref{estimation1}). Alternatively we can use the following elementary inequality
\[
-m^2 + \sqrt{ \mathpzc{x}^2 + m^2} \sqrt{ \mathpzc{y}^2 + m^2} -  \lambda  \mathpzc{x} \mathpzc{y}  
\geq  {\textstyle\frac{m^2 \big( \mathpzc{x}^2 + \mathpzc{y}^2 - 2 \lambda \mathpzc{x} \mathpzc{y}  \big)}{\mathpzc{x}^2+\mathpzc{y}^2 + 2m^2}},
\]
valid for real $\mathpzc{x},\mathpzc{y}$, $m, \lambda$ in the range  $-\infty < \mathpzc{x},\mathpzc{y},m < +\infty$,
and $-1 \leq \lambda \leq 1$. In order to prove this inequality, we multiply both sides by the denominator.
Then, after transferring the terms without the root to the right-hand side and squaring,
we reduce the above inequality to the following inequality for the polynomial  
\begin{multline*}
Q(\mathpzc{x},\mathpzc{y}) =
( \mathpzc{x}^2+\mathpzc{y}^2 + 2m^2)^2 ( \mathpzc{x}^2 + m^2 ) ( \mathpzc{y}^2 + m^2 )  
\\
- \Big[m^2 (\mathpzc{x}^2 + \mathpzc{y}^2 - 2 \lambda \mathpzc{x} \mathpzc{y} ) + m^2 ( \mathpzc{x}^2+\mathpzc{y}^2 + 2m^2) + \lambda \mathpzc{x} \mathpzc{y}
( \mathpzc{x}^2+\mathpzc{y}^2 + 2m^2) \Big]^2 \geq 0
\end{multline*}
in the real variables $\mathpzc{x},\mathpzc{y}$ with the parameters $m, \lambda$. We prove the positivity of $Q$ along the sraight lines
$\mathpzc{y} = a \mathpzc{x}$, by considering the polynomial $P(\mathpzc{x}) = Q(\mathpzc{x}, a\mathpzc{x})$ of the single
real variable $\mathpzc{x}$ with the real parameters $a,m, \lambda$. Writing explicitly $P(\mathpzc{x})$ we see that for
all $\mathpzc{x}$,$a$ and in the assumed domain $-1 \leq \lambda \leq 1$ of $\lambda$:
\begin{multline*}
P(\mathpzc{x}) = (1-\lambda^2)a^2(a^2+1)^2 \mathpzc{x}^8 \,\,\,\,
 + \,\,\,\, m^2(a^6 -4a^5\lambda + 7a^4 - 8a^3\lambda +7a^2-4a\lambda+1)\mathpzc{x}^6
\\
+ m^4(1-4a^3\lambda+6a^2-4a\lambda +1) \mathpzc{x}^4
\\
\geq
(1-\lambda^2)a^2(a^2+1)^2 \mathpzc{x}^8 \,\,\,\,
 + \,\,\,\, m^2(a^6 -4a^5 + 7a^4 - 8a^3 +7a^2-4a+1)\mathpzc{x}^6
\\
+ m^4(1-4a^3+6a^2-4a +1) \mathpzc{x}^4
\\
=
(1-\lambda^2)a^2(a^2+1)^2 \mathpzc{x}^8 \,\, + \,\,  m^2(a-1)^4(a^2+1)\mathpzc{x}^6 \,\,  + \,\, m^4(a-1)^4\mathpzc{x}^4 \geq 0,
\end{multline*}
so that $P(\mathpzc{x})$ is of degree $8$ in which there are only even degrees present, and such that all
its coefficients are positive in the full range  $-\infty < a,m < + \infty$, $-1 \leq \lambda \leq 1$, of the parameters.
Next we use the last elementary inequality in the subrange $0 < \mathpzc{x},\mathpzc{y}$ with the interpretation
\[
\mathpzc{x} = |\boldsymbol{\p}_1|, \,\,\,\, \mathpzc{y}=|\boldsymbol{\p}_2|,
\,\,\,\, \lambda  \mathpzc{x} \mathpzc{y} = \langle \boldsymbol{\p}_1, \boldsymbol{\p}_2 \rangle
\]
with $\lambda$ equal to the cosine of the angle between $\boldsymbol{\p}_1$ and $\boldsymbol{\p}_2$, immediately obtaining the inequality
(\ref{estimation1}).

From the estimation (\ref{estimation1}) it follows that the function 
\[
{\textstyle\frac{|\boldsymbol{\p}_1-\boldsymbol{\p}_2|^2}{-(p_1-p_2)^2}}, \,\,\, p_1, p_2 \in \mathscr{O}_{{}_{m,0,0,0}},
\]
is locally integrable (even with its absolute value majorized everywhere by a single polynomial) of polynomial growth in the variables  $(\boldsymbol{\p}_1, \boldsymbol{\p}_2)$ and also
\[
\sqrt{{\textstyle\frac{|\boldsymbol{\p}_1-\boldsymbol{\p}_2|^2}{-(p_1-p_2)^2}}}
=
{\textstyle\frac{|\boldsymbol{\p}_1-\boldsymbol{\p}_2|}{\sqrt{-(p_1-p_2)^2}}},
 \,\,\, p_1, p_2 \in \mathscr{O}_{{}_{m,0,0,0}},
\] 
is locally integrable (even with its absolute value majorized everywhere by a single polynomial) of polynomial growth in the variables  $(\boldsymbol{\p}_1, \boldsymbol{\p}_2)$.

Because, for $p_1, p_2 \in \mathscr{O}_{{}_{m,0,0,0}}$, 
\[
{\textstyle\frac{(p_1 -p_2)_0 (p_1 -p_2)_0}{(p_1 -p_2)^2}}
= -\Big({\textstyle\frac{\sqrt{|\boldsymbol{\p}_1|^2+m^2}-\sqrt{|\boldsymbol{\p}_2|^2+m^2}}{\sqrt{-(p_1-p_2)^2}}} \Big)^2
= {\textstyle\frac{(p_1-p_2)^2+|\boldsymbol{\p}_1-\boldsymbol{\p}_2|^2}{(p_1-p_2)^2}}
=
1 + {\textstyle\frac{|\boldsymbol{\p}_1-\boldsymbol{\p}_2|^2}{(p_1-p_2)^2}},
\]
then by (\ref{estimation1}) it follows that the $\mu,\nu=0,0$ component of the function 
(\ref{ScaleInv(p1-p2)}), with $p_1, p_2 \in \mathscr{O}_{{}_{m,0,0,0}}$, and regarded as a function of 
the variables  $(\boldsymbol{\p}_1, \boldsymbol{\p}_2)$, is locally integrable 
(even with its absolute value majorized everywhere by a single polynomial) of polynomial growth.
We have, thus, also shown that the function
\[
{\textstyle\frac{\sqrt{|\boldsymbol{\p}_1|^2+m^2}-\sqrt{|\boldsymbol{\p}_2|^2+m^2}}{\sqrt{-(p_1-p_2)^2}}} 
\]
of the variables  $(\boldsymbol{\p}_1, \boldsymbol{\p}_2)$ is also locally integrable 
(even with its absolute value majorized everywhere by a single polynomial) of polynomial growth.
Therefore, also the other components $\mu,\nu$ of the function 
(\ref{ScaleInv(p1-p2)}), with $p_1, p_2 \in \mathscr{O}_{{}_{m,0,0,0}}$, and regarded as a function of 
the variables  $(\boldsymbol{\p}_1, \boldsymbol{\p}_2)$, are locally integrable 
(even with its absolute value majorized everywhere by a single polynomial) of polynomial growth, because
\[
\begin{split}
{\textstyle\frac{(p_1 -p_2)_0 (p_1 -p_2)_k}{(p_1 -p_2)^2}} =
-{\textstyle\frac{\sqrt{|\boldsymbol{\p}_1|^2+m^2}-\sqrt{|\boldsymbol{\p}_2|^2+m^2}}{\sqrt{-(p_1-p_2)^2}}} 
{\textstyle\frac{|\boldsymbol{\p}_1-\boldsymbol{\p}_2|}{\sqrt{-(p_1-p_2)^2}}}
{\textstyle\frac{(\boldsymbol{\p}_1-\boldsymbol{\p}_2)_k}{|\boldsymbol{\p}_1-\boldsymbol{\p}_2|}}
\\
{\textstyle\frac{(p_1 -p_2)_i (p_1 -p_2)_k}{(p_1 -p_2)^2}} =
{\textstyle\frac{|\boldsymbol{\p}_1-\boldsymbol{\p}_2|^2}{(p_1-p_2)^2}}
{\textstyle\frac{(\boldsymbol{\p}_1-\boldsymbol{\p}_2)_i}{|\boldsymbol{\p}_1-\boldsymbol{\p}_2|}}
{\textstyle\frac{(\boldsymbol{\p}_1-\boldsymbol{\p}_2)_k}{|\boldsymbol{\p}_1-\boldsymbol{\p}_2|}},
\end{split}
\]
for $\mu=0$, and $\mu=k=1,2,3$ or  $\mu=i=1,2,3$ and $\mu=k=1,2,3$.  

Therefore, in the integrand 
\[
{\textstyle\frac{\xi_1(s_1, \boldsymbol{\p}_1)\xi_2(s_2, \boldsymbol{\p}_2) u_{s_1}^{+}(\boldsymbol{\p}_1)^+\gamma^\nu u_{s_2}^{+}(\boldsymbol{\p}_2)}{[(p_1- p_2)^2 ]^{k+1}}} \,\, 
(-1)^{k+1}
\Big({\textstyle\frac{(p_1-p_2)_\mu (p_1-p_2)_\nu}{(p_1-p_2)^2}} - g_{\mu \nu} \Big) 
\big[\widetilde{\Pi^{{}^{\textrm{av}}}}(p_1 - p_2)\big]^k
\widetilde{\phi}(p_1 - p_2)
\]
 of (\ref{+-int}), which in case $k=1$ becomes equal 
\[
{\textstyle\frac{\xi_1(s_1, \boldsymbol{\p}_1)\xi_2(s_2, \boldsymbol{\p}_2) u_{s_1}^{+}(\boldsymbol{\p}_1)^+\gamma^\nu u_{s_2}^{+}(\boldsymbol{\p}_2)}
{[(p_1- p_2)^2 ]^{2}}} \,\, 
\Big({\textstyle\frac{(p_1-p_2)_\mu (p_1-p_2)_\nu}{(p_1-p_2)^2}} - g_{\mu \nu} \Big) 
\widetilde{\Pi^{{}^{\textrm{av}}}}(p_1 - p_2)
\widetilde{\phi}(p_1 - p_2)
\]
the factor 
\[
\Big({\textstyle\frac{(p_1-p_2)_\mu (p_1-p_2)_\nu}{(p_1-p_2)^2}} - g_{\mu \nu} \Big) 
\]
is locally integrable 
(even with its absolute value majorized everywhere by a single polynomial) of polynomial growth in $(\boldsymbol{\p}_1, \boldsymbol{\p}_2)$, and in the factor
\[
\widetilde{\Pi^{{}^{\textrm{av}}}}(p_1 - p_2) = [(p_1 - p_2)^2]^2 g_0(p_1 - p_2)
\]
also $g_0(p_1 - p_2)$, as the function of  $(\boldsymbol{\p}_1, \boldsymbol{\p}_2)$, is majorized by a polynomial. We can therefore cancel 
$[(p_1- p_2)^2 ]^{2}$ in the denominator coming from the photon propagator with that $[(p_1 - p_2)^2]^2$ in 
$\widetilde{\Pi^{{}^{\textrm{av}}}}(p_1 - p_2)$, and after this we are left with the following integrand
\[
\xi_1(s_1, \boldsymbol{\p}_1)\xi_2(s_2, \boldsymbol{\p}_2) u_{s_1}^{+}(\boldsymbol{\p}_1)^+\gamma^\nu u_{s_2}^{+}(\boldsymbol{\p}_2)
\widetilde{\phi}(p_1 - p_2)
 \,\, 
\Big({\textstyle\frac{(p_1-p_2)_\mu (p_1-p_2)_\nu}{(p_1-p_2)^2}} - g_{\mu \nu} \Big). 
g_0(p_1 - p_2)
\]

Because
\[
\xi_1(s_1, \boldsymbol{\p}_1)\xi_2(s_2, \boldsymbol{\p}_2) u_{s_1}^{+}(\boldsymbol{\p}_1)^+\gamma^\nu u_{s_2}^{+}(\boldsymbol{\p}_2)
\widetilde{\phi}(p_1 - p_2)
\]
is a Schwartz function of  $(\boldsymbol{\p}_1, \boldsymbol{\p}_2)$, then we are left with an absolutely integrable
(and locally integrable) integrand and the integral  (\ref{+-int}), also in case $k=1$,
is absolutely convergent. 

Because $(-p_1+p_2)^2 = (p_1-p_2)^2$, the proof of convergence of the integral  (\ref{-+int}), also in case $k=1$,
is identical.

Thus, the Lemma is now fully proved. 
\qed

Proof of the absolute convergence of the evaluation integrals of the kernels of the
higher order contributions with arbitrary many ``self-energy'' loop insertions is analogous,
provided the self-energy distribution and its advanced and retarded parts $\Sigma, \Sigma_{\textrm{ret}}, \Sigma_{\textrm{av}}$,
are computed with the ``natural'' normalization in the splitting and $m\neq 0$.

\vspace*{0.5cm}

\begin{center}
{\scriptsize MASSLESS CHARGED FIELD}
\end{center}

Let us consider for example the kernels (\ref{++})-(\ref{-+}) of the analogue odd $n=2k+1$-order contributions with vacuum polarization loop
insertions for spinor QED with massless Dirac field, \emph{i.e.} with $m=0$. In this case the Fourier transform
of the causal combinations of the product of pairings whose retarded part enters into the
vacuum polarization distribution has the jump singularity $\theta(p^2)$ on the cone in momentum space, and the normalization point
in the computation of this retarded part cannot be chosen at zero. In particular we cannot naively put $m=0$ in the formula for the vacuum
polarization in the massive spinor QED in order to compute the vacuum polarization for the massless spinor QED.
Therefore, in QED with massless spinor
field we compute the vacuum polarization using the normalization point $p'$ shifted from zero. In particular
for the formula valid in the positive energy cone and in the open domain including it, we use the normalization point
$p'$ somewhere within the positive energy cone in the momentum space. The result is given by (\ref{FTPim=0}) in which
$p'^2>0$ so $p'$ cannot be put equal to zero. The absolute value of this function (\ref{FTPim=0}) is everywhere
bounded by a polynomial,
but it has the jump singularity $\theta(p^2)$ at the cone $p^2=0$. These properties are preserved
by the splitting, as is easily seen by the dispersion formula (\ref{dispesionFm=0}) for the splitting in case $m=0$
and singularity degree of $d$ equal $\omega=2$ (which is the case for $\Pi^{\mu\nu}$). Thus, also
$\widetilde{\Pi^{\textrm{av}}_{\mu \nu}}$ is everywhere majorized by a polynomial, and is given by the analogous formula
with the analogous scalar factor $\widetilde{\Pi^{\textrm{av}}}$ with the jump-type
singularity $\theta(p^2)$ at the cone $p^2=0$ in the momentum space. This jump cannot be removed by using the
freedom in the splitting, which in the momentum space is determined up to the polynomial
of degree equal to the singularity degree of the splitted distribution (in the considered case equal $\omega =2$).

There are also further replacements in the massless charged field case, namely
the FT of the basic solutions $u_{s_i}^{\pm}$ of the massive Dirac equation are replaced by the scale invariant
FT of the basic solutions of the massless Dirac field, and a less obvious replacement: the functions
$\xi_1$, $\xi_2$ belong now, for the massless spinor field, to the nuclear space $\mathcal{S}^{0}(\mathbb{R}^3)$
of all those Schwartz functions with all derivatives vanishing at zero. Only in this case massless field composes
a well-defined free field as an operator-valued distribution in the white noise sense. The analysis presented above,
and applied in the $m=0$ case, with the mentioned replacements shows that 
 the $\epsilon \rightarrow 0$ limit of the last Lemma does not exist in spinor QED with the massless charged spinor field,
and this is the case for all possible choices in the splitting. 
The same holds for other QED's with massless charged fields.

Alternatively one can give a very short argument proving this result. Because the FT of the vacuum polarization in QED
with massless charged field is not smooth at the cone $p^2=0$, having the jump $\theta(p^2)$ there, and this singularity
cannot be repaired by any choice of the
splitting (addition of any polynomial in momenta of second degree), then we are confronted with the valuation of the
distribution
\[
{\textstyle\frac{1}{[\upsilon + i \epsilon]^{k+1}}}
\overset{\epsilon\rightarrow 0}{\longrightarrow}
{\textstyle\frac{1}{\upsilon^{k+1}}}
-
{\textstyle\frac{i\pi(-1)^{k}}{k!}}\delta^{(k)}(\upsilon)
\]
in single real variable $\upsilon =(p_1\pm p_2)^2$ at the ``test function'' which has the jump-type and $\sim \tfrac{1}{\sqrt{\upsilon}}$-type
singularity at $\upsilon=(p_1\pm p_2)^2=0$, which, as we know from the distribution theory,
is not well-defined, or alternatively: there is no sensible way of definition of the product of the theta function $\theta(\upsilon)$-distribution
(or the $\tfrac{1}{\sqrt{\upsilon}}$-function-type-distribution)
and the derivatives of the Dirac delta distribution $\delta^{(k)}(\upsilon)$.

Analogous results hold for other QED's with massive and massless charged fields. This is because the commutation functions of the
various type of charged free fields are equal to the action of invariant linear differential operators on one and the same 
commutation function of the scalar free field. 

Thus, by the last existence/nonexistence Lemma and
Lemma \ref{S*Xi}, and Theorems 3.6 and 3.9  of  \cite{obataJFA} (or their generalization to the Fermi case
or general Fock space) 
we obtain our main theorem for the existence of the adiabatic limit of interacting fields in QED with massive charged field
and nonexistence of this limit if the charged field is massless. 

\vspace*{0.2cm}

Summing up:

\vspace{0.2cm}

\begin{twr}
For causal perturbative QED on the Minkowski space-time with the Hida operators as the creation-annihilation 
operators and with massive charged field, the higher order contributions to
interacting fields in the adiabatic limit $g\rightarrow 1$ are well-defined as sums of generalized integral kernel operators
with vector valued kernels in the sense of Obata, and this is the case only for the ``natural''
choice in the Epstein-Glaser splitting in the construction of the scattering operator.
\label{ExistenceIntFields.g=1.m>0}
\end{twr}

\vspace{0.1cm}

But:

\vspace{0.1cm}

\begin{twr}
For causal perturbative QED on the Minkowski space-time with the Hida operators as 
the creation-annihilation operators and with massless charged field, the higher order contributions to
interacting fields  in the adiabatic limit $g\rightarrow 1$ are not well-defined, 
even as sums of generalized integral kernel operators in the sense of Obata, and for
no choice in the Epstein-Glaser splitting in the construction of the scattering operator.
\label{NonExistenceIntFields.g=1.m=0}
\end{twr}

\subsection{Comparison with the approach based on Wightman's operator distributions}\label{Comparison}

As we have seen, in the adiabatic limit $g \rightarrow 1$ the higher order contributions 
$\mathbb{A}_{{}_{\textrm{int}}}^{(n)}(g=1)$ to interacting fields  $\mathbb{A}_{{}_{\textrm{int}}}$ in QED preserve 
in general the meaning of the generalized operators -- finite sums of integral kernel operators, \emph{i.e.} continuous maps
\[
\oplus_{1}^{4} \mathscr{E} \ni \phi \xrightarrow{\textrm{continously}} \mathbb{A}_{{}_{\textrm{int}}}^{(n)}(g=1, \phi)
\in \mathscr{L}\big((E), (E)^*\big) \,\,\,\,\,\,\,\,\,\,\,\, \textrm{case A)}
\] 
and only for some exceptional contributions $\mathbb{A}_{{}_{\textrm{int}}}^{(n)}(g=1)$, or some of their subcontributions, we have
\[
\oplus_{1}^{4} \mathscr{E} \ni \phi \xrightarrow{\textrm{continously}} \mathbb{A}_{{}_{\textrm{int}}}^{(n)}(g=1, \phi)
\in \mathscr{L}\big((E), (E)\big). \,\,\,\,\,\,\,\,\,\,\,\, \textrm{case B)}
\] 
Only in case B) the contribution $\mathbb{A}_{{}_{\textrm{int}}}^{(n)}(g=1)$ can be understood as operator valued distributions
also in the Wightman sense \cite{wig}. Contributions of class A) which are not of class B) cannot be understood as operator valued distributions, 
and in particular cannot be accounted for within the approach based on operator valued distributions in the Wightman sense.
Let us recall that the Hida space $(E)$ contains the so called fundamental domain $\mathcal{D}_0$
used in \cite{wig}, which consists of all images of the vacuum state under the polynomial expressions in 
$\boldsymbol{\psi}(f_1), \boldsymbol{\psi}^\sharp(f_2), A(f_3)$, $\ldots$, $\boldsymbol{\psi}(f_{n-1}), \boldsymbol{\psi}^\sharp(f_{n-1}), A(f_n)$,
for $n\in  \mathbb{N}$ and $f_k$ ranging over the Schwartz test functions (if we restrict the arguments $f_k$ of $A$
to the subspace $\mathcal{S}^{00}$ of the Schwartz space). Contributions of class A) which are not of class B) transform some of the Fock states
in $(E)$ into nonnormalizable states which do not belong to the Fock space, but only to the space $(E)^*$ dual to the Hida space $(E)$.
In general, the states of $\mathcal{D}_0 \subset (E)$ are transformed by the contributions of class A) into nonnormalizable states of $(E)^*$
and cannot represent any operator valued distributions in the Wightman sense. An example of the contribution
of type B) is the first order contribution $A_{{}_{\textrm{int}}}^{(1)}(g=1)$ to the interacting e.m. potential field.

Let us consider an example of type A) which does not belong to the class B). We encounter such contributions already at the first
order approximation in the adiabatic limit $g \rightarrow 1$. Namely, the first order contribution $\psi_{{}_{\textrm{int}}}^{(1)}(g=1)$
in the adiabatic limit belongs to class A) but not to class B), and cannot be subsumed within the approach based on  Wightman operator
distributions. As we have seen  $\psi_{{}_{\textrm{int}}}^{(1)}(g=1)$ is a finite sum of well-defined integral kernel operators 
of class A), which evaluated at a test function $\phi \in \oplus_{1}^{4} \mathscr{E}$ is equal to 
\begin{align*}
\psi_{{}_{\textrm{int}}}^{(1)}(g=1;\phi) &
=
\sum\limits_{\nu',s}
\int \ud^3 \boldsymbol{\p}' \ud^3\boldsymbol{\p} &
{\textstyle\frac{(m+\slashed{p} +\slashed{p'})\gamma^{\nu'}u_s(\boldsymbol{\p})\widetilde{\phi}(-|\boldsymbol{\p}'|-p_0(\boldsymbol{\p}), -\boldsymbol{\p}'- \boldsymbol{\p})}{(\boldsymbol{\p}'|(\langle \boldsymbol{\p}' | \boldsymbol{\p} \rangle -|\boldsymbol{\p}'|p_0(\boldsymbol{\p}) )}}
& \,\,\, a_{\nu'}(\boldsymbol{\p}') \, d_s(\boldsymbol{\p})
\\
 & + \sum\limits_{\nu',s}
\int \ud^3 \boldsymbol{\p}' \ud^3\boldsymbol{\p} & \ldots \,\,\,\,\,\,\,\,\,\,\,\,\,\,\,\,\,\,\,\,\,\,\,\,\,\,\,\,\,\,\,\,\,\,\,\,   
& \,\,\,a_{\nu'}(\boldsymbol{\p}')^+ \, d_s(\boldsymbol{\p})
\\
 & + \sum\limits_{\nu',s}
\int \ud^3 \boldsymbol{\p}' \ud^3\boldsymbol{\p} & \ldots \,\,\,\,\,\,\,\,\,\,\,\,\,\,\,\,\,\,\,\,\,\,\,\,\,\,\,\,\,\,\,\,\,\,\,\, 
& \,\,\, a_{\nu'}(\boldsymbol{\p}') \, d_s(\boldsymbol{\p})^+
\\
 & + \sum\limits_{\nu',s}
\int \ud^3 \boldsymbol{\p}' \ud^3\boldsymbol{\p} & \ldots \,\,\,\,\,\,\,\,\,\,\,\,\,\,\,\,\,\,\,\,\,\,\,\,\,\,\,\,\,\,\,\,\,\,\,\, 
& \,\,\, a_{\nu'}(\boldsymbol{\p}')^+ \, d_s(\boldsymbol{\p})^+
\end{align*}
where dots denote the kernels
\[
\kappa_{\mathpzc{l},\mathpzc{m}}(\phi)(\nu',\boldsymbol{\p}', s,\boldsymbol{\p})
= \pm {\textstyle\frac{(m+ \pm \slashed{p} \pm \slashed{p'})\gamma^{\nu'}u_s(\boldsymbol{\p})\widetilde{\phi}(\pm |\boldsymbol{\p}'|\pm p_0(\boldsymbol{\p}), \pm \boldsymbol{\p}'\pm \boldsymbol{\p})}{(\boldsymbol{\p}'|(\langle \boldsymbol{\p}' | \boldsymbol{\p} \rangle -|\boldsymbol{\p}'|p_0(\boldsymbol{\p}) )}},
\,\,\, \mathpzc{l}+\mathpzc{m} = 2
\]
with the respective $\pm$ sings in front of the whole expression and in front of the components $p'_{0}(\boldsymbol{\p}') = |\boldsymbol{\p}'|$,
$\boldsymbol{\p}'$, $p_{0}(\boldsymbol{\p}) = \sqrt{|\boldsymbol{\p}|^2 + m^2}$, $\boldsymbol{\p}$ of the momenta in the denominator, 
correspondingly to the annihilation or the creation operators. Using the elementary estimation
\[
\left | \textstyle{\frac{1}{(\boldsymbol{\p}'|(\langle \boldsymbol{\p}' | \boldsymbol{\p} \rangle -|\boldsymbol{\p}'|p_0(\boldsymbol{\p}) )}} \right |
> \textstyle{\frac{1}{|\boldsymbol{\p}'|^2(|p_0(\boldsymbol{\p}) + |\boldsymbol{\p}|)}}
\]
we see that the kernels $\kappa_{\mathpzc{l},\mathpzc{m}}(\phi)(\nu',\boldsymbol{\p}', s,\boldsymbol{\p})$, regarded as two-particle functions of spin-momenta variables $(\nu',\boldsymbol{\p}', s,\boldsymbol{\p})$ do not belong to the tensor product of single particle Schwartz spaces or even to  the two-particle Hilbert
spaces having their $L^2(\mathbb{R}^{3\times 4} \times \mathbb{R}^{3\times 2})$-norms IR divergent. In particular $\psi_{{}_{\textrm{int}}}^{(1)}(g=1;\phi)$
acting on a finite number particle state with smooth Schwartz functions in each spin-momentum variable, lying in the domain $\mathcal{D}_0$, 
gives a nonnormalizable state which, moreover, is not smooth in  $(\boldsymbol{\p}', \boldsymbol{\p})$. Thus,  $\psi_{{}_{\textrm{int}}}^{(1)}(g=1)$
 cannot represent any operator valued distribution in the Wightman sense. But, as we have seen in Subsection \ref{Proof}, 
$\psi_{{}_{\textrm{int}}}^{(1)}(g=1)$ is a finite sum of well-defined integral kernel operators in the sense of \cite{obataJFA}.

Analogous analysis of the kernels of higher order contributions to interacting fields in the adiabatic limit $g \rightarrow 1$ shows that in general 
higher order contributions to interacting fields in the adiabatic limit $g \rightarrow 1$ are not well-defined operator 
valued distributions in the Wightman sense and do not belong to class B) but only to class A), and this is the case only if the normalization in the splitting 
of causal distributions in the computation of the scattering operator is ``natural''. The adiabatic limit $g \rightarrow 1$
does not exist in the theory (I)-(V) based on Wightman distributions, so that in particular Theorems of Subsection \ref{Proof}
cannot be proved within the approach based on Wightman distributions, contrary to what we have in the approach (I)-(V) based on the 
integral kernel operators in the sense of \cite{obataJFA}.

\section*{Acknowledgements}

The author would like to express his deep gratitude to Professor D. Kazakov  and Professor I. Volovich 
for the very helpful discussions. I am especially grateful to Prof. D. Kazakov for bringing the problem, 
summarized in Theorem \ref{NonExistenceIntFields.g=1.m=0}, to my attention.
He also would like to thank for the excellent conditions for work at JINR, Dubna.
He would like to thank Professor M. Je\.zabek 
for the excellent conditions for work at INP PAS in Krak\'ow, Poland
and would like to thank Professor A. Staruszkiewicz and 
Professor M. Je\.zabek for the warm encouragement. 
The author expresses his gratitude to the Reviewer for his suggestions.

\end{document}